\documentstyle[eqsecnum,aps,epsfig]{revtex}
\begin{document}
\draft
\def\edth{{\check\partial}}
\def\antiedth{{\hat\partial}}
\title{
The evolution of circular, non-equatorial orbits of Kerr black holes
due to gravitational-wave emission
}
\author{Scott A.\ Hughes}
\address{
Theoretical Astrophysics, California Institute of Technology,
Pasadena, CA 91125\\
Department of Physics, University of Illinois at Urbana-Champaign,
Urbana, IL 61801}
\maketitle
\begin{abstract}
A major focus of much current research in gravitation theory is on
understanding how radiation reaction drives the evolution of a binary
system, particularly in the extreme mass ratio limit.  Such research
is of direct relevance to gravitational-wave sources for space-based
detectors (such as LISA).  We present here a study of the radiative
evolution of circular ({\it i.e.}, constant Boyer-Lindquist coordinate
radius), non-equatorial Kerr black hole orbits.  Recent theorems have
shown that, at least in an adiabatic evolution, such orbits evolve
from one circular configuration into another, changing only their
radius and inclination angle.  This constrains the system's evolution
in such a way that the change in its Carter constant can be deduced
from knowledge of gravitational wave fluxes propagating to infinity
and down the black hole's horizon.  Thus, in this particular case, a
local radiation reaction force is not needed.  In accordance with
post-Newtonian weak-field predictions, we find that inclined orbits
radiatively evolve to larger inclination angles (although the
post-Newtonian prediction overestimates the rate of this evolution in
the strong field by a factor $\lesssim 3$).  We also find that the
gravitational waveforms emitted by these orbits are rather
complicated, particularly when the hole is rapidly spinning, as the
radiation is influenced by many harmonics of the orbital frequencies.
\end{abstract}

\section{Introduction}
\label{sec:intro}

\subsection{Motivation: the relativistic two-body problem in
the extreme mass ratio limit}
\label{subsec:motivation}

An outstanding problem in general relativity is the evolution of
binary systems with strong gravity and compact bodies.  Although
post-Newtonian approximations very successfully describe the evolution
of such binaries when the bodies are widely separated
{\cite{postnewtonian}}, no approximation schemes work well when the
bodies are close: in general, there is no small parameter that can be
used to develop any approximations.  Numerical relativity will be
needed to accurately evolve and understand the dynamics of compact
binaries in the strong-field regime.  Despite much effort and progress
{\cite{numericalrelativity}}, numerical relativity is still several
years away from being able to solve the most interesting strong-field
problems such as the final inspiral and merger of binary black hole
and binary neutron star systems.  The full relativistic two-body
problem is thus likely to remain unsolved into the near future.

There is one limit in which the two-body problem can be solved to very
high accuracy with tools available now: the limit in which the mass of
one body, $m_1 \equiv \mu$, is much less than the mass of the other, a
black hole with $m_2 \equiv M$.  The body of mass $\mu$ can be treated
as perturbing the ``background'' black hole spacetime generated by the
mass $M$.  The evolution of the system can then be studied with
perturbation techniques such as the Teukolsky equation
{\cite{teuk72}}.

This limit is in fact of great interest since extreme mass ratio
systems ({\it e.g.}, a compact body of mass $1-10\,M_\odot$ orbiting a
black hole of mass $10^{5-7}\,M_\odot$) are among the most important
candidate sources for space-based gravitational-wave detectors such as
the Laser Interferometer Space Antenna (LISA) {\cite{lisa}}.  Recent
estimates place the number of such extreme mass ratio inspirals,
occurring at distances $D \lesssim 1$ Gpc, in the range of 1/year to
1/month {\cite{sigurdssonrees,sigurdsson}}.  LISA should be able to
measure the gravitational waves emitted by these inspirals with
amplitude signal-to-noise ratios around $10 - 100$
{\cite{finnthorne}}.

During its last year of inspiral, the compact body orbits in the very
strong-field region near the massive black hole's event horizon,
typically spiraling due to gravitational-wave emission from $r
\lesssim 10 M$ to the innermost stable circular orbit.  (We use
units with $G = c = 1$.)  The orbit is dynamically unstable there, so
the body quickly plunges into the hole, cutting off the signal.
During this year, the system emits roughly $10^5$ gravitational-wave
cycles, mostly in LISA's most sensitive frequency band (depending upon
the mass of the black hole).  If it is possible to accurately track
the waves' phase during this year (using, for instance, the technique
of matched filtering with templates), it will be possible to make very
accurate measurements of the black hole's characteristics, and perhaps
to make detailed tests of general relativity (for example, by
``mapping'' the massive object's spacetime to test whether it is in
fact a black hole or a more exotic compact object
{\cite{fintanmeasure}}).

Making such accurate measurements will require a high-precision means
of modeling gravitational waves from extreme mass ratio binaries.  One
can divide in two the influences which drive such systems' evolutions:
the radiation reaction force (backreaction due to gravitational-wave
emission causing the orbiting body to lose orbital energy and spiral
in), and environmental influences (effects due to the systems'
astrophysical environment).  It is possible that environmental
influences for the most important sources will be negligible.  Perhaps
the most important influence is the interaction of the orbiting body
with material accreting onto the massive black hole: black holes in
the target mass range $10^{5-7}M_\odot$ reside at the core of galaxies
where they accrete gas from their environment.  In the majority of
cases, accretion occurs at rather low rates (several orders of
magnitude less than the Eddington rate {\cite{narayannodrag}}).  For
these ``normal'' galaxies, much evidence
{\cite{accretionstats,narayannodrag}} suggests that the gas accretes
via an advection dominated accretion flow (ADAF).  Narayan has shown
that the influence of an ADAF upon an inspiraling compact body will be
far less important than radiation reaction {\cite{narayannodrag}}: the
timescale for ADAF drag to change an orbit's characteristics is many
orders of magnitude longer than the radiation reaction timescale.  If
accretion drag remains the most important enviromental influence on
extreme mass ratio binaries, then radiation reaction is likely to be
the only important element needed to construct precise models of their
evolution.  Because in general we expect orbits to be inclined and
rather eccentric {\cite{sigurdsson}}, techniques must be constructed
for analyzing the radiative evolution of generic Kerr orbits.

If the mass ratio is extreme, one can use a linear approximation to
study the system's evolution.  Split the spacetime metric into a
``background'' black hole piece plus a perturbation: $g_{\alpha\beta}
= g^{\rm Kerr}_{\alpha\beta}(M, a) + h_{\alpha\beta}(\mu)$.  The
system's evolution is governed by the properties of the full spacetime
$g_{\alpha\beta}$, but it is convenient to regard the motion of the
small body $dx^\mu/d\tau$ as a geodesic of $g^{\rm
Kerr}_{\alpha\beta}$ plus corrections from an instantaneous radiation
reaction force $f^\mu_{\rm RR}(\tau)$:
\begin{equation}
{dx\over d\tau}^\mu = \left.{dx\over d\tau}^\mu\right|_{\rm geodesic} +
\int f^\mu_{\rm RR}(\tau)\,d\tau\;.
\label{eq:radreactionforcedef}
\end{equation}
The force $f^\mu_{\rm RR}(\tau)$ (where $\tau$ is proper time measured
by the orbiting body) encapsulates the manner in which the motion of
the body deviates from motion in the background spacetime.  In
particular, it embodies the effects of radiation reaction, causing the
inspiral of the small body toward the black hole as the system's
orbital energy and angular momentum are bled off by gravitational-wave
emission.  Detailed quantitative understanding of this force is needed
to precisely model the evolution of extreme mass ratio binaries.

There currently exists a well-defined prescription for calculating the
radiation reaction force.  Quinn and Wald {\cite{quinnwald}} and Mino
{\it et al.}\ {\cite{minoforce}} have independently and by several
different techniques derived a rather general expression for the
force, depending upon a ``tail'' which is integrated over the past
worldline of the orbiting body's motion.  This tail reflects the fact
that, because of scatter from spacetime curvature, the domain of
dependence of radiation for an event lies inside the light cone.
Various groups are currently working on implementations of the force
(see, {\it e.g.}, Refs.\ {\cite{warreneanna,wisemanscalar}} for a
approaches based on direct evalulation of the Quinn-Wald-Mino
quantities on the worldline, and
{\cite{amosforce_old,amosforce,liorstatic,liorscalar}} for an approach
based on multipole decomposition).  It is likely to be some time
before practical implementations for astrophysically realistic cases
will be available.

\subsection{Radiation reaction without radiation reaction forces}
\label{subsec:rrworrforces}

One can parameterize Kerr orbits by their three constants of motion:
energy $E$, $z$-component of angular momentum $L_z$, and Carter
constant $Q = p_\theta^2 + \cos^2\theta\left[a^2(1 - E^2) +
\csc^2\theta L_z^2\right]$ {\cite{mtw}}.  A given set $(E,L_z,Q)$ can
be thought of as a point in the space of all possible Kerr orbits.
When orbits evolve due to radiation reaction, they generate a
trajectory $[E(t),L_z(t),Q(t)]$ through this space.  One can then
regard the major goal of radiation reaction research as understanding
all physically allowed trajectories through this orbital phase space.

Consider the limit in which the system's evolution is adiabatic: the
timescale $\tau_{\rm RR}$ for the orbit's parameters to change is much
longer than an orbital period $T$.  In this limit, the system spends
many periods near any point on its phase space trajectory: each point
on the trajectory is itself very nearly a geodesic orbit.  [If the
evolution is not adiabatic, the system moves along this trajectory too
rapidly for this to be the case: it does not remain near any set of
constants $(E,L_z,Q)$ long enough to approximate an orbit.]  One can
then regard the system as slowly passing from one geodesic
configuration to another.

The mechanism pushing the system from one geodesic to the next is a
radiation reaction force.  In this limit, the effect of this force can
be understood as a slow change to the ``constants'' of the orbital
motion.  Might it not be possible to {\it indirectly} infer the change
in these constants and thereby deduce the radiative evolution without
actually computing the radiation reaction force?  For example, the
energy carried by gravitational waves to infinity and down the event
horizon is well understood (cf.\ Refs.\ {\cite{isaacson}} and
{\cite{hawkinghartle}}).  Might it be possible to deduce from the
gravitational-wave flux the change of all three orbital constants and
thereby deduce, in the adiabatic limit, the system's radiative
evolution?

In general, it is {\it not} possible to deduce the rate at which all
three constants so change: the energy and the angular momentum can be
read from the gravitational-wave flux, but usually the Carter constant
cannot be.  This is because the energy and angular momentum are
scalars which are linearly constructed from an orbit's momentum
$p^\mu$, whereas the Carter constant is a scalar which is
quadratically constructed from the momentum.

Consider a particle orbiting a Kerr black hole.  At some instant, the
particle has momentum $p^\mu_{\rm B}$.  This particle radiates for
some time and falls into a new orbit with instantaneous momentum
$p^\mu_{\rm A}$.  The change in the particle's momentum is $\delta
p^\mu = p^\mu_{\rm A} - p^\mu_{\rm B}$.

The Kerr metric admits a timelike Killing vector $T_\mu$, an azimuthal
Killing vector $\Phi_\mu$, and a Killing tensor $Q_{\mu\nu}$ (see
{\cite{chandra}} for an explicit representation of this tensor in
Boyer-Lindquist coordinates).  The energy of the orbiting particle is
given by $E = -T_\mu p^\mu$.  Thus, the particle has energy $E_{\rm B}
= -T_\mu p^\mu_{\rm B}$ before it radiates, and energy $E_{\rm A} =
-T_\mu p^\mu_{\rm A}$ after it radiates.  The change in the energy is
carried away by the radiation:
\begin{equation}
\delta E = E_{\rm B} - E_{\rm A} = -T_\mu (p^\mu_{\rm B} - p^\mu_{\rm A})
= T_\mu \delta p^\mu\;.
\label{eq:radiationenergy}
\end{equation}
If we consider a ``graviton limit''\footnote{For the purposes of this
illustrative calculation, it is far simpler to regard the radiation as
a stream of particles for which we can identify an associated
4-momentum.  In the limit in which the radiation is clearly wavelike,
we must be far more careful to make sure that we confine our analysis
to the wave zone, and to average over several wavelengths.  A more
careful analysis in this limit would likely modify Eqs.\
(\ref{eq:deltacarter}) and (\ref{eq:carterchange}).} of the emitted
radiation, then $\delta p^\mu_{\rm graviton} = -\delta p^\mu$ is the
4-momentum carried by the radiation itself.  The quantity $\delta E$
is the energy that observers at infinity measure the radiation to
carry (or, that it adds to the black hole's mass after it falls down
the event horizon).  Since $\delta E = -T_\mu\delta p^\mu_{\rm
graviton}$ depends only on properties of the gravitons radiated to
infinity or down the horizon, one can deduce that the particle's
energy is changed from $E_{\rm B}$ to $E_{\rm B} - \delta E = E_{\rm
A}$ directly out of the gravitational-wave flux.  Using $\Phi_\mu$
rather than $T_\mu$, we see that the change in the particle's
($z$-component of) angular momentum can also be read from the flux.

Consider now the particle's Carter constant.  It is related to the
particle's instantaneous momentum via the Killing tensor: $Q =
Q_{\mu\nu} p^\mu p^\nu$.  Thus, before radiating, the particle has
$Q_{\rm B} = Q_{\mu\nu} p^\mu_{\rm B} p^\nu_{\rm B}$, and after has
$Q_{\rm A} = Q_{\mu\nu} p^\mu_{\rm A} p^\nu_{\rm A}$.  Writing $Q_{\rm
A}$ using $p^\mu_{\rm A} = p^\mu_{\rm B} + {\delta} p^\mu$, and then
evaluating $Q_{\rm B} - Q_{\rm A}$ yields
\begin{equation}
\delta Q = Q_{\rm B} - Q_{\rm A} = 
2 Q_{\mu\nu} p^{\mu}_{\rm B} \delta p^\nu
+ Q_{\mu\nu} \delta p^\mu \delta p^\nu\;.
\label{eq:deltacarter}
\end{equation}
Because this depends explicitly on the local, instantaneous momentum
of the particle, this quantity {\it cannot} be read from the radiation
flux.  Indeed, dividing by $\delta\tau$, taking the limit $\delta \tau
\to 0$, and recognizing that in the limit $\delta p^\mu/\delta\tau$ is
the radiation reaction force $f^\mu_{\rm RR}$, we find that the rate
of change of the Carter constant is given by
\begin{equation}
{\dot Q} = 2 Q_{\mu\nu} p^\mu f^\nu_{\rm RR} \;.
\label{eq:carterchange}
\end{equation}
{\it In general, the radiation reaction force is needed to compute the
rate at which the Carter constant changes even in the adiabatic
limit.}

In general, ``radiation reaction without radiation reaction forces''
does not work.  There are, however, special cases where it does work.
Consider first equatorial orbits.  The Carter constant for equatorial
orbits is zero, and, since there is no means by which these orbits can
be raised out of the equatorial plane (the Kerr metric is reflection
symmetric about its equator), $Q$ remains zero at all times.  Thus,
the system's evolution is entirely given by the two quantities $\dot
E$ and $\dot L_z$.  This is not surprising.  Equatorial orbits can be
described with two parameters: a radial measure $p$ (such as the
semi-latus rectum; see {\cite{dan1}}), and an eccentricity $e$.
Orbital evolution is given by the rates at which these parameters
change.  Connecting the evolution of these two orbital quantities to
the evolution of the two ``conserved'' quantities $E$ and $L_z$ fully
specifies the system's evolution.  Thus, equatorial orbit evolution
can be fixed by examining radiation flux.  In particular, since
Schwarzschild black holes have no preferred orientation, any
Schwarzschild orbit is equatorial and can be evolved by studying
radiation flux.  Detailed studies of such orbits' evolution (and the
waveforms they produce) can be found in
{\cite{eric1,eric2,eric3,dan1,minoetal97}}.

Kerr black holes have a preferred orientation defined by their spin
axis, so equatorial and non-equatorial orbits are quite different.
Various evolutionary studies of such orbits have been done: see
{\cite{det78}} for a first analysis of waveforms and energy fluxes
from circular equatorial orbits; {\cite{shibata94}} for a study of
waveforms and energy fluxes from eccentric equatorial orbits;
{\cite{dan2}} for an analysis of the stability of circular equatorial
orbits; and {\cite{finnthorne}} for an examination of circular
equatorial orbits with an emphasis on measurability by LISA.

Circular, non-equatorial Kerr orbits can also be analyzed without
radiation reaction forces.  (``Circular orbit'' means ``orbit of
constant Boyer-Lindquist coordinate radius''.  The properties of such
orbits in the limit $a = M$ are discussed at length in
{\cite{wilkins}}.)  Such an orbit has non-zero Carter constant, but it
turns out that, in the adiabatic limit, its evolution is entirely
determined by the radiated energy and angular momentum.  This is
because, as has recently been proved
{\cite{danamos,fintan2,minothesis}}, circular orbits remain circular
as they adiabatically evolve due to radiation reaction.  That is, to
very high accuracy the system evolves from one circular configuration
to another.  By requiring that the circularity condition hold at all
times, we can write down a relationship ${\dot Q} = {\dot Q}({\dot E},
{\dot L_z})$ [cf.\ Eq.\ (\ref{eq:Qdotrdot})].  Thus, measurement of
the fluxes ${\dot E}$ and ${\dot L_z}$ is enough to entirely specify
the evolution of the system.  This is not surprising: inclined
circular systems can be described by two parameters, a radius $r$ and
an inclination angle $\iota$.  Despite the fact that such orbits have
three non-trivial conserved quantities, one would guess that
information about the evolution of two of these quantities should
suffice to specify the system's evolution.

The remainder of this paper discusses the evolution of inclined,
circular orbits of Kerr black holes.  A first analysis of such
evolution was given by Shibata {\cite{shibata93}} before the
``circular goes to circular'' theorems were proved.  Noting that in
the $a = 0$ limit the Carter constant contains information about the
$x$ and $y$ components of the angular momentum, Shibata argues that
even in the case $a \ne 0$ the evolution of $Q$ should be driven by
$L_x$ and $L_y$ carried in gravitational radiation
{\cite{lxlycomment}}.  Neglecting certain terms since their
evolutionary timescales will be much longer than other terms, Shibata
gives an expression for Carter constant evolution in terms of
quantities that can be measured in the radiation flux.  Although we
have not checked this explicitly, we suspect that Shibata's
prescription is adequate for describing how $Q$ changes in the weak
field [where the timescale separation he describes should be quite
accurate, and also where radiation flux down the horizon is
unimportant (see {\cite{lxlycomment}})], but would not work well in
the strong field.

\subsection{The Sasaki-Nakamura-Teukolsky formalism}
\label{subsec:SNT}

The formalism used in this paper to compute waveforms and fluxes is
based on the Teukolsky equation [Eq.\ (\ref{eq:teukolsky})].  The
Teukolsky equation describes the (linearized) evolution of
perturbations to the Kerr spacetime.  In particular, it gives the
evolution of the complex Weyl curvature scalar $\psi_4 =
-C_{\alpha\beta\gamma\delta} n^\alpha {\bar m}^\beta n^\gamma {\bar
m}^\delta$ (where $C_{\alpha\beta\gamma\delta}$ is the Weyl curvature
tensor and $n^\alpha$, ${\bar m}^\beta$ are legs of the Newman-Penrose
null tetrad; see, {\it e.g.}, {\cite{chandra}}).  All information
about the radiation flux at infinity and down the horizon can be
extracted from $\psi_4$.

The Teukolsky equation has a source term which depends on an integral
over the world line of the orbiting particle.  In this paper, we work
in the frequency domain, thus requiring a Fourier transform of this
source.  For this Fourier transform to be formally valid, we must know
the full worldline from $t = -\infty$ to $\infty$.  Since, in fact, we
do not yet know the radiative corrections, we approximate the
worldline as a Kerr geodesic orbit.  This approximation is valid in
the adiabatic limit, and is in fact how the assumption of adiabatic
motion mathematically manifests itself: we use the zeroth order,
geodesic motion to find the first order radiative correction.

For each set of constants $(E,L_z,Q)$ there is in fact an entire
family of orbits.  Each member of the family is distinguished by the
particle's initial position at the beginning of a period.  In
constructing the source function, we assume that all of these orbits
are equivalent in the sense that they generate the same radiative
trajectory through $(E,L_z,Q)$ parameter space.  This should be a
valid assumption in the adiabatic limit.  To see this, pick a
parameter space trajectory with some initial conditions, and divide it
into segments of length $T$ (a single period).  In the adiabatic
limit, each segment has identical constants, but has different initial
conditions and so corresponds to a different member of the orbital
family.  Over many periods, the orbiting particle averages all members
of the family.  Thus, in the adiabatic limit, the initial conditions
of the trajectory make no difference: if we had chosen different
initial conditions, the orbiting particle would have sampled all
members of the family anyway.  This averaging lets us assume that we
may pick one representative member of the orbital family and obtain
reliable results.

To actually solve the Teukolsky equation, we integrate its source over
a Green's function constructed from the equation's homogeneous
solutions.  Here, we run into a well-known technical difficulty:
because it has a long-ranged potential, the asymptotic forms of these
homogeneous solutions are somewhat ill-behaved.  In a numerical
computation, we need to be able to calculate the Teukolsky solution at
some field point $r$ by integrating from an asymptotic regime where
its behavior is simple.  Because of the long-ranged potential, it is
very difficult to properly set the phase of the asymptotic solution:
as $r\to\infty$, it has an outgoing piece $\propto e^{i\omega r^*}$
whose amplitude grows at a rate $r^4$ times an ingoing piece $\propto
e^{-i\omega r^*}$ [$r^*$ is the Kerr ``tortoise coordinate'', cf.\
Eq.\ (\ref{eq:rstarofr})].  The ingoing piece is completely lost in a
numerical calculation.

To get around this problem, we use the Sasaki-Nakamura equation, Eq.\
(\ref{eq:sasakinakamura}).  Solutions of the Sasaki-Nakamura equation
are related to solutions of the Teukolsky equation by a simple
transformation [Eq.\ (\ref{eq:XtoR})]; and, since the Sasaki-Nakamura
equation has a short-ranged potential, its asymptotic solutions are
very well-behaved.  We thus integrate the Sasaki-Nakamura equation,
perform a transformation to the Teukolsky solution, use that solution
to construct a Green's function, and then integrate the Green's
function over the Teukolsky source.  From this final integration, we
construct the Weyl scalar $\psi_4$ and obtain all needed details about
the radiation flux.

\subsection{Overview and organization of this paper}
\label{subsec:overview}

Throughout this paper, an overdot denotes $d/dt$ and a prime denotes
$d/dr$.  An overbar denotes complex conjugation.  Unless otherwise
specified, $t$, $r$, $\theta$, and $\phi$ refer to the Boyer-Lindquist
coordinates.  The superscript ``rad'' refers to a quantity carried by
radiation.  Hence, ${\dot E}^{\rm rad}$ is the flux of energy carried
by radiation.  By contrast, ${\dot E}$ is the change rate of an
orbit's energy.  We assume that ${\dot E} + {\dot E}^{\rm rad} = 0 =
{\dot L_z} + {\dot L_z}^{\rm rad}$.

Section {\ref{sec:geodesic}} gives an overview of the properties of
circular Kerr geodesic orbits.  We review the equations governing
these geodesics, and then review the properties of equatorial orbits
in Sec.\ {\ref{subsec:eqcircorbits}}.  In Sec.\
{\ref{subsec:noneqcircorbits}} we discuss non-equatorial orbits,
giving relations that $E$, $L_z$, $Q$, and the radius $r$ must satisfy
for the orbit to be circular.  Frequencies of motion for
non-equatorial orbits are calculated in Sec.\
{\ref{subsec:noneqfrequencies}}.  There are two important frequencies
for circular, non-equatorial orbits: an azimuthal frequency
$\Omega_\phi$ (connected to the time for an orbiting particle to pass
through its range of $\phi$), and a polar frequency $\Omega_\theta$
(related to the time for an orbiting particle to pass through its
range of $\theta$).  These frequencies are equal when $a = 0$, and
approach one another in the weak-field limit.  In the strong field,
and particularly for rapidly spinning black holes, they can be quite
different.

In Sec.\ {\ref{sec:radreaction1}}, we begin analyzing the effects of
radiation reaction.  By requiring that circular orbits remain circular
as the system evolves, we derive in Sec.\
{\ref{subsec:circremainscirc}} an expression relating $\dot Q$ and
$\dot r$ to $\dot E$ and $\dot L_z$.  We show in Sec.\
{\ref{subsec:adiabatic}} that requiring the system to evolve
adiabatically strongly constrains the applicability of this analysis
to astrophysical systems.  For orbits in the strong field, the
analyses of this paper can only be considered valid if the mass ratio
of the system is quite extreme [cf.\ Eq.\ (\ref{eq:muoverM3})].

In Sec.\ {\ref{sec:radreaction2}}, we describe in detail the
Sasaki-Nakamura-Teukolsky formalism used to compute all radiation
reaction quantities.  Although this formalism has been used by several
other authors in the past, we have found that the literature contains
several critical errors (particularly for non-zero black hole spin).
Thus, we discuss this formalism in some detail (hopefully without
introducing any errors of our own).  Section {\ref{subsec:teukolsky}}
reviews the Teukolsky equation {\cite{teuk72}}, discussing the
asymptotic behavior of its homogeneous solutions, using those
solutions to construct a Green's function, and then using the Green's
function to construct $\psi_4$ and extract the waveforms and fluxes.
(Extended discussion of the separated $\theta$ dependence and
efficient algorithms for computing that dependence are given in
Appendix {\ref{app:spheroid}}.)  In Sec.\
{\ref{subsec:sasakinakamura}}, we introduce the Sasaki-Nakamura
transformation and equation, and show how one can obtain the
homogeneous Teukolsky solutions from the well-behaved Sasaki-Nakamura
solution.  In Sec.\ {\ref{subsec:sourceterm}}, we develop the source
term of the Teukolsky equation, and use it in Sec.\
{\ref{subsec:evalcoeffs}} to complete the description of $\psi_4$.

Section {\ref{sec:implement}} describes the numerical implementation
used in this study.  In Sec.\ {\ref{subsec:codeimplementation}} we
sketch the code used to compute all radiation reaction quantities.  We
describe in Sec.\ {\ref{subsec:codevalidation}} several tests
(``sanity checks'') that this code was required to pass: in the weak
field, the code was required to reproduce results that are well known
from previous post-Newtonian analyses; and, in the strong-field
Schwarzschild limit, radiation reaction quantities for inclined orbits
were checked to see that they obeyed a simple relation to quantities
computed for the equatorial plane.  (In addition, many quantities were
spot-checked against results for equatorial orbits that were kindly
provided by D.\ Kennefick and L.\ S.\ Finn; they agreed in all cases.)

Results for strong-field Kerr orbits are given in Sec.\
{\ref{sec:results}}.  We discuss the waveforms and energy fluxes for a
few representative orbits in Sec.\ {\ref{subsec:radiation}}.  We find
that as the black hole's spin increases, the effect of high harmonics
of the orbit's fundamental frequencies becomes more important.
Mathematically, this is due to the fact that the transmissivity of the
Teukolsky potential to radiation increases as the spin parameter $a$
is dialed up.  More physically, it is probably due to the fact that
for large spins and strong-fields, the frequencies $\Omega_\phi$ and
$\Omega_\theta$ become quite different, and so the motion of the
source becomes almost aperiodic.  The orbital motion, which is the
source of the radiation, thus requires many harmonics of these
frequencies to be accurately described.  In Sec.\
{\ref{subsec:radreactseq}} we describe a radiation reaction sequence
for strong-field orbits near a black hole with spin parameter $a = 0.8
M$.  We examine a set of many orbits, parameterized by the radius $r$
and inclination angle $\iota$.  At each parameter space point
$(r,\iota)$, we compute the direction $({\dot r},{\dot\iota})$ in
which radiation reaction drives the system.  These directions would be
the tangent vectors to the trajectories $[r(t),\iota(t)]$ that
evolving circular orbits would follow as they evolved.  An interesting
result is that the change in inclination angle $\dot\iota$ found in
the strong field is rather smaller (by a factor $\lesssim 3$) than
what one would predict based on extrapolating post-Newtonian theory.

The spin choice $a = 0.8 M$ is a reasonable value for a black hole
whose angular momentum has been buffered by magnetohydrodynamic
extraction of spin energy, such as are described by some models of
quasar engines {\cite{sterlcomment,moderskietal}}.  In general,
whenever a black hole is coupled to an accretion disk by magnetic
fields, one expects there to be a buffering torque that can
significantly impact the hole's spin {\cite{rogercomment}}.  However,
if the hole's evolution is driven by thin-disk accretion (particularly
by photon capture in the thin-disk limit as described in
{\cite{kipspinbuffer}}), or the hole is produced by the merger of two
black holes (such as a result of galaxy collisions), the spin can be
much larger: $a = 0.998 M$ is the prediction of photon-buffered
thin-disk accretion.  An analysis of radiation reaction on circular
orbits in this large spin regime will be presented in a separate paper
{\cite{paperII}}.

Some concluding discussion is given in Sec.\ {\ref{sec:conclusion}}.
In particular, we note that the assumption of adiabaticity might not
be reasonable in the real world.  This assumption would not be needed
if it were possible to compute the instantaneous radiation reaction
force $f^\mu_{\rm RR}$.  This illustrates the importance of such a
force, and provides further impetus to workers implementing the
radiation reaction force.

\section{Circular, geodesic Kerr orbits}
\label{sec:geodesic}

In this section, we discuss bound circular orbits of Kerr black holes,
without incorporating radiation reaction.

Kerr geodesics are governed by the following four equations
{\cite{mtw}}:
\begin{mathletters}
\begin{eqnarray}
\Sigma^2\left({dr\over d\tau}\right)^2 && =\left[E(r^2+a^2)
- a L_z\right]^2- \Delta\left[r^2 + (L_z - a E)^2 +
Q\right] \equiv R\;,\label{eq:rdot}\\
\Sigma^2\left({d\theta\over d\tau}\right)^2 && = Q - \cot^2\theta L_z^2
-a^2\cos^2\theta(1 - E^2)\equiv\Theta^2\;,\label{eq:thetadot}\\
\Sigma\left({d\phi\over d\tau}\right) && =
\csc^2\theta L_z + aE\left({r^2+a^2\over\Delta} - 1\right) -
{a^2L_z\over\Delta}\;,\label{eq:phidot}\\
\Sigma\left({dt\over d\tau}\right) && =
E\left[{(r^2+a^2)^2\over\Delta} - a^2\sin^2\theta\right] +
aL_z\left(1 - {r^2+a^2\over\Delta}\right)\;.\label{eq:tdot}
\label{eq:geodesiceqns}
\end{eqnarray}
\end{mathletters}
The quantities $E$, $L_z$, and $Q$ (``energy'', ``$z$-component of
angular momentum'', and ``Carter constant'') specify a family of
geodesic orbits.  They are conserved along any orbit in this family.
Particular members of the family are specified by initial conditions.
In Eqs.\ (\ref{eq:rdot}--\ref{eq:tdot}), $\Sigma\equiv r^2 +
a^2\cos^2\theta$ and $\Delta= r^2 - 2 M r + a^2$.  Eqs.\
(\ref{eq:rdot}) and (\ref{eq:thetadot}) have been divided by $\mu^2$
and Eqs.\ (\ref{eq:phidot}) and (\ref{eq:tdot}) by $\mu$ (where $\mu$
is the mass of the orbiting particle).  Because of this, $E$, $L_z$,
and $Q$ as used throughout this paper are the specific energy, angular
momentum and Carter constant: $E = E^{\rm usual}/\mu$, $L_z = L_z^{\rm
usual}/\mu$, and $Q = Q^{\rm usual}/\mu^2$.

The right-hand side of Eq.\ (\ref{eq:thetadot}) goes to zero as the
orbiting particle's position approaches the turning points of its
$\theta$ motion, $\theta_{\rm max}$ and $\theta_{\rm min}$.  This
causes problems in a numerical implementation [cf.\ the source term of
Sec.\ {\ref{subsec:sourceterm}}, which could be written as an integral
over $(d\theta/dt)^{-1}$].  To deal with this problem, we transform to
a coordinate that is much better behaved at the turning points.
First, define the variable $z = \cos^2\theta$.  Equation
(\ref{eq:thetadot}) becomes
\begin{eqnarray}
{d\theta\over d\tau} &=& \pm {\sqrt{z^2\left[a^2(1 - E^2)\right] -
z\left[Q + L_z^2 + a^2(1 - E^2)\right] + Q}\over
(r^2 + a^2 z)\sqrt{1 - z}}\;,\nonumber\\
&=& \pm {\sqrt{\beta(z_+ - z)(z_- - z)}\over(r^2 + a^2 z)\sqrt{1 - z}}\;.
\label{eq:thetadot2}
\end{eqnarray}
The plus sign corresponds to motion from $\theta_{\rm min}$ to
$\theta_{\rm max}$, and vice versa for the minus sign.  We have
defined $\beta = a^2(1 - E^2)$, and $z_\pm$ are the two roots of the
quadratic in the top line of Eq.\ (\ref{eq:thetadot2}).

Next, define the variable $\chi$ via $z = z_- \cos^2\chi$.  The
parameter $\chi$ ranges from $0$ to $2\pi$.  As $\chi$ varies from $0$
to $2 \pi$, $\theta$ goes from $\theta_{\rm min}$ to $\theta_{\rm
max}$ (at $\chi = \pi$) then back to $\theta_{\rm min}$.  Examining
$dz/d\theta$ and $dz/d\chi$ we see that
\begin{eqnarray}
{d\chi\over d\theta} &=& \sqrt{1 - z \over z_- - z}\;,\qquad
0 \le \chi \le \pi\;;\nonumber\\
&=& -\sqrt{1 - z \over z_- - z}\;,\qquad
\pi \le \chi \le 2\pi\;.
\label{eq:dchidtheta}
\end{eqnarray}
Combining Eqs.\ (\ref{eq:thetadot2}) and (\ref{eq:dchidtheta}), we obtain
the geodesic equation for $\chi$:
\begin{equation}
{d\chi\over d\tau} = {\sqrt{\beta(z_+ - z)}\over r^2 + a^2 z}\;.
\label{eq:chidot}
\end{equation}
This form is perfectly well behaved over the entire orbit.  Although
the interest here is in circular orbits, circularity was not assumed
at any point in this derivation.  Equation (\ref{eq:chidot}) could be
useful in the study of generic Kerr orbits.

We specialize to circular orbits from this point onward.  These orbits
satisfy $R = 0 = R'$.  The first condition guarantees that $dr/d\tau =
0$, so that the radius does not change.  The second states that the
orbit is eternally at a turning point of its radial motion.  In
addition, the condition $R'' < 0$ must be met for the orbit to be
stable.

An orbit can be fixed by specifying $r$ and $L_z$: from the conditions
$R = 0 = R'$, $Q$ and $E$ are determined.  The Carter constant and
angular momentum determine the amount by which the orbit is inclined
from the equatorial plane.  As in {\cite{fintan2,fintan1}}, we will
use the following inclination angle:
\begin{equation}
\cos\iota = {L_z\over\sqrt{L_z^2 + Q}}\;.
\label{eq:iota}
\end{equation}
This inclination angle is a constant of the motion and is very easy to
compute.  One could define other inclination angles [{\it e.g.}, the
arctangent of $(d\theta/d\tau)/(d\phi/d\tau)$ evaluated at $\theta =
\pi/2$, the angle at which infinitely distant observers would
see the particle cross the equatorial plane].  We shall stick with
$\iota$ as defined above since it is adequate for this analysis.

\subsection{Equatorial orbits}
\label{subsec:eqcircorbits}

To begin, consider $Q=0$.  By Eq.\ (\ref{eq:iota}) these orbits have
$\iota = 0^\circ$ or $180^\circ$, lying in the equatorial plane.  By
Eq.\ (\ref{eq:thetadot}), $d\theta/d\tau = 0$, so they remain in the
equatorial plane at all times.  The conditions $R = 0 = R'$ yield the
following formulae for $E(r)$ and $L_z(r)$:
\begin{mathletters}
\begin{eqnarray}
E^{\rm pro} &&= {{1 - 2 v^2 + q v^3}\over
\sqrt{1 - 3 v^2 + 2 q v^3}}\;,
\label{eq:Eeqpro}\\
L_z^{\rm pro} &&= r v{{1 - 2 q v^3 + q^2 v^4}\over
\sqrt{1 - 3 v^2 + 2 q v^3}}\;,
\label{eq:Lzeqpro}\\
E^{\rm ret} &&= {{1 - 2 v^2 - q v^3}\over
\sqrt{1 - 3 v^2 - 2 q v^3}}\;,
\label{eq:Eeqret}\\
L_z^{\rm ret} &&= -r v{{1 + 2 q v^3 + q^2 v^4}\over
\sqrt{1 - 3 v^2 - 2 q v^3}}\;.
\label{eq:Lzeqret}
\end{eqnarray}
\end{mathletters}
Here, $v\equiv\sqrt{M/r}$ and $q\equiv a/M$.  The superscripts ``pro''
and ``ret'' correspond to prograde and retrograde orbits,
respectively\footnote{Throughout this paper, we distinguish between
prograde and retrograde orbits by the sign of $L_z$.  Another common
convention is to switch the sign of the black hole's spin $a$.  This
is not useful here since the characteristics of the orbit should
smoothly vary from prograde to retrograde as the orbit's inclination
varies from $0^\circ$ to $180^\circ$.}.

As seen by observers at infinity, the particle in this equatorial
orbit moves with azimuthal frequency
\begin{equation}
\Omega_\phi = {d\phi\over dt} = {d\phi/d\tau\over dt/d\tau}
= \pm {M^{1/2}\over{r^{3/2} \pm a M^{1/2}}}\;,
\label{eq:equatorial_omegaphi}
\end{equation}
where the upper sign refers to prograde and the lower to retrograde
orbits.

\subsection{Non-equatorial orbits}
\label{subsec:noneqcircorbits}

Non-equatorial orbits have $Q \ne 0$.  We will use the following
algorithm to ensure that we find all stable orbits at some particular
radius $r$:

\begin{enumerate}
\item The most stable orbit is the prograde equatorial orbit, so it
makes a useful starting point.  Pick a value of $r$ and use Eqs.\
(\ref{eq:Eeqpro}) and (\ref{eq:Lzeqpro}) to calculate this orbit's
energy and angular momentum.  Since it is equatorial, $Q=0$.

\item Decrease the angular momentum $L_z$, holding $r$ fixed.
Solve the system of equations $R = 0 = R'$ to find $Q$ and $E$.  These
equations admit simple analytic solutions for $Q(r,L_z)$ and
$E(r,L_z)$ [Eqs.\ (\ref{eq:EofLzr}) and (\ref{eq:QofLzr})].

\item Repeatedly decrement the angular momentum and solve again for
$Q(r,L_z)$ and $E(r,L_z)$ until either the angular momentum reaches
$L_z^{\rm ret}$ (indicating that we have reached the least-bound
retrograde equatorial orbit) or else $R'' = 0$.  Orbits with $R''=0$
are marginally bound; orbits with still lower values of $L_z$ are not
stable, and are thus not of interest.  Radiation reaction causes such
orbits to catastrophically plunge into the black hole.

\end{enumerate}

As mentioned above, when $r$ and $L_z$ are specified, the equations $R
= 0 = R'$ admit a simple solution for $E$ and $Q$:
\begin{eqnarray}
E(r,L_z) &&= {{a^2L_z^2(r-M) + r\Delta^2}
\over{a L_z M\left(r^2 - a^2\right) \pm \Delta\sqrt{r^5(r-3M)
+ a^4r(r+M)+a^2r^2(L_z^2-2Mr+2r^2)]}}}\;,
\label{eq:EofLzr}\\
Q(r,L_z) &&= {\left[(a^2 + r^2) E(r,L_z) - a L_z\right]^2\over \Delta}
- \left[r^2 + a^2 E(r,L_z)^2 - 2 a E(r,L_z) L_z + L_z^2\right]\;.
\label{eq:QofLzr}
\end{eqnarray}
There are two roots for $E$.  Only one of these roots is physical; the
other typically gives an energy less than the energy of the most
strongly bound orbit (the prograde equatorial orbit).  In this paper,
we shall focus exclusively on orbits for which the plus sign in the
denominator of Eq.\ (\ref{eq:EofLzr}) is physical.  The minus sign
turns out to be physical only for strong-field orbits of very rapidly
rotating holes; an analysis of such orbits will be presented in a
separate paper {\cite{paperII}}.  At any rate, since we have fixed the
choice of root in Eq.\ ({\ref{eq:EofLzr}}), circular orbits are
entirely determined by choosing $r$ and $L_z$ (and checking that $R''
\le 0$).

\subsection{Frequencies of non-equatorial orbits}
\label{subsec:noneqfrequencies}

As the particle orbits, its motions in $\theta$ and $\phi$ are
separately periodic.  The periods for these two motions are generally
different.  In this section, we derive expressions for the periods
$T_\theta$ and $T_\phi$, and show that if we perform our analysis in a
certain frame we need only worry about $T_\theta$.  The analysis here
is very similar to that of Wilkins {\cite{wilkins}, but does not
specialize to $a=M$.

\subsubsection{Period of $\theta$ motion}

We first calculate the time (as seen by observers at infinity) for the
particle to move from $\chi = 0$ to $\chi$.  Combining Eqs.\
(\ref{eq:tdot}) and (\ref{eq:chidot}),
\begin{equation}
{dt\over d\chi} = {{\gamma + a^2 E z}\over\sqrt{\beta(z_+ - z)}}\;,
\label{eq:dtdchi}
\end{equation}
where
\begin{equation}
\gamma\equiv E\left[{(r^2+a^2)^2\over\Delta}-a^2\right]
+a L_z\left(1 - {r^2+a^2\over\Delta}\right)\;.
\label{eq:gammadef}
\end{equation}
The time it takes to go from $0$ to $\chi$ is then
\begin{eqnarray}
t_0(\chi) &=& \int_0^\chi d\chi'\,{{\gamma + a^2 E z(\chi')}
\over{\sqrt{\beta[z_+ - z(\chi')]}}}\;,\nonumber\\
&=&  {\gamma\over\sqrt{\beta z_+}}
\left[K\left(\sqrt{z_-/z_+}\right) -
F\left(\pi/2 - \chi,\sqrt{z_-/z_+}\right)\right]\nonumber\\
& + & a^2 E \sqrt{z_+\over\beta}
\left[E\left(\pi/2 - \chi,\sqrt{z_-/z_+}\right) -
F\left(\pi/2 - \chi,\sqrt{z_-/z_+}\right) -
E\left(\sqrt{z_-/z_+}\right) + K\left(\sqrt{z_-/z_+}\right)\right]\;.
\label{eq:t0}
\end{eqnarray}
On the last line, $F(\varphi,k)$ is the incomplete elliptic integral
of the first kind, $K(k)$ is the complete elliptic integral of the
first kind; $E(\varphi,k)$ and $E(k)$ are respectively the incomplete
and complete elliptical integrals of the second kind (using the
notation of {\cite{numrec}}).

Equation (\ref{eq:t0}) only applies to the interval
$0\le\chi\le\pi/2$.  It is straightforward to generalize to the
interval $\pi/2 \le \chi
\le \pi$:
\begin{eqnarray}
t(\chi) &=& t_0(\chi)\;, \qquad 0 \le \chi \le \pi/2\;,
\nonumber\\
&=& t_0(\pi/2) + t_0(\chi - \pi/2)\;, \qquad \pi/2 \le \chi \le \pi\;.
\label{eq:tofchi}
\end{eqnarray}
Similar formulae can be written down for $\pi\le\chi\le 2\pi$; it
turns out that they are not needed.  We could add any constant to
$t(\chi)$; this would specify a different member of the orbital family
corresponding to $(E,L_z,Q)$.

The particle moves through one fourth of its $\theta$ range as $\chi$
varies from $0$ to $\pi/2$.  Hence, $T_\theta$, the period of the
particle's $\theta$ motion, is given by
\begin{equation}
T_\theta = 4 t_0(\pi/2) = {4\gamma\over\sqrt{\beta z_+}}
K\left(\sqrt{z_-/z_+}\right) + 4a^2E\sqrt{z_+\over\beta}
\left[K\left(\sqrt{z_-/z_+}\right)- E\left(\sqrt{z_-/z_+}\right)\right]\;.
\label{eq:Ttheta2}
\end{equation}
The corresponding frequency of $\theta$ motion is $2\pi/T_\theta$.

\subsubsection{Period of $\phi$ motion}

In this subsection we calculate the angle $\phi$ accumulated as the
particle moves from $0$ to $\chi$, and use it to calculate the
azimuthal period $T_\phi$.

Applying the transformations $z=\cos^2\theta = z_-\cos^2\chi$ to Eq.\
(\ref{eq:phidot}) and combining the result with Eq.\ (\ref{eq:chidot})
yields
\begin{equation}
{d\phi\over d\chi}={1\over\sqrt{\beta (z_+ - z)}}
\left({L_z\over1-z} + \delta\right)\;,
\label{eq:dphidchi}
\end{equation}
where
\begin{equation}
\delta = aE\left({r^2+a^2\over\Delta}-1\right)-
{a^2L_z\over\Delta}\;.
\label{eq:deltadef}
\end{equation}
Integrating Eq.\ (\ref{eq:dphidchi}) from $0$ to $\chi$ gives
the amount of $\phi$ accumulated as the particle orbits:
\begin{eqnarray}
\phi(\chi) &=& \phi_0(\chi)\;, \qquad 0 \le \chi \le \pi/2\;,
\nonumber\\
&=& \phi_0(\pi/2) + \phi_0(\chi - \pi/2)\;, \qquad \pi/2 \le \chi \le \pi\;;
\label{eq:phiofchi}
\end{eqnarray}
where
\begin{eqnarray}
\phi_0(\chi) &=&
{1\over\sqrt{\beta z_+}}
\left\{L_z\left[\Pi\left(\pi/2,-z_-,\sqrt{z_-/z_+}\right) -
\Pi\left(\pi/2 - \chi,-z_-,\sqrt{z_-/z_+}\right)\right]
\right.\nonumber\\
&+&\left. \delta\left[K\left(\sqrt{z_-/z_+}\right) -
F\left(\pi/2 - \chi,\sqrt{z_-/z_+}\right)\right]\right\}\;.
\label{eq:phi0}
\end{eqnarray}
Here $\Pi(\varphi,n,k)$ is the incomplete elliptical integral of the
third kind, again using the notation of {\cite{numrec}}.  In a period
$T_\theta$, the particle moves through an azimuthal angle $\Phi$ given
by
\begin{equation}
\Phi = 4 \phi_0(\pi/2) = {4\over\sqrt{\beta z_+}}
\left[L_z\Pi\left(\pi/2,-z_-,\sqrt{z_-/z_+}\right) -
\delta K\left(\sqrt{z_-/z_+}\right)\right]\;.
\label{eq:Phi}
\end{equation}

Unless $\Phi$ equals $2\pi$ (which is only the case for $a = 0$), the
periods of $\theta$ and $\phi$ motion will be incommensurate.  This is
potentially problematic, since it is not clear which period is the
fundamental one to use for describing the orbits and gravitational
radiation.  However, as shown by Cutler, Kennefick and Poisson (Sec.\
IID of {\cite{dan1}}), we can expand the $\phi$ motion in a Fourier
series of $\Omega_\theta$ harmonics:
\begin{equation}
{d\phi\over dt} = \sum_{n=0}^\infty a_n e^{in\Omega_\theta t}\;.
\label{eq:decompose}
\end{equation}
First, integrate this up to obtain
\begin{equation}
\phi(t) = a_0 t + \sum_{n=1}^\infty b_n e^{in\Omega_\theta t}\;,
\end{equation}
where $b_n = ia_n/n\Omega_\theta$.  Next, average this over a time
$T_\theta$.  The sum goes to zero by periodicity, so
\begin{equation}
a_0 \equiv \Omega_\phi = {\Phi\over T_\theta} =
{\Phi\over2\pi}\Omega_\theta\;.
\end{equation}
Analyze the system in a coordinate system that is rotating at angular
velocity $\Omega_\phi$, so that $\phi' = \phi - \Omega_\phi t$.  In
this coordinate system, the only frequency that matters in the
analysis of azimuthal motion is $\Omega_\theta$ (and its harmonics):
\begin{equation}
\phi'(t) = \sum_{n=1}^\infty b_n e^{in\Omega_\theta t}\;.
\end{equation}
Thus $\Omega_\theta$ and its harmonics are the only frequencies that
are needed in order to analyze the motion of the orbiting particle in
this coordinate system.  The only timescale we need be concerned about
is $T_\theta$.

\section{Radiation reaction: initial considerations}
\label{sec:radreaction1}

In this section, we begin to consider radiation reaction and how it
modifies the orbits from their purely geodesic form.  The requirement
that circular orbits remain circular allows us to write down a simple
condition for $\dot Q$ and $\dot r$ given $\dot E$ and $\dot L_z$.
This analysis is presented in Sec.\ {\ref{subsec:circremainscirc}}.

As discussed in the Introduction, we assume that the system evolves in
an adiabatic manner: the change in any orbital parameter $q$ over one
period $T_\theta$ must be much less than $q$.  By insisting that the
evolution be adiabatic everywhere, we derive conditions on the mass
ratio $\mu/M$.  As we show in Sec.\ {\ref{subsec:adiabatic}}, this
leads to stringent conditions when the particle is near the black
hole, but essentially irrelevant conditions when the particle is in
the weak field.  When discussing astrophysical gravitational-wave
sources, our results can be considered relevant only when these
conditions are met.

\subsection{Circular remains circular}
\label{subsec:circremainscirc}

As discussed in Sec.\ {\ref{subsec:rrworrforces}}, it has recently
been shown {\cite{danamos,fintan2,minothesis}} that, under adiabatic
radiation reaction, circular orbits evolve from one circular
configuration to another (except possibly in the very strong-field
regime, as the orbiting particle begins to plunge into the black
hole).  An orbit which is circular remains circular.

For the orbit to remain circular at all times, the equations $\dot R =
0$ and $\dot R' = 0$ must hold, where
\begin{eqnarray}
{\dot R} &=& {\dot r} R' + 2 E {\dot E} r^4 + \left[2 a^2 E {\dot E} -
	2 L_z {\dot L_z} - {\dot Q}\right] r^2 +
	2\left[{\dot Q} + 2({\dot L_z} - a {\dot E})(L_z - a E)\right] M r -
	a^2 {\dot Q}\;,\nonumber\\
{\dot R'} &=& {\dot r} R'' + 8 E {\dot E} r^3 + 2\left[2 a^2 E {\dot E} -
	2 L_z {\dot L_z} - {\dot Q}\right] r +
	2\left[{\dot Q} + 2({\dot L_z} - a {\dot E})(L_z - a E)\right] M\;.
\label{eq:potentialdotdefs}
\end{eqnarray}
These quantities are found by taking the time derivative of $R$ and
$R'$ [where $R$ is defined in Eq.\ (\ref{eq:rdot})].

Notice that ${\dot R} = 0$ and ${\dot R'} = 0$ can be combined into a single
matrix equation,
\begin{equation}
A\cdot v_1 + B\cdot v_2 = 0\;,
\label{eq:matrixdef}
\end{equation}
where
\begin{eqnarray}
v_1 &=& \left({\matrix{{\dot E} \cr {\dot L_z}}}\right)\;,\nonumber\\
v_2 &=& \left({\matrix{{\dot Q} \cr {\dot r}}}\right)\;.
\end{eqnarray}
The matrices $A$ and $B$ can then be explicitly written out by
gathering the proper terms in $\dot E$, $\dot L_z$, $\dot Q$, and
$\dot r$.  The solution for $\dot Q$ and $\dot r$ in terms of $\dot E$
and $\dot L_z$ is now transparent: solving Eq.\ (\ref{eq:matrixdef})
for $v_2$ yields
\begin{equation}
v_2 = -B^{-1}\cdot A \cdot v_1 \equiv -C\cdot v_1\;.
\label{eq:matrixdef2}
\end{equation}
Thus, we may write
\begin{eqnarray}
{\dot Q} &=& -{c_{11}\over d} {\dot E} - {c_{12}\over d} {\dot L_z}\;,
\nonumber\\
{\dot r} &=& -{c_{21}\over d} {\dot E} - {c_{22}\over d} {\dot L_z}\;,
\label{eq:Qdotrdot}
\end{eqnarray}
where
\begin{mathletters}
\begin{eqnarray}
c_{11}(Q,E,L_z,r) \equiv&&
	-4E(1 - E^2)Mr^6 + 12EM^2r^5 - 2E[a^2(1 - E^2)
	+ 3(L_z^2 + Q)]Mr^4\nonumber\\
	&&+ 8[a^2E(2 - E^2) + E(L_z^2 + Q) - 2aL_z]M^2r^3
	\nonumber\\
	&&-2a[aE[6M^2 + L_z^2 + Q + a^2(1 - E^2)] - 6M^2 L_z]Mr^2
	\nonumber\\
	&&+4a^2E[Q + (L_z - aE)^2]M^2r - 4a(L_z - aE)[Q + (L_z - aE)^2]M^3\;,
	\label{eq:c11}\\
c_{12}(Q,E,L_z,r) \equiv&& -4L_z(1 - E^2)Mr^4 +
	16(1 - E^2)(L_z - a E)M^2 r^3\nonumber\\
	&&+2\left[L_z[a^2(1 - E^2) + L_z^2 + Q] - 6M^2(L_z - a E)\right]Mr^2
	\nonumber\\
	&&-4 L_z\left[Q + (L_z - a E)^2\right]M^2r +
	4 (L_z - a E)\left[Q + (L_z - a E)^2\right]M^3\;,
	\label{eq:c12}\\
c_{21}(Q,E,L_z,r) \equiv&& 2 E r^5 - 6 EM r^4 + 4 a^2 E r^3 +
	2 a (L_z - 2 a E)Mr^2 + 2 a^4 E r - 2a^3(L_z - a E)M\;,
	\label{eq:c21}\\
c_{22}(Q,E,L_z,r) \equiv&& 2aEMr^2 - 2a^2 L_z r + 2a^2(L_z - a E)M\;,
	\label{eq:c22}\\
d(Q,E,L_z,r) \equiv&& -2(1 - E^2)Mr^4 + 8(1 - E^2)M^2r^3
	+\left[Q + L_z^2 - 5 a^2(1 - E^2) - 6M^2\right]Mr^2\nonumber\\
	&&+2\left[a^2 (3 - E^2) + 2 a E L_z - (L_z^2 + Q)\right]M^2r
	\nonumber\\
	&&+2(L_z^2 + Q)M^3 - 4 a E L_zM^3 + a^2(2E^2M^2 - L_z^2 - Q)M
	- a^4(1 - E^2)M\;.
	\label{eq:denom}
\end{eqnarray}
\end{mathletters}
By determining $\dot E$ and $\dot L_z$, we determine $\dot Q$ and
$\dot r$, fully fixing the evolution of the particle's orbit.  In
particular, the rate of change of the inclination angle is
\begin{equation}
{\dot\iota} = -{d(\cos\iota)/dt\over{1 - \cos^2\iota}}\;,
\label{eq:iotadot}
\end{equation}
where
\begin{equation}
{d(\cos\iota)\over dt} = {1\over\sqrt{L_z^2 + Q}}\left[{\dot L_z} -
\left({L_z\over2}\right){2 L_z {\dot L_z} + {\dot Q}\over{L_z^2
+ Q}}\right]\;.
\label{eq:cosiotadot}
\end{equation}

\subsection{Adiabaticity}
\label{subsec:adiabatic}

In this section, we impose adiabaticity on the orbiting particle and
show that it constrains the mass ratio of the system.  We first impose
adiabaticity on motion in the weak field of the black hole ($r \gg
M$), and then impose adiabaticity on motion in the strong field ($r
\sim M$).

\subsubsection{Weak-field radiation reaction}
\label{weakfield}

In the weak-field, a post-Newtonian expansion suffices to calculate
the radiated energy and angular momentum.  Ryan {\cite{fintan2}} shows
that such a post-Newtonian expansion leads to the following results
for the change in the orbital radius and inclination angle:
\begin{eqnarray}
{\dot r}_{\rm weak} &=& -{64\over5}{\mu\over M}
	\left({M\over r}\right)^3\;,	\nonumber\\
{\dot \iota}_{\rm weak} &=& {244\over15}{\mu\over M^2}
	{a\over M}\left({M\over r}\right)^{11/2}\sin\iota
	\simeq{244\over15}{\mu\over M^2}{a\over M}
	\left({M\over r}\right)^{11/2}\iota\;.
\label{eq:pn_rdotiotadot}
\end{eqnarray}
In the weak-field, $T_\theta \simeq 2\pi {\sqrt{r^3/M}} + O(a)$
{\cite{fintan1}}.  Putting all of this together, we find to leading
order
\begin{eqnarray}
{{\dot r}_{\rm weak} T_\theta\over r} &\simeq&
	{128\pi\over5}{\mu\over M}
	\left({M\over r}\right)^{5/2}\;,
	\nonumber\\
{{\dot\iota}_{\rm weak} T_\theta\over\iota} &\simeq&
	{488\pi\over15}{\mu\over M}{a\over M}
	\left({M\over r}\right)^4\;.
	\nonumber\\
\label{eq:pn_rdotiotadot2}
\end{eqnarray}
Now impose adiabaticity: the condition ${\dot r}_{\rm weak}
T_\theta/r\ll 1$ leads to the condition
\begin{equation}
{\mu\over M} \ll {5\over128\pi}\left({r\over M}\right)^{5/2}\;,
\label{eq:muoverM1}
\end{equation}
and ${\dot\iota}_{\rm weak}T_\theta/\iota\ll 1$ leads to
\begin{equation}
{\mu\over M} \ll {15\over488\pi}\left({r\over M}\right)^{4}\;.
\label{eq:muoverM2}
\end{equation}
Since, by definition, $\mu/M \le 1/4$, Eqs.\ (\ref{eq:muoverM1}) and
(\ref{eq:muoverM2}) are {\it always} satisfied in the weak-field.
Weak-field radiation reaction is always adiabatic, regardless of the
mass ratio.  This is not surprising.

\subsubsection{Strong-field radiation reaction}
\label{strongfield}

In the strong-field, we do not know {\it a priori} $\dot E$ and $\dot
L_z$.  Nonetheless, we would like to understand what constraints
adiabaticity places on the mass ratio, so we must at least estimate
$\dot E$ and $\dot L_z$.  Ryan {\cite{fintan2}} gives the following
quadrupole-order formulae, valid in the weak field:
\begin{eqnarray}
{\dot E}^{\rm quad} &=& -{32\over5}\left({\mu\over M}\right)^2
	\left({M\over r}\right)^5\left[1 - {73\over12}{a\over M}
	\left({M\over r}\right)^{3/2}\cos\iota\right]\;,\nonumber\\
{\dot L_z}^{\rm quad} &=& -{32M\over5}\left({\mu\over M}\right)^2
	\left({M\over r}\right)^{7/2}\left[\cos\iota + {61\over24}
	{a\over M}\left({M\over r}\right)^{3/2}\left(1 -
	3\cos^2\iota\right)\right]\;.
\label{eq:quadedotlzdot}
\end{eqnarray}
(These are the rates of change of the {\it physical} energy and
angular momentum, not the specific energy and angular momentum used in
the rest of this paper.)  We will take the {\it ansatz} that these
formulae are valid up to factors of order unity even in the strong
field.  (Detailed analysis shows that this {\it ansatz} is reasonable;
cf.\ Tables {\ref{tab:r7_a0.95_rrnumbers}} --
{\ref{tab:r100_a0.05_rrnumbers}}.)

Consider prograde, equatorial orbits with $a = M$, $r = (1 +
\epsilon)M$: a particle orbiting just barely outside the event horizon
of an extreme Kerr black hole.  In this limit,
\begin{equation}
T_\theta = M \left[{4\pi\over\epsilon}\sqrt{2\over3} +
	{22\pi\over3}\sqrt{2\over3} + O(\epsilon)\right]\;.
\end{equation}
Solving Eq.\ (\ref{eq:Qdotrdot}) using the quadrupole formulae
(\ref{eq:quadedotlzdot}) with $a=M$, $r = (1+\epsilon)M$ and expanding
in $\epsilon$ gives
\begin{equation}
{\dot r} \simeq -{584\over5\sqrt{3}\epsilon} +
	{13657\over15\sqrt{3}} + O(\epsilon)\;.
\end{equation}
Combining these results and imposing the adiabatic condition ${\dot r}
T_\theta/r \ll 1$, we obtain
\begin{equation}
{\mu\over M} \ll {15\epsilon^2\over2336\sqrt{2}\pi} + O(\epsilon^3)\;.
\label{eq:muoverM3}
\end{equation}
This is a very stringent requirement.  Since our calculation will
assume that adiabaticity holds over the entire evolution, Eq.\
(\ref{eq:muoverM3}) tells us that our results in the strong field will
only be astrophysically meaningful when applied to systems for which
the mass ratio is very extreme.  (Cutler, Kennefick, and Poisson
showed a similar condition holds as an orbiting point particle
approaches the innermost stable circular orbit of the Schwarzschild
spacetime {\cite{dan1}}.)

\section{Radiation reaction: the Teukolsky and Sasaki-Nakamura
equations}
\label{sec:radreaction2}

\subsection{The Teukolsky equation}
\label{subsec:teukolsky}

We use a formalism based on the Teukolsky equation to study radiation
reaction.  The Teukolsky equation describes the behavior of the Weyl
curvature component $\psi_4$, which encapsulates all information about
the gravitational radiation flux at infinity and at the event horizon.
Teukolsky showed {\cite{teuk72}} that the multipolar decomposition
\begin{equation}
\psi_4 = {1\over(r - i a \cos\theta)^4}\int_{-\infty}^{\infty}d\omega
\sum_{lm} R_{lm\omega}(r) _{-2}S^{a\omega}_{lm}(\theta)
e^{im\phi} e^{-i\omega t}
\label{eq:psi4decomp}
\end{equation}
separates the evolution equation for $\psi_4$.  The function
$_{-2}S^{a\omega}_{lm}(\theta)$ is a spin-weighted spheroidal
harmonic; it is discussed in Appendix {\ref{app:spheroid}}.  The
radial function $R_{lm\omega}(r)$ obeys the Teukolsky equation:
\begin{equation}
\Delta^2 {d\over dr}\left({1\over\Delta}
{dR_{lm\omega}\over dr}\right) - V(r) R_{lm\omega}(r) =
-{\cal T}_{lm\omega}(r)\;.
\label{eq:teukolsky}
\end{equation}
The potential is
\begin{equation}
V(r) = -{K^2 + 4i(r-M)K\over\Delta} + 8i\omega r + \lambda\;,
\label{eq:teukpotential}
\end{equation}
where $K = (r^2+a^2)\omega - m a$, and\footnote{For general spin
fields, $\lambda = {\cal E}_{lm} - 2am\omega + a^2\omega^2 - s(s+1)$;
however, we have specialized to $s=-2$.} $\lambda\equiv {\cal E}_{lm}
- 2 a m \omega + a^2 \omega^2 - 2$.  The number ${\cal E}_{lm}$ is the
eigenvalue of the spheroidal harmonic; see Appendix
{\ref{app:spheroid}} for details.

The homogeneous Teukolsky equation admits two independent solutions,
$R^H_{lm\omega}$ and $R^\infty_{lm\omega}$, with the following
asymptotic values:
\begin{eqnarray}
R^H_{lm\omega} &=& B^{\rm hole}_{lm\omega}\Delta^2 e^{-i p_{m\omega} r^*},
\qquad r\to r_+\nonumber\\
R^H_{lm\omega} &=& B^{\rm out}_{lm\omega}r^3 e^{i\omega r^*} +
{B^{\rm in}_{lm\omega}\over r} e^{-i\omega r^*},
\qquad r\to\infty\; ;
\label{eq:RHasymptotic}
\end{eqnarray}
and
\begin{eqnarray}
R^\infty_{lm\omega} &=& D^{\rm out}_{lm\omega}e^{i p_{m\omega} r^*} +
D^{\rm in}_{lm\omega} \Delta^2 e^{-i p_{m\omega} r^*},
\qquad r\to r_+\nonumber\\
R^\infty_{lm\omega} &=& D^{\infty}_{lm\omega}r^3 e^{i\omega r^*},
\qquad r\to\infty\; .
\label{eq:Rinftyasymptotic}
\end{eqnarray}
We have introduced $p_{m\omega} = \omega - m\omega_+$, where $\omega_+
\equiv a/2Mr_+$ is the ``angular velocity of the horizon'' (the
angular velocity with which inertial observers at the horizon are seen
to rotate due to frame dragging {\cite{membrane}}), and the ``tortoise
coordinate''
\begin{equation}
r^*(r) = r + {2M r_+\over r_+ - r_-}\ln{r-r_+\over 2 M}
 - {2M r_-\over r_+ - r_-}\ln{r-r_-\over 2 M}\;,
\label{eq:rstarofr}
\end{equation}
which is derived from the rule $dr^*/dr = (r^2 + a^2)/\Delta$.  (In
these relations, $r_\pm = M \pm \sqrt{M^2 - a^2}$; recall that $r_+$
is the coordinate of the event horizon.)  With these solutions and
using the theory of Green's functions {\cite{arfken}}, the general
solution of the Teukolsky equation can be written
\begin{equation}
R_{lm\omega}(r) = Z^H_{lm\omega}(r)R^\infty_{lm\omega}(r) +
Z^\infty_{lm\omega}(r)R^H_{lm\omega}(r)\;,
\label{eq:fullradteuksoln}
\end{equation}
where
\begin{eqnarray}
Z^H_{lm\omega}(r) &=& {1\over2i\omega B^{\rm in}_{lm\omega}
D^{\infty}_{lm\omega}}\int_{r_+}^r dr'
{R^H_{lm\omega}(r'){\cal T}_{lm\omega}(r')\over\Delta(r')^2}\;,\nonumber\\
Z^\infty_{lm\omega}(r) &=& {1\over2i\omega B^{\rm in}_{lm\omega}
D^{\infty}_{lm\omega}}\int_r^\infty dr'
{R^\infty_{lm\omega}(r'){\cal T}_{lm\omega}(r')\over\Delta(r')^2}\;.
\label{eq:ZHZinfty}
\end{eqnarray}

By construction, $Z^H_{lm\omega}(r\to r_+) =
Z^\infty_{lm\omega}(r\to\infty) = 0$.  Defining $Z^H_{lm\omega}\equiv
Z^H_{lm\omega}(r\to\infty)$, $Z^\infty_{lm\omega} \equiv
Z^\infty_{lm\omega}(r\to r_+)$, the asymptotic radial solutions are
\begin{eqnarray}
R_{lm\omega}(r\to\infty) &=& Z^H_{lm\omega} D^{\infty}_{lm\omega}
r^3 e^{i\omega r^*}\;,\nonumber\\
R_{lm\omega}(r\to r_+) &=& Z^\infty_{lm\omega} B^{\rm hole}_{lm\omega}
\Delta^2 e^{-i p_{m\omega} r^*}\;.
\label{eq:Rfullasymptotic}
\end{eqnarray}
This solution is purely ingoing at the horizon and purely outgoing at
infinity, which is physically correct.  It is convenient to absorb the
factors $D^{\infty}_{lm\omega}$ and $B^{\rm hole}_{lm\omega}$ into
$Z^H_{lm\omega}$ and $Z^\infty_{lm\omega}$ respectively, and rewrite
Eq.\ (\ref{eq:ZHZinfty}):
\begin{eqnarray}
Z^H_{lm\omega}(r) &=& {1\over2i\omega B^{\rm in}_{lm\omega}}
\int_{r_+}^r dr'
{R^H_{lm\omega}(r'){\cal T}_{lm\omega}(r')\over\Delta(r')^2}\;,\nonumber\\
Z^\infty_{lm\omega}(r) &=& {B^{\rm hole}_{lm\omega}\over
2i\omega B^{\rm in}_{lm\omega} D^{\infty}_{lm\omega}}\int_r^\infty dr'
{R^\infty_{lm\omega}(r'){\cal T}_{lm\omega}(r')\over\Delta(r')^2}\;.
\label{eq2:ZHZinfty}
\end{eqnarray}

From these asymptotic solutions, we construct the energy and angular
momentum flux due to gravitational radiation that goes to infinity and
down the event horizon.  First, note that the particle's motion is
describable as a set of harmonics of the frequencies $\Omega_\theta$
and $\Omega_\phi$, and define
\begin{equation}
\omega_{mk} = m\Omega_\phi + k\Omega_\theta\;.
\label{eq:omegamkdef}
\end{equation}
Then, decompose the $\omega$ dependence of $Z^H_{lm\omega}$ and
$Z^\infty_{lm\omega}$ as
\begin{eqnarray}
Z^H_{lm\omega} &=& \sum_k Z^H_{lmk}
\delta(\omega - \omega_{mk})\;,\nonumber\\
Z^\infty_{lm\omega} &=& \sum_k Z^\infty_{lmk}
\delta(\omega - \omega_{mk})\;.
\label{eq:ZHZinftydecompose}
\end{eqnarray}
The coefficients $Z^{H,\infty}_{lmk}$ fully determine the energy and
angular momentum fluxes.

\subsubsection{Fluxes as $r\to\infty$}
\label{subsec:fluxatinfty}

As $r\to\infty$,
\begin{equation}
\psi_4(r\to\infty) \to {1\over2}\left({\ddot h}_+
- i{\ddot h}_\times\right)\;.
\label{eq:psi4limit}
\end{equation}
The energy flux in gravitational waves, from the Isaacson
stress-energy tensor \cite{isaacson}, is
\begin{equation}
\left({dE\over dA dt}\right)^{\rm rad}_{r\to\infty} =
{1\over16\pi}\left\langle\left(\partial h_+\over\partial t\right)^2 +
\left(\partial h_\times\over\partial t\right)^2\right\rangle\;,
\label{eq:dEdAdt}
\end{equation}
where the angle brackets denote averaging over several wavelengths.
Combining Eqs.\ (\ref{eq:psi4decomp}), (\ref{eq:Rfullasymptotic}),
(\ref{eq:ZHZinftydecompose}), and (\ref{eq:dEdAdt}), we obtain
\begin{eqnarray}
\left({dE\over dt}\right)^{\rm rad}_{r\to\infty} &=&
\sum_{lmk} {{|Z^H_{lmk}|^2}\over4\pi\omega_{mk}^2}\;,\nonumber\\
\left({dL_z\over dt}\right)^{\rm rad}_{r\to\infty} &=&
\sum_{lmk} {{m |Z^H_{lmk}|^2}\over4\pi\omega_{mk}^3}\;.
\label{eq:fluxatinfinity}
\end{eqnarray}

\subsubsection{Fluxes as $r\to r_+$}
\label{subsec:fluxathorizon}

The energy and angular momentum flux at the horizon can be calculated
by measuring the rate at which the event horizon's area increases as
radiation falls into it, following the prescription of
{\cite{hawkinghartle}} as described in {\cite{teukpress}}.  The result
is
\begin{eqnarray}
\left({dE\over dt}\right)^{\rm rad}_{r\to r_+} &=&
\sum_{lmk} \alpha_{lmk}
{|Z^\infty_{lmk}|^2\over4\pi\omega_{mk}^2}\;,\nonumber\\
\left({dL_z\over dt}\right)^{\rm rad}_{r\to r_+} &=&
\sum_{lmk} \alpha_{lmk}
{m |Z^\infty_{lmk}|^2\over4\pi\omega_{mk}^3}\;.
\label{eq:fluxathorizon}
\end{eqnarray}
The coefficient $\alpha_{lmk}$ is found by transforming from
Kinnerley's null tetrad (used to construct $\psi_4$) to the
Hawking-Hartle null tetrad {\cite{hawkinghartle}} (which is
well-behaved at the Kerr event horizon); see {\cite{teukpress}} for
details.  It is given by
\begin{equation}
\alpha_{lmk} = {256 (2Mr_+)^5 p_{mk}(p_{mk}^2 +
4\varepsilon^2)(p_{mk}^2 + 16\varepsilon^2)\omega_{mk}^3
\over |C_{lmk}|^2}\;,
\label{eq:horizonalphadef}
\end{equation}
with $\varepsilon = \sqrt{M^2 - a^2}/4Mr_+$, and
\begin{eqnarray}
|C_{lmk}|^2 &=& \left[(\lambda+2)^2 + 4 a \omega_{mk} -
4 a^2 \omega_{mk}^2\right]\left(\lambda^2 + 36 m a \omega_{mk}
 - 36 a^2\omega_{mk}^2\right)\nonumber\\
& & + \left(2\lambda+3\right) \left(96 a^2\omega_{mk}^2 -
48 m a \omega_{mk}\right) + 144 \omega_{mk}^2(M^2 - a^2)\;.
\end{eqnarray}

In order to calculate all of the fluxes and from them deduce the
evolution of the particle's orbit, we need a method to calculate the
coefficients $Z^H_{lmk}$ and $Z^\infty_{lmk}$.  This, in turn,
requires us to calculate the coefficients $B^{\rm in}_{lmk}$, $B^{\rm
hole}_{lmk}$, and $D^\infty_{lmk}$.  To do so, we use the
Sasaki-Nakamura equation.

\subsection{The Sasaki-Nakamura equation}
\label{subsec:sasakinakamura}

In principle, one should be able to calculate all the necessary
coefficients directly from the Teukolsky equation.  Consider in
particular $B^{\rm in}_{lmk}$.  From Eq.\ (\ref{eq:RHasymptotic}), we
know that we should be able to start with a purely ingoing pulse of
radiation at the event horizon with unit amplitude; we should then be
able to use the Teukolsky equation to integrate out very far, and read
off the ratio $B^{\rm in}_{lmk}/B^{\rm hole}_{lmk}$ (as well as
$B^{\rm out}_{lmk}/B^{\rm hole}_{lmk}$).  This approach does not work
well in a practical numerical implementation.  The reason is that by
Eq.\ (\ref{eq:RHasymptotic}) the outgoing solution grows with a
coefficient $r^4$ relative to the ingoing coefficient --- it
completely swamps the ingoing solution, making it impossible to read
off $B^{\rm in}_{lmk}$ with any kind of accuracy.

The fundamental reason for this difficulty is that the Teukolsky
equation's potential $V(r)$ is long ranged.  A solution to this
difficulty was given by Sasaki and Nakamura {\cite{sasaknak}}, who
discovered a transformation that takes the Teukolsky function $R(r)$,
governed by an equation with long-ranged potential, to the
Sasaki-Nakamura function $X(r)$, governed by an equation with
short-ranged potential.

The Sasaki-Nakamura equation is
\begin{equation}
{d^2 X_{lmk}\over dr^{*2}} -
F(r){d X_{lmk}\over dr^*} - U(r) X_{lmk} = 0\;.
\label{eq:sasakinakamura}
\end{equation}
The functions $F(r)$ and $U(r)$ are shown explicitly in Appendix
{\ref{app:sasaknak}}.  Like the Teukolsky equation, the
Sasaki-Nakamura equation admits two solutions, whose asymptotic forms
are
\begin{eqnarray}
X^H_{lmk} &=& e^{-ip_{mk}r^*}, \qquad r\to r_+\;,\nonumber\\
X^H_{lmk} &=& A^{\rm out}_{lmk} {\bar P}(r) e^{i\omega_{mk} r^*} +
A^{\rm in}_{lmk} P(r) e^{-i\omega_{mk} r^*},\qquad r\to\infty\;;
\label{eq:XHasymptotic}
\end{eqnarray}
and
\begin{eqnarray}
X^\infty_{lmk} &=& C^{\rm out}_{lmk} e^{ip_{mk}r^*} +
C^{\rm in}_{lmk} e^{-ip_{mk}r^*},\qquad r\to r_+\;,\nonumber\\
X^\infty_{lmk} &=& {\bar P(r)} e^{i\omega_{mk}r^*},\qquad r\to\infty\;.
\label{eq:Xinftyasymptotic}
\end{eqnarray}
The function
\begin{equation}
P(r) = 1 + {{\cal A}\over\omega r} + {{\cal B}\over(\omega r)^2} +
{{\cal C}\over(\omega r)^3}
\label{eq:Pdef}
\end{equation}
allows us to more accurately describe the behavior of $X^{H,\infty}$
near infinity.  Inserting $X^{H,\infty}$ into the Sasaki-Nakamura
equation (\ref{eq:sasakinakamura}), and taking the limit $r\to\infty$,
we read off ${\cal A}$, ${\cal B}$, and ${\cal C}$:
\begin{mathletters}
\begin{eqnarray}
{\cal A} = && -{i\over2}(\lambda + 2 + 2am\omega)\;,
	\label{Acof}\\
{\cal B} = && -{1\over8}\left[(\lambda + 2)^2 - (\lambda + 2)(2 -
	4am\omega) - 4[am\omega + 3iM\omega -
	am\omega(am\omega + 2IM\omega)]\right]\;,
	\label{Bcof}\\
{\cal C} = && -{i\over6}\left[4am\omega + {\cal B}(\lambda - 4 +
	2am\omega + 8iM\omega) + 12(M\omega)^2 -
	2{\cal A}\lambda M\omega -
	(a\omega)^2(\lambda - 3 + m^2 + 2 am\omega)\right]\;.
	\label{Ccof}
\end{eqnarray}
\end{mathletters}
In the limit $a \to 0$, Eqs.\ (\ref{Acof})--(\ref{Ccof}) reduce to
the results given in \cite{eric1}.

One transforms from the Sasaki-Nakamura function to the Teukolsky
function with the rule
\begin{equation}
R^{H,\infty}_{lmk} = {1\over\eta}\left[\left(\alpha +
{\beta_{,r}\over\Delta}\right)\chi^{H,\infty}_{lmk} -
{\beta\over\Delta}\chi^{H,\infty}_{lmk,r}\right]\;,
\label{eq:XtoR}
\end{equation}
where $\chi^{H,\infty}_{lmk} = X^{H,\infty}_{lmk}\Delta/\sqrt{r^2 +
a^2}$.  From this rule follow relations for the various coefficients:
\begin{mathletters}
\begin{eqnarray}
B^{\rm in}_{lmk} &=& -{A^{\rm in}_{lmk}\over4\omega_{mk}^2}\;,
\label{eq:AintoBin}
\\
D^{\infty}_{lmk} &=& -{4\omega_{mk}^2\over c_0}\;,
\label{eq:Dinfty}
\\
B^{\rm hole}_{lmk} &=& {1\over d_{lm\omega}}\;,\quad\mbox{where}
\label{eq:Bhole}
\\
d_{lm\omega} &=& \sqrt{2Mr_+}
\left[(8 - 24 i M\omega - 16M^2\omega^2)r_+^2\right. \nonumber\\
& & +\left.(12iam - 16 M + 16amM\omega + 24iM^2\omega)r_+
- 4a^2m^2 - 12iamM + 8 M^2\right]\;.
\label{eq:dlmw}
\end{eqnarray}
\end{mathletters}
The functions $\alpha$, $\beta$, and $\eta$ and the coefficient $c_0$
are given in Appendix\ \ref{app:sasaknak}.  The quantity $B^{\rm
hole}_{lmk}$ is taken from Ref.\ \cite{minoetal97}.

The only coefficient which must be calculated numerically is $A^{\rm
in}_{lmk}$.  This is quite reasonable: from Eq.\
(\ref{eq:XHasymptotic}), we begin with a unit amplitude, purely
ingoing pulse of radiation near the event horizon and integrate
outward to read off $A^{\rm in}$.  As $r \to \infty$, the outgoing
piece of $X^H_{lmk}$ remains of constant amplitude, so $A^{\rm
in}_{lmk}$ is easy to extract.  This prescription is sufficient to
calculate the solutions to the homogeneous Teukolsky equation.

\subsection{The source term}
\label{subsec:sourceterm}

The next part of the analysis is to compute the source of the
Teukolsky equation.  This term is given by {\cite{breuer}}
\begin{equation}
{\cal T}_{lm\omega}(r) = 4\int d\Omega\,dt\,{\Sigma\over\rho^4}
\left(B_2' + B_2^{*\prime}\right)_{-2}S^{a\omega}_{lm}(\theta)e^{-im\phi}
e^{i\omega t}\;,
\label{eq:source1}
\end{equation}
where the functions $B_2'$ and $B_2^{*\prime}$ are
\begin{eqnarray}
B_2' &=& -{\rho^8\bar\rho\over2}L_{-1}\left[\rho^{-4}
L_0\left(\rho^{-2}\bar\rho^{-1} T_{nn}\right)\right] +
{\Delta^2\rho^8\bar\rho\over2\sqrt{2}}L_{-1}\left[\rho^{-4}
\bar\rho^2 J_+\left(\rho^{-2}\bar\rho^{-2}\Delta^{-1}
T_{n\bar m}\right)\right]\;,
\label{eq:B_2'}
\\
B_2^{*\prime} &=& {\Delta^2\rho^8\bar\rho\over2\sqrt{2}} J_+
\left[\rho^{-4}\bar\rho^2 \Delta^{-1} L_{-1}\left(\rho^{-2}
\bar\rho^{-2} T_{n\bar m}\right)\right] -
{\Delta^2\rho^8\bar\rho\over4}J_+\left[\rho^{-4} J_+
\left(\rho^{-2}\bar\rho T_{\bar m\bar m}\right)\right]\;.
\label{eq:B_2^*'}
\end{eqnarray}
Here, $\rho = -1/(r - ia\cos\theta)$, $\bar\rho = -1/(r +
ia\cos\theta)$ (note that $\rho$ and $\bar\rho$ have the opposite sign
from that used in, for example, {\cite{minoetal97}}; this is the sign
convention used in, for example, {\cite{teuk72}}).  The differential
operators $J_+$ and $L_{s}$ are
\begin{eqnarray}
J_+ &=& \partial_r + {iK(r)\over\Delta}\;,\nonumber\\
L_s &=& \partial_\theta + m\csc\theta - a\omega\sin\theta
+ s\cot\theta\;.
\label{eq:J+_and_Ls}
\end{eqnarray}

The quantities $T_{nn}$, $T_{n\bar m}$ and $T_{\bar m\bar m}$ are
projections of the stress-energy tensor for the orbiting point
particle onto the Newman-Penrose null-tetrad legs ${\bf n}$,
${\bf{\bar m}}$: $T_{nn} = T^{\alpha\beta} n_{\alpha} n_{\beta}$, {\it
etc.}  For a point particle moving in the Kerr spacetime,
\begin{eqnarray}
T^{\alpha\beta}(x) &=& \mu\int d\tau\, u^\alpha u^\beta
\delta^{(4)}\left[x - z(\tau)\right]\nonumber\\
&=& \mu\int d\tau\, u^\alpha u^\beta\left(-g\right)^{1/2}
\delta\left[t - t(\tau)\right]\delta\left[r - r(\tau)\right]
\delta\left[\theta - \theta(\tau)\right]
\delta\left[\phi - \phi(\tau)\right]\;.
\label{eq:stresstensor1}
\end{eqnarray}
Here, $x$ is an arbitrary spacetime point, $\tau$ is proper time
measured by the moving particle, $z(\tau)$ is the particle's
worldline, and $(-g)^{1/2} = \Sigma\,\sin\theta$ is the factor which
converts coordinate volume to proper volume ($g$ is the determinant of
the covariant components of the Kerr metric.)

As discussed in the Introduction and in Sec.\
{\ref{subsec:adiabatic}}, the true world line of the particle is given
by a Kerr geodesic plus radiation reaction corrections.  Since we do
not yet know these corrections, we use the zeroth order geodesic
motion to write down this stress-energy tensor.  Performing the
integral, we find
\begin{equation}
T^{\alpha\beta}(r,\theta,\phi,t) = \mu\,{u^\alpha u^\beta\over
\Sigma\,\sin\theta\,{\dot t}}\,\delta(r - r_0)
\delta\left[\theta - \theta(t)\right]
\delta\left[\phi - \phi(t)\right]\;.
\label{eq:stresstensor}
\end{equation}
This is specialized to circular orbits of radius $r_0$.

Next, project this stress-energy tensor onto the Newman-Penrose null
tetrad.  From {\cite{chandra}},
\begin{eqnarray}
n_\alpha &=& {1\over2}\left({\Delta\over\Sigma},1,0,
-{a\Delta\sin^2\theta\over\Sigma}\right)\;,\nonumber\\
\bar m_\alpha &=& {\rho\over\sqrt{2}}\left(ia\sin\theta,0,
\Sigma,-i(r^2+a^2)\sin\theta\right)\;.
\label{eq:NPelements}
\end{eqnarray}
It is useful to write the projected stress-energy tensor components in
the form
\begin{equation}
T_{ab} = {C_{ab}\over\sin\theta}\delta(r - r_0)
\delta\left[\theta - \theta(t)\right]
\delta\left[\phi - \phi(t)\right]\;.
\label{eq:Tab}
\end{equation}
Setting $\dot r = 0$, performing the projection, and using Eqs.\
(\ref{eq:thetadot})--(\ref{eq:tdot}) yields
\begin{eqnarray}
C_{nn} &=& {\mu\over4\Sigma^3{\dot t}}
\left[E(r^2 + a^2) - a L_z\right]^2\;,\nonumber\\
C_{n\bar m} &=& {\mu\rho\over2\sqrt{2}\Sigma^2{\dot t}}
\left[E(r^2 + a^2) - a L_z\right]
\left[i\sin\theta\left(a E - {L_z\over\sin^2\theta}\right) +
\Theta\right]\;,\nonumber\\
C_{\bar m\bar m} &=& {\mu\rho^2\over2\Sigma{\dot t}}
\left[i\sin\theta\left(a E - {L_z\over\sin^2\theta}\right) +
\Theta\right]^2\;.
\label{eq:Cab}
\end{eqnarray}
The function $\Theta$ is given in Eq.\ (\ref{eq:thetadot}).  As
written, its sign is ambiguous, depending on whether the particle is
ascending ($\theta$ decreasing) or descending ($\theta$ increasing).
We take $\Theta$ to be positive, and define
\begin{eqnarray}
C^{\pm}_{nn} &=& {\mu\over4\Sigma^3{\dot t}}
\left[E(r^2 + a^2) - a L_z\right]^2 = C_{nn}\;,\nonumber\\
C^{\pm}_{n\bar m} &=& {\mu\rho\over2\sqrt{2}\Sigma^2{\dot t}}
\left[E(r^2 + a^2) - a L_z\right]
\left[i\sin\theta\left(a E - {L_z\over\sin^2\theta}\right) \pm
\Theta\right]\;,\nonumber\\
C^{\pm}_{\bar m\bar m} &=& {\mu\rho^2\over2\Sigma{\dot t}}
\left[i\sin\theta\left(a E - {L_z\over\sin^2\theta}\right) \pm
\Theta\right]^2\;.
\label{eq:pmCab}
\end{eqnarray}
These quantities will be used later in order to break this ambiguity
when computing the source function.

Next use Eqs.\ (\ref{eq:B_2'}), (\ref{eq:B_2^*'}), (\ref{eq:Tab}), and
(\ref{eq:Cab}) to expand Eq.\ (\ref{eq:source1}), and then repeatedly
integrate by parts so that no $\theta$ derivatives are taken of any
delta functions.  The following identity {\cite{chandra,dan2}}
simplifies this integration:
\begin{equation}
\int_0^\pi h(\theta)L_s\left[g(\theta)\right]\sin\theta d\theta =
-\int_0^\pi g(\theta)L^\dag_{1 - s}\left[h(\theta)\right]
\sin\theta d\theta\;,
\label{eq:Ls_identity}
\end{equation}
where
\begin{equation}
L^\dag_s = \partial_\theta - m\csc\theta +
a\omega\sin\theta + s\cot\theta\;.
\label{eq:Ldags}
\end{equation}
With moderate effort, we find
\begin{eqnarray}
{\cal T}_{lm\omega}(r) = -4\int dt\,e^{i[\omega t - m\phi(t)]}
&&\left\{
{\rho^{-2}\bar\rho^{-1}\over2}\,C_{nn}\,\delta(r - r_0)\,L^\dag_1
\left[\rho^{-4} L^\dag_2\left(\rho^3 S\right)\right]\right.
\nonumber\\
&&\left. +
{\Delta^2\rho^{-1}\bar\rho^2\over\sqrt{2}}\,J_+
\left[\rho^{-2}\bar\rho^{-2}\Delta^{-1}C_{n\bar m}\,\delta(r - r_0)\right]
\left[ia\sin\theta(\rho - \bar\rho)S + L^\dag_2 S\right] \right.
\nonumber\\
&&\left. +
{\Delta\rho^{-2}\bar\rho^{-2}\over2\sqrt{2}}\,C_{n\bar m}\,\delta(r - r_0)
L^\dag_2\left[\rho^3\,S\,\partial_r\left(\rho^{-4}\bar\rho^2\right)\right]
\right.\nonumber\\
&&\left. +
{\Delta^2\rho^3\over4}\,S\,J_+\left[\rho^{-4}J_+\left(\rho^{-2}\bar\rho\,
C_{\bar m\bar m}\,\delta(r - r_0)\right)\right]\right\}\;.
\label{eq:source2}
\end{eqnarray}
All functions of $\theta$ are evaluated at $\theta(t)$, the particle's
$\theta$ coordinate at $t$.  We have written $S$ as shorthand for
$_{-2}S^{a\omega}_{lm}\left[\theta(t)\right]$.

Following \cite{minoetal97}, it is very convenient to write this in the
form
\begin{equation}
{\cal T}_{lm\omega}(r) = \int dt\,\Delta^2
\left\{\left[
A_{nn0} + A_{n\bar m0} + A_{\bar m\bar m0}\right]\delta(r - r_0) +
\partial_r\left(\left[A_{n\bar m1} +
A_{\bar m\bar m1}\right]\delta(r - r_0)\right) +
\partial^2_r\left[A_{\bar m\bar m2}\,\delta(r - r_0)\right]\right\}\;.
\label{eq:source3}
\end{equation}
From Eq.\ (\ref{eq:source2}), it is not too difficult to work out
$A_{abi}$:
\begin{mathletters}
\begin{eqnarray}
A_{nn0} &=& -{2\rho^{-3}\bar\rho^{-1}C_{nn}\over\Delta^2}
\left[L^\dag_1L^\dag_2 S + 2 i a \rho\sin\theta\,L^\dag_2 S\right]\;,
\label{eq:Ann0}
\\
A_{n\bar m0} &=& -{2\sqrt{2}\rho^{-3}C_{n\bar m}\over\Delta}
\left[\left({iK\over\Delta} - \rho - \bar\rho\right)L^\dag_2 S +
\left({iK\over\Delta} + \rho + \bar\rho\right)ia\sin\theta\,
S(\rho - \bar\rho)\right]\;,
\label{eq:Anm0}
\\
A_{\bar m\bar m 0} &=& S\rho^{-3}\bar\rho
\left[\left({K\over\Delta}\right)^2 + 2 i \rho {K\over\Delta} +
i \partial_r\left({K\over\Delta}\right)\right]\;,
\label{eq:Amm0}
\\
A_{n\bar m1} &=& -{2\sqrt{2}\rho^{-3}C_{n\bar m}\over\Delta}
\left[L^\dag_2 S + ia\rho\sin\theta(\rho - \bar\rho)S\right]\;,
\label{eq:Anm1}
\\
A_{\bar m\bar m1} &=& 2S\rho^{-3}\bar\rho\,C_{\bar m\bar m}
\left(\rho - {iK\over\Delta}\right)\;,
\label{eq:Amm1}
\\
A_{\bar m\bar m2} &=& -S\rho^{-3}\bar\rho\,C_{\bar m\bar m}\;.
\label{eq:Amm2}
\end{eqnarray}
\end{mathletters}
We will refer to quantities $A^{\pm}_{nn0}$, $A^{\pm}_{n\bar m0}$,
$A^{\pm}_{\bar m\bar m0}$, {\it etc.}, which are the various $A_{abi}$
written using Eq.\ (\ref{eq:pmCab}) rather than Eq.\ (\ref{eq:Cab}).
Equations (\ref{eq:Cab}), (\ref{eq:source2}), and
(\ref{eq:Ann0})--(\ref{eq:Amm2}) completely specify the source for
circular orbits.

\subsection{Evaluation of the coefficients $Z^H_{lmk}$ and $Z^\infty_{lmk}$}
\label{subsec:evalcoeffs}

The next step is to substitute this source into Eq.\
(\ref{eq2:ZHZinfty}) and integrate for the coefficients
$Z^H_{lm\omega}$, $Z^\infty_{lm\omega}$.  Thanks to the
delta functions, these integrals are trivial; the result is
\begin{eqnarray}
Z^H_{lm\omega} &=& {1\over2i\omega B^{\rm in}_{lmk}}
\int_{-\infty}^{\infty}dt\,e^{i[\omega t - m\phi(t)]}
\biggl\{
R^H_{lm\omega}(r_0)
\left[A_{nn0} + A_{n\bar m0} + A_{\bar m\bar m0}\right]
\nonumber\\
&& \qquad\qquad\qquad\qquad\qquad\qquad\qquad\qquad
- {dR^H_{lm\omega}\over dr}\biggr|_{r_0}
\left[A_{n\bar m1} + A_{\bar m\bar m1}\right]
+ {d^2R^H_{lm\omega}\over dr^2}\biggr|_{r_0}\,
A_{\bar m\bar m2}\biggr\}\;,
\nonumber\\
Z^\infty_{lm\omega} &=& -{c_0\over8i\omega^3d_{lmk}
B^{\rm in}_{lmk}}
\int_{-\infty}^{\infty}dt\,e^{i[\omega t - m\phi(t)]}
\biggl\{
R^\infty_{lm\omega}(r_0)
\left[A_{nn0} + A_{n\bar m0} + A_{\bar m\bar m0}\right]
\nonumber\\
&&\qquad\qquad\qquad\qquad\qquad\qquad\qquad\qquad
- {dR^\infty_{lm\omega}\over dr}\biggr|_{r_0}
\left[A_{n\bar m1} + A_{\bar m\bar m1}\right]
+ {d^2R^\infty_{lm\omega}\over dr^2}\biggr|_{r_0}\,
A_{\bar m\bar m2}\biggr\}\;.
\label{eq3:ZHZInfty}
\end{eqnarray}

For a numerical implementation, it is not useful to leave this in the
form of an integral over infinite domain.  For simplicity, write the
above integrals as
\begin{equation}
Z^{H,\infty}_{lm\omega} = C^{H,\infty} \int_{-\infty}^{\infty} dt\,
e^{i[\omega t - m\phi(t)]} I^{H,\infty}[r_0,\theta(t)]\;.
\label{eq:def:IHinfty}
\end{equation}
To bring out the harmonic structure of the source, we would like to
write the integrand in the form $J^{H,\infty}_{mk}(r_0)
\exp[i(\omega - m\Omega_\phi - k\Omega_\theta)t]$.  First, define
\begin{equation}
J^{H,\infty}_m[r_0,\theta(t)] = I^{H,\infty}[r_0,\theta(t)]
e^{im[\Omega_\phi t - \phi(t)]}\;.
\label{eq:def:JmHinfty}
\end{equation}
From the discussion in Sec.\ {\ref{subsec:noneqfrequencies}}, we
know that
\begin{equation}
J^{H,\infty}_m[r_0,\theta(t)] = \sum_{k=0}^\infty
J^{H,\infty}_{mk}(r_0) e^{-ik\Omega_\theta t(\theta)}\;,
\end{equation}
where
\begin{equation}
J^{H,\infty}_{mk}(r_0) = {1\over T_\theta}\int_0^{T_\theta} dt\,
J^{H,\infty}_m[r_0,\theta(t)]\,e^{ik\Omega_\theta t}\;.
\label{eq:def:JmkHinfty}
\end{equation}
Substituting for $I^{H,\infty}[r_0,\theta(t)]$ in Eq.\
(\ref{eq:def:IHinfty}) gives
\begin{eqnarray}
Z^{H,\infty}_{lm\omega} &=& C^{H,\infty} \sum_{k=0}^\infty
\int_{-\infty}^{\infty} dt\,
e^{i[\omega - m\Omega_\phi - k\Omega_\theta]t}
J^{H,\infty}_{mk}(r_0)\;,\nonumber\\
&=& 2\pi C^{H,\infty} \sum_{k=0}^\infty \delta(\omega - \omega_{mk})
J^{H,\infty}_{mk}(r_0)\;.
\end{eqnarray}
Then use Eq.\ (\ref{eq:ZHZinftydecompose}) to find
\begin{equation}
Z^{H,\infty}_{lmk} = 2\pi C^{H,\infty} \sum_{k=0}^\infty
J^{H,\infty}_{mk}(r_0)\;.
\end{equation}

We must next evaluate the number $J^{H,\infty}_{mk}(r_0)$.  By Eqs.\
(\ref{eq:def:JmHinfty}) and (\ref{eq:def:JmkHinfty}),
\begin{equation}
J^{H,\infty}_{mk}(r_0) = {1\over T_\theta}\int_0^{T_\theta}
dt\,e^{i[\omega_{mk} t - m \phi(t)]} I^{H,\infty}[r_0,\theta(t)]\;.
\label{eq:JmkHinfty}
\end{equation}
Because the dependence of $I^{H,\infty}_{mk}$ on $t$ is implicit, it
is useful to change the integration variable to $\chi$.  The integrand
then picks up a factor $(d\chi/dt)^{-1}$.  This is well-behaved over
the entire domain of the integral [cf.\ Eq.\ (\ref{eq:dtdchi})].
Changing to $\theta$, for example, would not work well since the
factor $(d\theta/dt)^{-1}$ is singular at the turning points
$\theta_{\rm max/min}$.

Writing $dt = d\chi\,(d\chi/dt)^{-1}$ and combining all of the results
of this section, gives
\begin{eqnarray}
Z^{H}_{lmk} &=& {\pi\over i\omega_{mk} T_\theta B^{\rm in}_{lmk}}
\sum_{\pm}
\int_0^{\pi} d\chi\,{\gamma + a^2 E z(\chi) \over
\sqrt{\beta[z_+ - z(\chi)]}}\,e^{\pm i\left[\omega_{mk} t(\chi)
- m \phi(\chi)\right]}\;_{\pm}I^{H}_{lmk}[r_0,z(\chi)]\;,
\label{eq:ZHfinal}\\
Z^{\infty}_{lmk} &=&
-{\pi c_0\over 4i\omega_{mk}^3 d_{lm\omega} T_\theta B^{\rm in}_{lmk}}
\sum_{\pm}
\int_0^{\pi} d\chi\,{\gamma + a^2 E z(\chi) \over
\sqrt{\beta[z_+ - z(\chi)]}}\,e^{\pm i\left[\omega_{mk} t(\chi)
- m \phi(\chi)\right]}\;_{\pm}I^\infty_{lmk}[r_0,z(\chi)]\;,
\label{eq:Zinftyfinal}
\end{eqnarray}
where
\begin{equation}
_{\pm}I^{H,\infty}_{lmk}[r_0,z(\chi)] =
R^{H,\infty}_{lm\omega}(r_0)
\left[A^{\pm}_{nn0} + A^{\pm}_{n\bar m0} + A^{\pm}_{\bar m\bar m0}\right]
- {dR^{H,\infty}_{lm\omega}\over dr}\biggr|_{r_0}
\left[A^{\pm}_{n\bar m1} + A^{\pm}_{\bar m\bar m1}\right]
+ {d^2R^{H,\infty}_{lm\omega}\over dr^2}\biggr|_{r_0}\,
A^{\pm}_{\bar m\bar m2}\;.
\label{eq:I_def}
\end{equation}

The coefficients $Z^{H,\infty}_{lmk}$ obey a useful symmetry between
harmonics $(m,k)$ and $(-m,-k)$.  From Eqs.\
(\ref{eq:ZHfinal})--(\ref{eq:I_def}) and by inspection of the
Teukolsky equation [Eq.\ (\ref{eq:teukolsky})], it is apparent that
$Z^{H,\infty}_{l-m-k} = e^{i\alpha} {\bar Z}^{H,\infty}_{lmk}$, where
$\alpha$ is some phase factor (as yet undetermined).  It is simple to
determine this phase factor in the Schwarzschild limit.  Because of
spherical symmetry, waveforms emitted by a particle orbiting at some
angle $\iota$ to the equator of a Schwarzschild hole and observed in
the equatorial plane are equivalent to those emitted by a particle
orbiting in the equatorial plane and observed at an angle $-\iota$.
Equating the expressions for the waveforms in these two cases yields
$e^{i\alpha} = (-1)^{k + l}$, so that
\begin{equation}
Z^{H,\infty}_{l-m-k} = (-1)^{k + l} {\bar Z}^{H,\infty}_{lmk}\;.
\label{eq:zedsymmetry}
\end{equation}
This factor in fact arises from the rotation properties of the
spin-weighted spherical harmonics.  Because we generate spheroidal
harmonics as a sum of spherical harmonics, Eq.\ (\ref{eq:zedsymmetry})
holds for Kerr black holes as well.  Hence, we need only consider $k
\ge 0$.

\section{Numerical implementation and validation}
\label{sec:implement}

In this section, we first describe the structure and methods of the
numerical code that was used to study circular orbit radiation
reaction.  We next present results of certain ``sanity checks'' that
were run to make sure that, in the proper limits, the code reproduces
well-known results.

\subsection{Code implementation}
\label{subsec:codeimplementation}

The numerical code can be broken into two pieces: a set of
``harmonic'' routines which calculate the coefficients $Z^H_{lmk}$ and
$Z^\infty_{lmk}$ (and thence the energy and angular momentum fluxes),
and a ``driver'' routine which repeatedly calls the harmonic routines
and determines when they have ``converged'' (as defined below) to give
the fluxes of energy and angular momentum emitted by that particular
orbit.

The driving algorithm has the following structure:
\begin{enumerate}

\item Choose the orbital radius $r$ and angular momentum $L_z$.
From this, compute the orbital energy $E$ and Carter constant $Q$ via
Eqs.\ (\ref{eq:EofLzr}) and (\ref{eq:QofLzr}), and also the inclination
angle $\iota$ via Eq.\ (\ref{eq:iota}).

\item Loop on the harmonic index $l$, starting with $l = 2$,
until the $l$-convergence criterion discussed below is satisfied.

\item For each value of $l$, loop on the index $m \in [-l, \dots, l]$.

\item For each value of $m$, loop on the index $k$ until the
$k$-convergence criterion discussed below is satisfied.  Because of
the symmetry condition Eq.\ (\ref{eq:zedsymmetry}), consider only $k
\ge 0$.

\item Compute the frequency $\omega_{mk} = m\Omega_\phi +
k\Omega_\theta$ using Eqs.\ (\ref{eq:Ttheta2}) and (\ref{eq:Phi}).

\item Compute the spheroidal harmonic expansion coefficients
$b^{a\omega}_j$ [cf.\ Eq.\ (\ref{eq:spheroidexpand})].

\item Compute the coefficients $Z^{H,\infty}_{lmk}$.  This is
described separately below.

\item Compute the energy and angular momentum fluxes to infinity
and down the hole via Eqs.\ (\ref{eq:fluxatinfinity}) and
(\ref{eq:fluxathorizon}).

\item Check the convergence of the $k$ loop.  The convergence test
implemented here is to check whether ${\dot E}^{\rm rad}_{lmk} <
\epsilon_k \max_k {\dot E}^{\rm rad}_{lmk}$ (where ${\dot E}^{\rm
rad}_{lmk} = {\dot E}^H_{lmk} + {\dot E}^\infty_{lmk}$, and $\max_k
{\dot E}^{\rm rad}_{lmk}$ is the largest value of ${\dot E}^{\rm
rad}_{lmk}$ over all $k$ values).  The value of $\epsilon_k$ is
discussed in greater detail in Sec. {\ref{sec:results}}.  When this
condition is met for three successive $k$ values, the $k$-loop is
ended.  Otherwise, increase $k$ by one and repeat the $k$-loop.

\item If $m = l$, terminate the $m$-loop, else increase $m$ by one
and repeat.

\item Check the convergence of the $l$ loop.  The convergence test
implemented here is to check whether ${\dot E}^{\rm rad}_l <
\epsilon_l \max_l {\dot E}^{\rm rad}_l$ (where ${\dot E}^{\rm rad}_l =
\sum_{mk} {\dot E}^{\rm rad}_{lmk}$, and $\max_l {\dot E}^{\rm rad}_l$
is the largest value of ${\dot E}^{\rm rad}_l$ over all $l$ values).
We have typically used $\epsilon_l = 10 \times \epsilon_k$.  When this
condition is met for three successive $l$ values, the $l$-loop is
ended.  Otherwise, increase $l$ by one and repeat.

\item Compute the total change in the angular momentum and energy
of the particle:
\begin{equation}
{\dot E} = -\sum_l {\dot E}^{\rm rad}_l\;,\qquad
{\dot L_z} = -\sum_l {\dot L_{z;l}}^{\rm rad}\;.
\end{equation}

\item Compute the total change in the particle's radius, Carter
constant and inclination angle using Eqs.\
(\ref{eq:Qdotrdot})--(\ref{eq:iotadot}).

\item Compute the gravitational waveform:
\begin{equation}
h_+(\theta,\phi,t) - i h_\times(\theta,\phi,t) =
\sum_{lmk} {1\over\omega_{mk}^2} Z^H_{lmk}
{_{-2}}S^{a\omega_{mk}}_{lm}(\theta) e^{i(m\phi - \omega t)}\;.
\label{eq:waveform}
\end{equation}
\end{enumerate}
For a given orbit with some radius and inclination angle ($r,\iota$),
this algorithm gives the orbit's gravitational waveform and the
direction (${\dot r},{\dot\iota}$) in which radiation reaction drives
it to a new orbit.

The ``harmonic'' routines which calculate the coefficients
$Z^{H,\infty}_{lmk}$ are sufficiently involved that they merit separate
discussion.  This algorithm assumes that the
orbital characteristics $E$, $L_z$, $Q$, and $r$ are known, as are the
harmonic indices $l$, $m$, $k$ (and hence the frequency
$\omega_{mk}$), and the spheroidal harmonic expansion coefficients
$b^{a\omega}_j$.

\begin{enumerate}

\item Integrate the Sasaki-Nakamura equation (\ref{eq:sasakinakamura})
inwards from ``infinity'' to $r$ using Bulirsch-Stoer integration (as
implemented in the routine {\tt bsstep()} driven by {\tt odeint()}
from {\cite{numrec}}, modified to integrate complex functions of real
arguments) to get $X^\infty_{lmk}$.  This integration is done in
Boyer-Lindquist coordinates.  The value at ``infinity'' is set using
the asymptotic form Eq.\ (\ref{eq:Xinftyasymptotic}).  It is of course
impossible to actually integrate numerically from $r = \infty$, so we
have implemented a variant of Richardson extrapolation {\cite{numrec}}
to accurately compute the value of $X^\infty_{lmk}$.  Let $r^\infty_i$
be the $i$th ``big number'' which will be used to approximate
infinity, and let $X^\infty_{i;lmk}$ be the value of $X^\infty_{lmk}$
computed using $r^\infty_i$ to set the $r\to\infty$ boundary
condition.  We set $r^\infty_1 \simeq 30/\omega_{mk}$, {\it i.e.},
roughly 5 mode wavelengths, and set $r^\infty_{i+1} = 2\times
r^\infty_i$.  We then construct a rational function approximation to
the sequence $X^\infty_{i;lmk}$ as a function of $x_i = 1/r^\infty_i$,
and use the approximation to extrapolate to the limit $x = 0$.  We
iterate until the difference in successive approximations is $\sim
10^{-7}$; this typically requires 5 -- 10 iterations.  We have found
this to be far more accurate and faster than simply setting ``$\infty
= \mbox{very large number}$'' in the code.

\item Integrate the Sasaki-Nakamura equation from the event horizon
to $r$ to get $X^H_{lmk}$, using the asymptotic form
Eq. (\ref{eq:XHasymptotic}) to set the value at the horizon.  We use
essentially the same numerical tricks and techniques as are used for
$X^\infty_{lmk}$.  In particular, we construct a rational function
approximation to a sequence of values $X^H_{i;lmk}$ computed using
$r^H_i = r_+ + x_i$, where $x_1 \sim 10^{-6}$ and $x_{i+1} = x_i/2$,
and then extrapolate to $x = 0$.  (This trick is needed since the
Sasaki-Nakamura equation is not well-behaved in Boyer-Lindquist
coordinates at the horizon.)  Roughly 4 -- 8 iterations are needed to
get a result accurate to $\sim 10^{-7}$.

\item Integrate the Sasaki-Nakamura equation from $r$ to infinity,
using $X^H_{lmk}$ as the value at $r$, and thereby compute the
coefficient $A^{\rm in}_{lmk}$.  The same numerical techniques used to
calculate $X^\infty_{lmk}$ are used here to integrate the equation and
to reach infinity.

\item Compute the homogeneous Teukolsky solutions $R^{H,\infty}_{lmk}$
using Eq.\ (\ref{eq:XtoR}), and $B^{\rm in}_{lmk}$ using Eq.\
(\ref{eq:AintoBin}).

\item Compute $Z^{H,\infty}_{lmk}$ using Eqs.\ (\ref{eq:ZHfinal}) and
(\ref{eq:Zinftyfinal}).

\end{enumerate}
This algorithm is then called by the ``driver'' algorithm for the step
at which $Z^{H,\infty}_{lmk}$ are needed.

\subsection{Code validation}
\label{subsec:codevalidation}

In order to verify that the algorithms described in Sec.\
{\ref{subsec:codeimplementation}} are working reliably, we ran a
series of tests to make sure that the code reproduces known results in
the proper limits.  First, we analyzed the weak-field limit and
verified that the code reproduces the results given in
{\cite{minoetal97}} (who analyzed gravitational radiation emitted by
point particles orbiting black holes using a post-Newtonian
expansion).  Second, we took the limit $a = 0$ and verified that the
radiation emitted by a particle in an inclined orbit about a
Schwarzschild black hole has the correct behavior.

Appendices G and I of {\cite{minoetal97}} give (very long and
detailed) post-Newtonian expansions of the energy flux to infinity and
down the horizon for a particle orbiting in the equatorial plane of a
Kerr black hole.  These formulae allow us to check that the fluxes at
infinity and down the horizon agree with known results as a function
of the orbit's radius $r$ and the black hole's spin $a$.  We have
compared both fluxes for a large number of cases and found excellent
agreement (to the degree expected) for all parameters.  Figure
{\ref{fig:fluxdownholel=3}} is a typical example, comparing for $l=3$
the numerically computed downhole flux for a co-rotating orbit at $r =
25 M$ as a function of black hole spin $a$.  The fits are quite good
(except for $l = 3$, $m = 2$) even in this relatively strong-field
region because in most cases the post-Newtonian expansion formulae of
{\cite{minoetal97}} are very robust, including many powers of $M/r$.
(There are not as many terms given for the case $l = 3$, $m = 2$, so
the fits are not as robust in that case.)

These tests demonstrate the code reliably produces radiation in the
Kerr equatorial plane.  The next check was whether the code reliably
produces radiation for orbits out of the equatorial plane.  A simple
test is to examine inclined orbits in the Schwarzschild limit.  In
this limit, the {\it total} flux radiated for a given $l$ mode must be
invariant as the plane of the orbit is tilted (due to spherical
symmetry); however, the {\it distribution} of the radiation among the
$k$ and $m$ indices changes.  The nature of this change can be deduced
by examining the Teukolsky equation source, Eq.\ (\ref{eq:source1}).
In the Schwarzschild limit, the spin-weighted spheroidal harmonic
${_{-2}}S^{a\omega}_{lm}(\theta)$ reduces to a spin-weighted spherical
harmonic ${_{-2}}Y_{lm}(\theta)$ (cf.\ Appendix
{\ref{subsec:spherical}}).  The behavior of the system under rotation
follows from the behavior of ${_{-2}}Y_{lm}(\theta)$ under rotation.

The rotation behavior of {\it zero}-weight spherical harmonics ({\it
i.e.}, the usual well-loved spherical harmonic) is well understood: if
one rotates about the $x$-axis of the coordinate system (so that the
$\phi$ angles remain unchanged) by some angle $\iota$, then spherical
harmonics in the new (primed) coordinate system are related to those
in the original (unprimed) coordinate system via
\begin{equation}
{_0}Y_{lm}(\theta') = \sum_{m' = -l}^l {\cal D}^l_{m'm}(\iota)
{_0}Y_{lm'}(\theta)\;.
\label{eq:wignerDdef}
\end{equation}
The function ${\cal D}^l_{m'm}(\iota)$ is the Wigner D-function; an
explicit expression for it is given in Eqs.\ (4.255)--(4.256) of
{\cite{arfken}} (with Arfken's angles $\alpha$ and $\gamma$ set to
zero, and $\beta = \iota$).  It represents one element of a matrix
that rotates spherical harmonics.  Because this matrix commutes with
the differential operator $\edth$ which lowers the spin weights of
spherical harmonics (cf.\ Appendix {\ref{subsec:spherical}}), Eq.\
({\ref{eq:wignerDdef}}) applies to {\it any} spin weight.

From this, it is a simple matter to show that the energy emitted by a
particle orbiting at angle $\iota$ into some set of harmonic indices
$(l,m,k)$ is related to that emitted into the indices $(l,m)$ by a
particle in equatorial orbit by
\begin{equation}
{{\dot E}_{lmk}(\iota)\over{\dot E}^{\rm eq}_{lm}} =
| {\cal D}^l_{(k-m),m}(\iota)|^2\;.
\label{eq:rotatededot}
\end{equation}

We have rather exhaustively compared the right- and left-hand sides of
Eq.\ (\ref{eq:rotatededot}).  In Figure {\ref{fig:compareinclined}},
we show a typical comparison, $l = 4$, $m = 2$, $k = 1$.  The
numerical points agree with the analytic D-function curve to a
fractional error $\sim 10^{-6} - 10^{-7}$; this kind of agreement was
found in all cases examined.  This indicates that the off-equator
capabilities of the code should be reliable.

In addition to these tests, we compared this code with unpublished
results from D.\ Kennefick (whose code was used for the analysis of
{\cite{dan2}}) and L.\ S.\ Finn (whose code is used for the analysis
of {\cite{finnthorne}}).  These two codes generate radiation for
equatorial Kerr orbits.  In all cases, we have found very good
agreement (typically agreeing to the number of digits available with
Finn's code, and fractional error $10^{-5} - 10^{-6}$ compared to
Kennefick's code).  This, plus the validation tests described above,
give us great confidence that the results found with this code are
reliable.

\section{Results}
\label{sec:results}

The major results of this analysis break into two pieces: the
gravitational waves (and associated energy spectra) produced by
particular orbits, and the sequence of orbits through which the system
passes as it evolves due to radiation reaction.  We consider these
results separately below.

\subsection{Radiation emitted at particular orbits}
\label{subsec:radiation}

These results were computed by implementing the algorithm discussed in
Sec.\ {\ref{subsec:codeimplementation}}, using the value $\epsilon_k =
10^{-7}$ (so that $\epsilon_l = 10^{-6}$).  The waveforms should
therefore have a fractional accuracy of about $10^{-5} - 10^{-6}$.  We
discuss in detail two particular strong-field orbits.

\subsubsection{$r = 7 M$, $a = 0.95 M$, $\iota = 62.43^\circ$}
\label{subsubsec:r7_a0.95}

The gravitational waveform produced by this orbit is shown in Figure
{\ref{fig:wave_r7_a0.95}}.  To achieve fractional accuracy $10^{-6}$
on the $l$-loop required summing to $l = 12$.  For each value of $l$,
there are of course $2l + 1$ values of $m$; and for each value of $m$,
we needed to compute $4$ to $20$ values of $k$ to achieve fractional
accuracy $10^{-7}$ on the $k$-loop.  The code uses roughly $2800$
separate harmonics for this orbit.  The fundamental orbital
frequencies have values $M \Omega_\phi = 0.05424$, $M \Omega_\theta =
0.04954$.

A notable feature of this waveform is the presence of many short
timescale features ({\it e.g.}, the small bump in $h_+$ near $t = 300
M$, and the spiky features in $h_\times$ between $t = 1100 M$ and $t =
1500 M$).  The presence of these features is consistent with the
breadth of this orbit's energy spectrum with respect to $k$ (cf.\
Fig.\ {\ref{fig:dEinfdt_r7_a0.95_l2}}).  High frequencies play a
rather important role in determining the radiation to infinity because
the Teukolsky potential [Eq.\ (\ref{eq:teukpotential})] is rather
transmissive to high frequency modes for large $a$ {\cite{chandra}}.
Thus, many sum and difference harmonics of the fundamental frequencies
are needed to accurately describe the motion.  The low-frequency
modulation of the waveform is due to frame-dragging induced precession
of the orbital plane --- Lense-Thirring precession.

The downhole energy spectrum for this orbit is plotted in Fig.\
{\ref{fig:dEHdt_r7_a0.95_l2}}.  The most notable feature is that the
energy radiated ``down the hole'' is negative: rather than the orbit
losing energy down the horizon, this indicates that the orbit gains
energy {\it from} the hole.  This seemingly bizarre phenomenon is due
to superradiant scattering {\cite{teukpress}}: the radiation extracts
energy from the ergosphere of the Kerr black hole and pumps that
energy into the orbit.  This transfers energy from the black hole's
spin to the particle's orbit, slowing the inspiral.  It is essentially
a manifestation of the Penrose process.

Calculations such as this have excited interest in the past in the
possibility of ``floating orbits'' {\cite{pressteuk}}: orbits which
absorb exactly as much energy as they lose to infinity, stopping their
secular inspiral due to radiation loss {\cite{spindownhole}}.
Floating orbits would appear to be incredibly interesting objects: as
sources of gravitational radiation, they would produce strong-field,
nearly stationary signals, offering the possibility of very high
precision studies of the Kerr ergosphere and the strong-field
spacetime of black holes.  Working with D.\ Kennefick, and checking
all results with his code {\cite{dan2}}, we have examined very
strong-field circular equatorial orbits of nearly maximal Kerr black
holes (for which superradiance is strongest, and so are most likely to
have floating orbits).  We find that in no cases does the energy
``flowing out of the horizon'' come close to that radiated to
infinity: summing over all multipoles, we find that at best ${\dot
E}^H \sim -0.1\,{\dot E}^\infty$.  Although we could not find this
result in the published literature, other authors who have written
similar codes (S.\ Detweiler and L.\ S.\ Finn in particular) have come
to the same conclusion {\cite{nofloatingorbits}}.

Table {\ref{tab:r7_a0.95_rrnumbers}} shows summed radiation reaction
quantities for this orbit.  They are compared with the values computed
from the post-Newtonian approximation, using formulae due to Ryan
{\cite{fintan2}}.  These numbers, coupled with the numbers for orbits
at $r = 7 M$, $a = 0.05 M$ (cf.\ Sec.\ {\ref{subsubsec:r7_a0.05}} and
Table {\ref{tab:r7_a0.05_rrnumbers}}) indicate that the post-Newtonian
formulae are not particularly useful in the strong-field.  As
predicted by Ryan, our results indicate that the inclination angle
increases --- the orbit tends to evolve toward anti-alignment with the
black hole's spin.  However, the quantitative details of the
post-Newtonian predictions are rather inaccurate. In particular, they
underestimate the rate at which the radius changes (by about a factor
of two in this case) and overestimate the rate at which the
inclination angle changes (in this case, by a factor of three).
(Interestingly, this table shows that the post-Newtonian predictions
for $\dot L_z$ and $\dot E$ aren't nearly as inaccurate as for $\dot
r$ and $\dot\iota$.  Small errors in $\dot L_z$ and $\dot E$ tend to
magnify to large errors in $\dot r$ and $\dot\iota$.)

\subsubsection{$r = 7 M$, $a = 0.05 M$, $\iota = 60.17^\circ$}
\label{subsubsec:r7_a0.05}

The gravitational waveform produced in this case is shown in Fig.\
{\ref{fig:wave_r7_a0.05}}.  Fractional accuracy of $10^{-6}$ on the
$l$-loop here required summing to $l = 10$.  As in the case $a = 0.95
M$, the code needed $4$ to $20$ values of $k$ for fractional accuracy
$10^{-7}$ on the $k$-loop.  In this case, the code needed roughly
$1300$ harmonics.  The fundamental orbital frequencies have values $M
\Omega_\phi = 0.05407$, $M
\Omega_\theta = 0.05378$.

Two features of this waveform are noteworthy, especially in comparison
with the waveform in Fig.\ {\ref{fig:wave_r7_a0.95}}.  First, note
that the low-frequency modulation of the waveform is much slower.
This is because the dragging of inertial frames is so much slower ---
the modulation is caused by the plane of the particle's orbit
precessing about the black hole.  At least at lowest order, this
Lense-Thirring precession frequency is proportional to the black
hole's spin.  Second, it is clear that the waveform is much
``simpler'' for small spin: far fewer short timescale features are
present (see the lowest panel of Fig.\ {\ref{fig:wave_r7_a0.05}}).
This is reflected in the narrowness of the spectra of energy radiated
to infinity (Fig.\ {\ref{fig:dEinfdt_r7_a0.05_l2}}).  When the black
hole's spin is small, the Teukolsky potential is not particularly
transmissive to high-frequency radiation; a large number of sum and
difference harmonics aren't needed.  The radiation at infinity is
largely determined by radiation emitted at frequencies $\omega \simeq
2\Omega_\phi$, $\omega\simeq\Omega_\theta+\Omega_\phi$, and
$\omega\simeq2\Omega_\theta$.  (For small spins,
$\Omega_\phi\simeq\Omega_\theta$, so these three frequencies are very
nearly equal.)

The downhole energy spectrum is plotted in Fig.\
{\ref{fig:dEHdt_r7_a0.05_l2}}.  Note that it is {\it nowhere}
negative: for $a = 0.05 M$, the particle does not extract any orbital
energy from the black hole's spin.  The ergosphere for such a slowly
rotating hole is small and unimportant, so radiation emitted toward
the hole tends to be absorbed by the event horizon rather than being
superradiantly scattered.

Table {\ref{tab:r7_a0.05_rrnumbers}} gives the summed radiation
reaction quantities, again comparing versus Ryan's post-Newtonian
results.  As in the case $a = 0.95 M$, we see that the post-Newtonian
results give the qualitatively correct result (orbital radius shrinks
and tilt increases), but are not quantitatively accurate in the strong
field.  Here, post-Newtonian theory underestimates the rate at which
the orbital radius changes by a factor of roughly three, and
overestimates the rate at which the inclination angle changes by about
$50\%$.

The behavior of $dE/dt$ summed over all $l$ as a function of $\omega$
is shown in Fig.\ {\ref{fig:dEsumdt_r7}}.  For both the downhole
energy and energy radiated to infinity, the greatest radiated flux
occurs at $\omega \simeq 0.1/M \simeq 2\Omega_\phi \simeq
2\Omega_\theta$.  Peaks in the spectrum occur near all integer
multiples of $\Omega_\phi$ and $\Omega_\theta$.  For $a = 0.95 M$,
these peaks are fairly broad.  This is because there is power at many
sum and difference harmonics of $\Omega_\phi$ and $\Omega_\theta$.
Since the fractional difference in these frequencies is about $10\%$,
all power spikes are distinctly separated from one another.  For $a =
0.05 M$, these peaks are quite narrow.  The fractional difference
between $\Omega_\phi$ and $\Omega_\theta$ is only about $0.5\%$ in
this case, so spikes for the various sum and difference harmonics are
{\it not} distinctly separated.

To demonstrate convergence to the post-Newtonian results, Tables
{\ref{tab:r100_a0.95_rrnumbers}} and {\ref{tab:r100_a0.05_rrnumbers}}
show summed radiation reaction quantities for orbits at $r = 100 M$,
spins $a = 0.95 M$ and $a = 0.05 M$, and inclination angles near
$60^\circ$.  The numerical results are in much better agreement with
the post-Newtonian results.  This is not surprising, since $r = 100 M$
is a relatively weak-field region of the spacetime.  It is however
reassuring; these tables are further evidence that the code's results
are trustworthy.

\subsection{Radiation reaction sequences}
\label{subsec:radreactseq}

Fig.\ {\ref{fig:innerseq_a0.8}} shows a section of the parameter space
in the strong-field of a Kerr black hole with spin $a = 0.8 M$.  As
discussed in the Introduction, this is a reasonable choice for
supermassive black holes whose spin has been buffered by
magnetohydrodynamic extraction of the holes' spin energy
{\cite{sterlcomment,moderskietal,rogercomment}}.  Each $(r,\iota)$
coordinate point in this figure represents an orbit.  The dotted line
separates stable from unstable orbits: orbits to the left of the curve
are unstable to perturbations and catastrophically plunge into the
black hole.  [This line is calculated by solving the system $R = R' =
R'' = 0$ for the constants parameterizing marginally stable orbits,
where $R$ is defined in Eq.\ ({\ref{eq:rdot}}).]

The effect of radiation reaction on these orbits is indicated in Fig.\
{\ref{fig:innerseq_a0.8}} by arrows at various points in the figure.
The tail of each arrow indicates a particular orbit; the arrow itself
is proportional to the vector $[(M/\mu)\dot r,(M^2/\mu)\dot\iota]$.
The arrow points in the direction to which radiation reaction pushes
the particle from orbit to orbit, and its length indicates how rapidly
radiation reaction acts.  As might be expected from the discussion in
Sec.\ {\ref{subsec:radiation}}, the arrows in this figure indicate
that the inclination angle does not change very quickly.  One can
regard the arrows as representing tangent vectors to a radiation
reaction trajectory $[r(t),\iota(t)]$.  The nearly horizontal aspect
of the arrows in this figure indicates that radiation reaction would
change $\iota$ by little during a physical inspiral.

The extremely long arrow near $r = 7M$, $\iota \simeq 120^\circ$ in
Fig.\ {\ref{fig:innerseq_a0.8}} highlights the effect radiation
reaction has on orbits which are close to the maximum inclination
angle for stable orbits.  That particular point is at $\iota =
119.194^\circ$; the maximum inclination angle at $r = 7M$ is $\iota =
119.670^\circ$.  Because it is extremely close to the marginally
stable orbit, a small ``push'' from radiation reaction has a very
large effect.  Orbits that are close to the stability threshold are
quickly pushed into the hole by gravitational-wave emission.

Figure {\ref{fig:iotadot_vs_iota}} plots $\dot\iota$ as a function of
$\iota$ for various black hole spins.  All curves are for orbits at $r
= 10 M$.  Notice that, for large spin, the distribution's peak is
pushed away from $\iota = 90^\circ$, where first order post-Newtonian
analysis predicts $\dot\iota$ is maximal [cf.\ Eq.\
(\ref{eq:pn_rdotiotadot}), which predicts
$\dot\iota\propto\sin\iota$].  One can see that the post-Newtonian
prediction is more accurate for small spin.  This is not surprising,
since Eq.\ (\ref{eq:pn_rdotiotadot}) is based on leading order
expansions in both $M/r$ and $a/M$.  Although not shown here, the peak
moves toward $90^\circ$ as $r/M$ increases, indicating that this shift
is truly a strong-field effect.

\section{Conclusion}
\label{sec:conclusion}

The results presented in this paper give the first strong-field
radiation reaction results in which the Carter constant evolves in a
non-trivial manner.  Circular, non-equatorial orbits are probably,
however, the only case in which one can evolve the Carter constant by
examining radiation fluxes --- in all other cases, Eq.\
(\ref{eq:Qdotrdot}), relating $\dot Q$ to $\dot E$ and $\dot L_z$,
will not hold.  More general prescriptions for evolving the Carter
constant will require calculation of an instantaneous radiation
reaction force.  Because the results presented here are constrained to
circular orbits, we cannot pretend that they are in any way generic.
However, they contain some very interesting features that may carry
over to more general --- eccentric and inclined --- orbits.  They also
should provide useful checks to future calculations that use a
radiation reaction force.

The calculations presented here indicates that $\dot\iota$ is
relatively small.  In particular, they show that the (dimensionless)
ratio $M\dot\iota/\dot r \ll 1$.  If this result holds in general, it
suggests an approximation in which one holds the inclination angle
fixed and allows the radius and eccentricity to radiatively evolve
{\cite{cutlercomment}}.  Such an approximation might give a first cut
of the trajectory a system follows through the phase space
$(r,e,\iota)$ of allowed orbits, perhaps serving as the first order
solution to an iterative scheme for finding such trajectories to
higher accuracy.  (It is worth noting, though, that preliminary
investigations of very strong-field orbits of rapidly rotating holes
indicate that the inclination angle changes rather more dramatically.
This result will be presented in a followup paper {\cite{paperII}}.)

The effect of spin on the gravitational waveforms (cf.\ the waves
plotted in Figs.\ {\ref{fig:wave_r7_a0.95}} and
{\ref{fig:wave_r7_a0.05}}, and their associated emission spectra,
Figs.\ {\ref{fig:dEinfdt_r7_a0.95_l2}} --
{\ref{fig:dEHdt_r7_a0.95_l2}} and {\ref{fig:dEinfdt_r7_a0.05_l2}} --
{\ref{fig:dEHdt_r7_a0.05_l2}}) is marked: large spin causes many
harmonics of the fundamental frequencies $\Omega_\phi$ and
$\Omega_\theta$ to influence the waveform.  This is reflected by the
relative breadth of emission spectra.  It seems likely that generic
orbits --- which will be further influenced by a third frequency,
$\Omega_r$ --- will have this property as well, and we thus expect to
see many harmonics of all three frequencies to influence the waveform
in the general case.  In terms of gravitational-wave observations,
this impact of the spin parameter could be nice: with such a marked
effect, the black hole's spin might be measurable to high accuracy.
On the other hand, because it has such a marked effect, data analysis
might require an unreasonably large number of waveform templates:
waveforms that are very ``interesting'' (in the sense of having a
rich, detailed structure) typically require many templates so that
signal-to-noise is not lost in measurement.  To understand how many
templates are needed would require constructing accurate radiation
reaction sequences through the phase space $(r,\iota)$ [$(r,e,\iota)$
in the general case], building the waveforms emitted along such
sequences, and then constructing the metric on the manifold of
waveform shapes as prescribed by Owen {\cite{owen}}.  If the number of
templates is very large, hierarchical search techniques will probably
be needed.  These analyses are beyond the scope of this paper
{\cite{nooverlaphere}}.

Finally, the two central approximations that go into the calculations
here --- adiabaticity and averaging over all members of the orbit
family --- could have a large effect on gravitational waveforms, and
hence on observations by space based detectors such as LISA.  First,
as noted in Eq.\ (\ref{eq:muoverM3}), the adiabatic assumption which
goes into these calculations requires that the mass ratio be rather
extreme in the strong field.  Astrophysical high-mass-ratio systems of
interest for gravitational-wave observations are likely to have $\mu/M
\sim 10^{-4} - 10^{-7}$.  This might not be extreme enough for
astrophysical systems to evolve adiabatically into the strong-field
regime.  However, it is impossible to relax the adiabatic assumption
without turning to an instantaneous radiation reaction force.  Second,
by averaging over all members of the orbit family, we wash away any
influence of initial conditions on the radiation reaction sequence ---
the position and momentum of the orbiting particle when observations
begin.  This averages the characteristics of the ``true'' waveform
over the time $T_{\rm return}$ it takes for the orbit to return (or at
the very least, come very close to) its initial conditions.  (``True''
waveform means the waveform constructed by evolving along a radiation
reaction sequence, not just the snapshots presented here at specific
moments on the sequence.)  For highly eccentric, inclined orbits, this
time could turn out to be very long.  If, in particular, it turns out
to be longer than the time it takes for radiation reaction to change
the system's orbital characteristics, such averaging would not be at
all accurate.  In this case, templates constructed for the analysis of
LISA-type gravitational-wave observations would require not only
information about the trajectory through the orbital phase space
$(r,e,\iota)$ but also about the initial conditions of all possible
orbits.  This could drastically increase the number of needed
templates.

In the end, we find that many of the questions raised here require the
instantaneous radiation reaction force $f^\mu_{\rm RR}$.  Although
``radiation reaction without radiation reaction force'' analyses such
as the one in this paper provide much insight and valuable information
about the waveforms emitted by high-mass-ratio systems, they are
essentially limited by the approximations that go into them, and
cannot answer some of the most important questions.  Because of the
eminent need, driven by future gravitational-wave observations, to
understand radiation reaction to very good accuracy in the
high-mass-ratio limit, programs to compute the radiation reaction
force should be given very high priority.

\acknowledgements

I am indebted to Daniel Kennefick for many useful conversations and
assistance, as well as for helping to validate my results with his
radiation reaction code; I likewise thank Sam Finn for providing data
from his code to help validate my results.  I am very grateful to Saul
Teukolsky for suggesting the spectral decomposition technique for
numerically calculating the spheroidal harmonics.  For many fruitful
conversations and advice, I particularly wish to thank those involved
in the Capra Ranch Radiation Reaction Analysis Program: Lior Burko,
Patrick Brady, \'Eanna Flanagan, Eric Poisson, and Alan Wiseman.
Finally for many useful conversations I thank Roger Blandford, Curt
Cutler, Steven Detweiler, Jeremy Heyl, Dustin Laurence, Yuri Levin,
Sterl Phinney, Masaru Shibata, Hideyuki Tagoshi, and Kip Thorne.  The
package {\sc Mathematica} was used to aid some of the calculations;
all of the numerical code was developed with tools from the Free
Software Foundation.  All plots were produced using the package SM.
This research was supported at Caltech by NSF Grants AST-9731698 and
AST-9618537, and NASA Grants NAG5-6840 and NAG5-7034; and at Illinois
by NSF Grant AST-9618524.

\appendix

\section{Calculation of spheroidal harmonics}
\label{app:spheroid}
\subsection{Spheroidal harmonics as a sum of spherical harmonics}
\label{subsec:spec_spheroid}

The separated $\theta$ dependence of the function $\psi_4$ is
governed by the equation
\begin{equation}
{1\over\sin\theta}{d\over d\theta}\left(\sin\theta
{d{_sS}^{a\omega}_{lm}\over d\theta}
\right) + \left[(a\omega)^2\cos^2\theta - 2 a\omega s\cos\theta -
\left({m^2 + 2 m s \cos\theta + s^2\over\sin^2\theta}\right) +
{\cal E}_{lm}\right] {_sS}^{a\omega}_{lm} = 0\;.
\label{eq:spheroideqn}
\end{equation}
For our purposes, we care only about the case $s=-2$.  Solutions to
this equation are the spin-weighted spheroidal harmonics.  When
$a\omega=0$, the solutions are the spin-weighted spherical harmonics;
in this case, ${\cal E}_{lm} = l(l+1)$.

This fact suggests that it may be useful to expand the spheroidal
harmonics in spherical harmonics.  This spectral decomposition takes
the following form:
\begin{equation}
_sS^{a\omega}_{lm}(\theta) = \sum_{j=l_{\rm min}}^\infty
b^{a\omega}_j {_sY}_{jm}(\theta)
\label{eq:spheroidexpand}
\end{equation}
Here, it is to be understood that $_sY_{lm}(\theta)$ denotes the
spin-weighted spherical harmonics without including the $\phi$
dependence: $_sY_{lm}^{\rm usual}(\theta,\phi) =
{_sY_{lm}}(\theta)e^{im\phi}$.  Also, $l_{\rm min} = \max(|s|,|m|)$.

It is convenient at this point to adopt Dirac-style notation, so that
\begin{eqnarray}
_sY_{jm}(\theta) \to |sjm\rangle\;,\qquad
_s{\bar Y}_{jm}(\theta) &\to& \langle sjm|\;,\nonumber\\
\int_0^\pi {_s}{\bar Y}_{lm}(\theta)\,f(\theta)\,_sY_{jm}(\theta)\,
\sin\theta\,d\theta &\to& \langle slm | f(\theta) | sjm \rangle\;.
\label{eq:diracconvention}
\end{eqnarray}
Substitute Eq.\ (\ref{eq:spheroidexpand}) into Eq.\
(\ref{eq:spheroideqn}), and use the fact that the functions $_sY_{jm}$
satisfy (\ref{eq:spheroideqn}) with $a\omega=0$ and ${\cal E}_{jm}=
j(j+1)$.  The result, in Dirac notation, is
\begin{equation}
\sum_{j=l_{\rm min}}^\infty b^{a\omega}_j
\left[(a\omega)^2\cos^2\theta - 2 a\omega s \cos\theta - j(j+1)\right]
|sjm\rangle = -{\cal E}_{lm}\sum_{j=l_{\rm min}}^\infty
b^{a\omega}_j |sjm\rangle\;.
\end{equation}
Next, multiply the above expression by $\langle slm |$.  The
various inner products are simply evaluated \cite{innerproducts}:
\begin{eqnarray}
\langle slm | \cos^2\theta | sjm\rangle &=& {1\over3}\delta_{jl} +
{2\over3}\sqrt{{2l+1\over2j+1}}\langle j,2,m,0 | l,m\rangle
\langle j,2,-s,0 | l,-s \rangle \equiv c^m_{j,l,2}\;,\nonumber\\
\langle slm | \cos\theta | sjm\rangle &=&
\sqrt{{2l+1\over2j+1}}\langle j,1,m,0 | l,m\rangle
\langle j,1,-s,0 | l,-s \rangle \equiv c^m_{j,l,1}\;,\nonumber\\
\langle slm | sjm \rangle &=& \delta_{jl}\;.
\end{eqnarray}
The numbers $\langle j,i,m,0 | l,n \rangle$ are Clebsch-Gordan
coefficients.  The fact that Clebsch-Gordan coefficients appear in
this expression greatly simplifies the sums: it tells us that
$c^m_{j,l,2} \ne 0$ only for $j\in [l-2,l-1,l,l+1,l+2]$, and
$c^m_{j,l,1} \ne 0$ only for $j\in [l-1,l,l+1]$.  Performing the
sums, we find
\begin{eqnarray}
b^{a\omega}_{l-2} (a\omega)^2 c^m_{l-2,l,2} &+&
b^{a\omega}_{l-1} \left[(a\omega)^2 c^m_{l-1,l,2} -
2 a\omega s\, c^m_{l-1,l,1}\right] +
b^{a\omega}_l \left[(a\omega)^2 c^m_{l,l,2} - 2 a\omega s\,
c^m_{l,l,1} - l(l+1)\right]\nonumber\\ &+&
b^{a\omega}_{l+1}
\left[(a\omega)^2 c^m_{l+1,l,2} - 2 a\omega s\, c^m_{l+1,l,1}\right] +
b^{a\omega}_{l+2} (a\omega)^2 c^m_{l+2,l,2} = -{\cal E}_{lm}
b^{a\omega}_l\;.
\label{eq:spheroidmatrix}
\end{eqnarray}
Equation (\ref{eq:spheroidmatrix}) can be rewritten as a matrix
equation, with the numbers $b^{a\omega}_l$ representing the
coefficients of the matrix's eigenvector, and ${\cal E}_{lm}$ the
matrix's eigenvalue.  The matrix so defined is clearly band-diagonal,
which means that solving for the eigenvalues and eigenvectors is
rather simple.  To do so, we used routines from {\cite{numrec}}.

\subsection{Numerical calculation of spin-weighted spherical harmonics}
\label{subsec:spherical}

The above section describes how to express the spin-weighted
spheroidal harmonics as a sum of spin-weighted spherical harmonics;
there remains the task of actually calculating the spherical
harmonics.  A favored method for stably and accurately calculating
spherical harmonics of spin-weight zero is with a recurrence relation
{\cite{numrec}}.  It is relatively simple to generalize such relations
to stably and accurately calculate spherical harmonics of non-zero
spin weight.  Our discussion here will be relevant to calculation of
spin weights $0$, $-1$, and $-2$; generalization to other spin weights
is straightforward.  See {\cite{goldberg}} for further discussion.

To begin, we note that the operator $\edth$ lowers the spin weight
of a function as follows:
\begin{eqnarray}
\edth {_sY}_{lm}(\theta) &\equiv&
-(\sin\theta)^{-s}\left[{\partial\over\partial\theta} +
{m\over\sin\theta}\right](\sin\theta)^s{_sY}_{lm}(\theta)
\nonumber\\ &=& -[(l+s)(l-s+1)]^{1/2} {_{(s-1)}Y}_{lm}(\theta)\;.
\label{eq:edthoperate}
\end{eqnarray}
The plan is to apply $\edth$ to ${_0Y}_{lm}$ and derive formulae for
the ${_{-1}Y}_{lm}$, and then to apply $\edth$ to ${_{-1}Y}_{lm}$ and
derive formulae for the ${_{-2}Y}_{lm}$.  Note also that there is
an operator $\antiedth$ which raises the spin weight of a function:
\begin{eqnarray}
\antiedth {_sY}_{lm}(\theta) &\equiv&
-(\sin\theta)^{s}\left[{\partial\over\partial\theta} -
{m\over\sin\theta}\right](\sin\theta)^{-s}{_sY}_{lm}(\theta)
\nonumber\\
&=& [(l-s)(l+s+1)]^{1/2} {_{(s+1)}Y}_{lm}(\theta)\;.
\label{eq:antiedthoperate}
\end{eqnarray}
This operator will come in handy when computing derivatives of the
spheroidal harmonics in Sec.\ {\ref{subsec:spheroidalderivs}} below.

The zero-weight spherical harmonics are written
\begin{equation}
{_0Y}_{lm}(\theta) = A(l,m) P_{lm}(\cos\theta)\;,
\label{eq:sphereweight0}
\end{equation}
where
\begin{equation}
A(l,m) = \sqrt{(2l+1)\over4\pi}\sqrt{(l-m)!\over(l+m)!}\;,
\label{eq:coeff_Alm}
\end{equation}
and the associated Legendre polynomial $P_{lm}(x)$ can be
accurately computed from the recurrence relation {\cite{numrec}}
\begin{equation}
P_{lm}(x) = {1\over l-m}\left[x(2l-1)P_{l-1,m} - (l+m-1)P_{l-2,m}\right]
\label{eq:plgndr}
\end{equation}
with starting values
\begin{eqnarray}
P_{mm}(x) &=& (-1)^m (2m-1)!! (1-x^2)^{m/2}\;,\nonumber\\
P_{m+1,m}(x) &=& x(2m+1) P_{mm}\;.
\label{eq:plgndr2}
\end{eqnarray}

Now, operate on ${_0Y}_{lm}$ with $\edth$.  Combining Eqs.\
(\ref{eq:edthoperate}) and (\ref{eq:sphereweight0}), and using the
notation $x\equiv\cos\theta$, we see that
\begin{equation}
{_{-1}Y}_{lm}(\theta) = -{A(l,m)\over\sqrt{l(l+1)}}
\left[\sqrt{1-x^2}\,{d\over dx} -
{m\over\sqrt{1-x^2}}\right]P_{lm}(x)\;.
\label{eq:sphereweightn1}
\end{equation}
We now need recurrence relations for the functions $\sqrt{1-x^2}\,
dP_{lm}/dx$ and $P_{lm}/\sqrt{1-x^2}$.  (We treat these combinations
as functions in and of themselves; this avoids problems as
$x\to\pm1$.)  These relations are easily derived from Eqs.\
(\ref{eq:plgndr}) and (\ref{eq:plgndr2}):
\begin{eqnarray}
{P_{lm}(x)\over\sqrt{1-x^2}}
&=& {1\over l-m}\left[x(2l-1){P_{l-1,m}\over\sqrt{1-x^2}}
- (l+m-1){P_{l-2,m}\over\sqrt{1-x^2}}\right]\,,\nonumber\\
{P_{mm}(x)\over\sqrt{1-x^2}} &=&
(-1)^m (2m-1)!! (1-x^2)^{(m-1)/2}\;,\nonumber\\
{P_{m+1,m}(x)\over\sqrt{1-x^2}} &=& x(2m+1)
{P_{mm}\over\sqrt{1-x^2}}\;;
\label{eq:plgndroversin}
\end{eqnarray}
\begin{eqnarray}
\sqrt{1-x^2}\,{dP_{lm}\over dx}
&=& {1\over l-m}\left[(2l-1)\left(\sqrt{1-x^2}\,P_{l-1,m} +
x\sqrt{1-x^2}\,{dP_{l-1,m}\over dx}\right)
- (l+m-1)\sqrt{1-x^2}\,{dP_{l-2,m}\over dx}\right]\,,\nonumber\\
\sqrt{1-x^2}\,{dP_{mm}\over dx} &=&
-mx(-1)^m (2m-1)!! (1-x^2)^{(m-1)/2}\;,\nonumber\\
\sqrt{1-x^2}\,{dP_{m+1,m}\over dx} &=& (2m+1)
\left[\sqrt{1-x^2}\,P_{mm} + x\sqrt{1-x^2}\,{dP_{mm}\over dx}\right]\;.
\label{eq:sindplgndr}
\end{eqnarray}

Next, operate on ${_{-1}Y}_{lm}$ with $\edth$.  Combine Eqs.\
(\ref{eq:edthoperate}) and (\ref{eq:sphereweightn1}) to obtain
\begin{equation}
{_{-2}Y}_{lm}(\theta) = {A(l,m)\over\sqrt{(l-1)l(l+1)(l+2)}}
\left[(1-x^2){d^2\over dx^2} - 2m {d\over dx} +
{m^2 - 2mx\over1-x^2}\right]P_{lm}(x)\;.
\label{eq:sphereweightn2}
\end{equation}
For these harmonics, we need recurrence relations for
$P_{lm}/(1-x^2)$, $dP_{lm}/dx$, and $(1-x^2)d^2P_{lm}/dx$.  Again,
these are straightforwardly derived from Eqs.\ (\ref{eq:plgndr}) and
(\ref{eq:plgndr2}):
\begin{eqnarray}
{P_{lm}(x)\over(1-x^2)}
&=& {1\over l-m}\left[x(2l-1){P_{l-1,m}\over(1-x^2)}
- (l+m-1){P_{l-2,m}\over(1-x^2)}\right]\,,\nonumber\\
{P_{mm}(x)\over(1-x^2)} &=&
(-1)^m (2m-1)!! (1-x^2)^{(m-2)/2}\;,\nonumber\\
{P_{m+1,m}(x)\over(1-x^2)} &=& x(2m+1)
{P_{mm}\over(1-x^2)}\;;
\label{eq:plgndroversinsqr}
\end{eqnarray}
\begin{eqnarray}
{dP_{lm}\over dx}
&=& {1\over l-m}\left[(2l-1)\left(P_{l-1,m} +
x{dP_{l-1,m}\over dx}\right)
- (l+m-1){dP_{l-2,m}\over dx}\right]\,,\nonumber\\
{dP_{mm}\over dx} &=&
-mx(-1)^m (2m-1)!! (1-x^2)^{(m-2)/2}\;,\nonumber\\
{dP_{m+1,m}\over dx} &=& (2m+1)
\left[P_{mm} + x{dP_{mm}\over dx}\right]\;;
\label{eq:dplgndr}
\end{eqnarray}
\begin{eqnarray}
(1-x^2)\,{d^2P_{lm}\over dx^2}
&=& {1\over l-m}\left[(2l-1)\left(2(1-x^2)\,{dP_{l-1,m}\over dx} +
x(1-x^2)\,{d^2P_{l-1,m}\over dx^2}\right)
- (l+m-1)(1-x^2)\,{d^2P_{l-2,m}\over dx^2}\right]\,,\nonumber\\
(1-x^2)\,{d^2P_{mm}\over dx^2} &=&
m(-1)^m (2m-1)!! (1-x^2)^{(m-2)/2}\left[x^2(m-2) -
(1-x^2)\right]\;,\nonumber\\
(1-x^2)\,{d^2P_{m+1,m}\over dx^2} &=& (2m+1)
\left[2(1-x^2)\,{dP_{mm}\over dx} +
x(1-x^2)\,{d^2P_{mm}\over dx^2}\right]\;.
\label{eq:sinsqrddplgndr}
\end{eqnarray}

This procedure can be continued as far as one's stamina allows; we
stop here since these functions are all that is needed for this paper.

\subsection{Some derivatives of the spheroidal harmonics}
\label{subsec:spheroidalderivs}

In the source term ${\cal T}_{lm\omega}$, we encounter the terms
$L_2^\dag S$ and $L_1^\dag L_2^\dag S$ [where $S$ is shorthand for
$_{-2}S^{a\omega_{mk}}_{lm}(\theta)$].  Using the spectral
decomposition described above, evaluating these derivatives is quite
straightforward.

Consider first $L_2^\dag S$.  Using the spectral decomposition, this
becomes
\begin{eqnarray}
L_2^\dag S &=& \sum_{k=l_{\rm min}}^\infty b_k L_2^\dag
{_{-2}}Y_{km}(\theta)\nonumber\\
&=& \sum_{k=l_{\rm min}}^\infty
b_k\left[\partial_\theta - {m\over\sin\theta} + 2\cot\theta\right]
{_{-2}}Y_{km}(\theta) + a\omega\sin\theta
\sum_{k=l_{\rm min}}^\infty b_k {_{-2}}Y_{km}(\theta)\;.
\label{eq:l2dag1}
\end{eqnarray}
The operator that appears in the first term on the second line is
$-\antiedth$ [cf.\ Eq.\ (\ref{eq:antiedthoperate})] for $s = -2$.
Thus,
\begin{equation}
L_2^\dag S = a\omega\sin\theta S -\sum_{k=l_{\rm min}}^\infty b_k
{\left[(k-1)(k+2)\right]}^{1/2}{_{-1}}Y_{km}(\theta)\;.
\label{eq:l2dagonS}
\end{equation}

Next consider $L_1^\dag L_2^\dag S$:
\begin{eqnarray}
L_1^\dag L_2^\dag S &=&
 a\omega L_1^\dag\sin\theta S -\sum_{k=l_{\rm min}}^\infty b_k
{\left[(k-1)(k+2)\right]}^{1/2}L_1^\dag{_{-1}}Y_{km}(\theta)\;,
\nonumber\\
&=& a\omega
\left[{\partial\over\partial\theta} - {m\over\sin\theta}
+\cot\theta + a\omega\sin\theta \right]
\sin\theta S\nonumber\\
 & &-\sum_{k=l_{\rm min}}^\infty b_k
{\left[(k-1)(k+2)\right]}^{1/2}
\left[{\partial\over\partial\theta} - {m\over\sin\theta}
+\cot\theta + a\omega\sin\theta\right]
{_{-1}}Y_{km}(\theta)\;,\nonumber\\
&=& a\omega
\left\{\sin^{-1}\theta\left[{\partial\over\partial\theta} -
{m\over\sin\theta} + a\omega\sin\theta \right]\sin\theta\right\}
\sin\theta S\nonumber\\
& & -\sum_{k=l_{\rm min}}^\infty b_k
{\left[(k-1)(k+2)\right]}^{1/2}
\left(\antiedth + a\omega \sin\theta\right)
{_{-1}}Y_{km}(\theta)\;,\nonumber\\
&=& a\omega\sin\theta
\left\{\sin^{-2}\theta\left[{\partial\over\partial\theta} -
{m\over\sin\theta} + a\omega\sin\theta \right]\sin^2\theta\right\}
S\nonumber\\
& & +\sum_{k=l_{\rm min}}^\infty b_k
{\left[(k-1)k(k+1)(k+2)\right]}^{1/2}
{_{0}}Y_{km}(\theta)
 -a\omega\sin\theta\sum_{k=l_{\rm min}}^\infty b_k
{\left[(k-1)(k+2)\right]}^{1/2}
{_{-1}}Y_{km}(\theta)\;,\nonumber\\
&=& a\omega\sin\theta
\left[{\partial\over\partial\theta} -
{m\over\sin\theta} + a\omega\sin\theta + 2\cot\theta\right]
S\nonumber\\
& & +\sum_{k=l_{\rm min}}^\infty b_k
{\left[(k-1)k(k+1)(k+2)\right]}^{1/2}
{_{0}}Y_{km}(\theta)
 +a\omega\sin\theta L_2^\dag S - (a\omega\sin\theta)^2 S\;.
\label{eq:l1dagl2dag1}
\end{eqnarray}
Thus, we finally end up with
\begin{equation}
L_1^\dag L_2^\dag S = \sum_{k=l_{\rm min}}^\infty b_k
{\left[(k-1)k(k+1)(k+2)\right]}^{1/2} {_{0}}Y_{km}(\theta)
+2a\omega\sin\theta L_2^\dag S - (a\omega\sin\theta)^2 S\;.
\label{eq:l1dagl2dagonS}
\end{equation}
Eqs.\ (\ref{eq:l2dagonS}) and (\ref{eq:l1dagl2dagonS}) are then
used in the source term evaluation.

\section{Functions that appear in the Sasaki-Nakamura equation}
\label{app:sasaknak}

The function $F(r)$ that appears in Eq.\ (\ref{eq:sasakinakamura}) is
given by
\begin{equation}
F(r) = {d\eta/dr\over\eta}{\Delta\over r^2 + a^2}\;,
\end{equation}
where
\begin{equation}
\eta(r) = c_0 + c_1/r + c_2/r^2 + c_3/r^3 + c_4/r^4\;,
\end{equation}
and
\begin{eqnarray}
c_0 &=& -12 i \omega M + \lambda(\lambda+2) -
12a\omega(a\omega - m)\;,\nonumber\\
c_1 &=& 8ia[3a\omega - \lambda(a\omega-m)]\;,\nonumber\\
c_2 &=& -24 i a M(a\omega - m) + 12 a^2[1 -
2(a\omega-m)^2]\;,\nonumber\\
c_3 &=& 24 i a^3(a\omega-m) - 24 Ma^2\;,\nonumber\\
c_4 &=& 12 a^4\;.
\end{eqnarray}

The function $U(r)$ that appears in Eq.\ (\ref{eq:sasakinakamura}) is
given by
\begin{equation}
U(r) = {\Delta U_1(r)\over(r^2+a^2)^2} + G(r)^2 +
{\Delta dG/dr\over r^2 + a^2} - F(r) G(r)\;,
\end{equation}
where
\begin{eqnarray}
G(r) &=& -{2(r-M)\over{r^2+a^2}} + {r\Delta\over(r^2+a^2)^2}\;,
\nonumber\\
U_1(r) &=& V(r) + {\Delta^2\over\beta}\left[{d\over
dr}\left(2\alpha + {d\beta/dr\over\Delta}\right) - {d\eta/dr\over\eta}
\left(\alpha + {d\beta/dr\over\Delta}\right)\right]\;,\nonumber\\
\alpha &=& -iK(r)\beta/\Delta^2 + 3i dK/dr + \lambda + 6\Delta/r^2\;,
\nonumber\\
\beta &=& 2\Delta[-iK(r) + r - M - 2\Delta/r]\;.
\label{eq:Ufunctions}
\end{eqnarray}
The functions $K(r)$ and $V(r)$ are from the Teukolsky potential, Eq.\
(\ref{eq:teukpotential}).

\newpage


\begin{table}
\caption
{Comparison of certain orbital quantities computed numerically and
computed using post-Newtonian expansions for orbits with $r = 7 M$, $a
= 0.95 M$, $\iota = 62.43^\circ$.  In this strong-field region, we do
not expect post-Newtonian expressions to be particularly valid.
Post-Newtonian theory underestimates $\dot r$ by about 2, and
overestimates $\dot\iota$ by about 3.}
\begin{tabular}{ccc}
Orbital quantity & Numerical value & Post-Newtonian value \\
\tableline
$(M/\mu)^2 {\dot E}$ & $-3.3885 \times 10^{-4}$
& $-3.2580 \times 10^{-4}$ \\
\tableline
$(M/\mu^2){\dot L_z}$ & $-3.3207 \times 10^{-3}$
& $-3.5926 \times 10^{-3}$\\
\tableline
$(M/\mu) \dot r$ & $-4.6574 \times 10^{-2}$
& $-2.7499 \times 10^{-2} $\\
\tableline
$(M^2/\mu) \dot\iota$ & $1.2073\times 10^{-4}$
& $3.0806 \times 10^{-4}$
\label{tab:r7_a0.95_rrnumbers}
\end{tabular}
\end{table}

\begin{table}
\caption
{Comparison of certain orbital quantities computed numerically and
computed using post-Newtonian expansions for orbits with $r = 7 M$, $a
= 0.05 M$, $\iota = 60.17^\circ$.  As in the $a = 0.95 M$ case, we do
not expect post-Newtonian expressions to work well here in the
strong-field.  Post-Newtonian theory underestimates $\dot r$ by about
3, and overestimates $\dot\iota$ by about $50\%$.}
\begin{tabular}{ccc}
Orbital quantity & Numerical value & Post-Newtonian value \\
\tableline
$(M/\mu)^2 {\dot E}$ & $-3.9524 \times 10^{-4}$
& $-3.7768 \times 10^{-4}$ \\
\tableline
$(M/\mu^2){\dot L_z}$ & $-3.6794 \times 10^{-3}$
& $-3.5210 \times 10^{-3}$\\
\tableline
$(M/\mu) \dot r$ & $-1.0964 \times 10^{-1}$
& $-3.6762 \times 10^{-2} $\\
\tableline
$(M^2/\mu) \dot\iota$ & $1.0875\times 10^{-5}$
& $1.5867 \times 10^{-5}$
\label{tab:r7_a0.05_rrnumbers}
\end{tabular}
\end{table}

\begin{table}
\caption
{Comparison of certain orbital quantities computed numerically and
computed using post-Newtonian expansions for orbits with $r = 100 M$,
$a = 0.95 M$, $\iota = 60.05^\circ$.  In this relatively weak-field
regime, we expect to see much better agreement with post-Newtonian
results.  The numbers in this table bear out this expectation.}
\begin{tabular}{ccc}
Orbital quantity & Numerical value & Post-Newtonian value \\
\tableline
$(M/\mu)^2 {\dot E}$ & $-6.2194 \times 10^{-10}$
& $-6.3815 \times 10^{-10}$ \\
\tableline
$(M/\mu^2){\dot L_z}$ & $-3.1175 \times 10^{-7}$
& $-3.1995 \times 10^{-7}$ \\
\tableline
$(M/\mu) \dot r$ & $-1.2610 \times 10^{-5}$
& $-1.2733 \times 10^{-5}$ \\
\tableline
$(M^2/\mu) \dot\iota$ & $1.2040\times 10^{-10}$
& $1.3389 \times 10^{-10}$
\label{tab:r100_a0.95_rrnumbers}
\end{tabular}
\end{table}

\begin{table}
\caption
{Comparison of certain orbital quantities computed numerically and
computed using post-Newtonian expansions for orbits with $r = 100 M$,
$a = 0.05 M$, $\iota = 60.00^\circ$.  As with the orbit at $r = 100M$,
$a = 0.95 M$, we find much better agreement with post-Newtonian
results.}
\begin{tabular}{ccc}
Orbital quantity & Numerical value & Post-Newtonian value \\
\tableline
$(M/\mu)^2 {\dot E}$ & $-6.2372 \times 10^{-10}$
& $-6.3990 \times 10^{-10}$ \\
\tableline
$(M/\mu^2){\dot L_z}$ & $-3.1191 \times 10^{-7}$
& $-3.2000 \times 10^{-7}$ \\
\tableline
$(M/\mu) \dot r$ & $-1.2676 \times 10^{-5}$
& $-1.2797 \times 10^{-5}$ \\
\tableline
$(M^2/\mu) \dot\iota$ & $6.6936\times 10^{-12}$
& $7.0439 \times 10^{-12}$
\label{tab:r100_a0.05_rrnumbers}
\end{tabular}
\end{table}


\begin{figure}[ht]
\begin{center}
\epsfig{file = 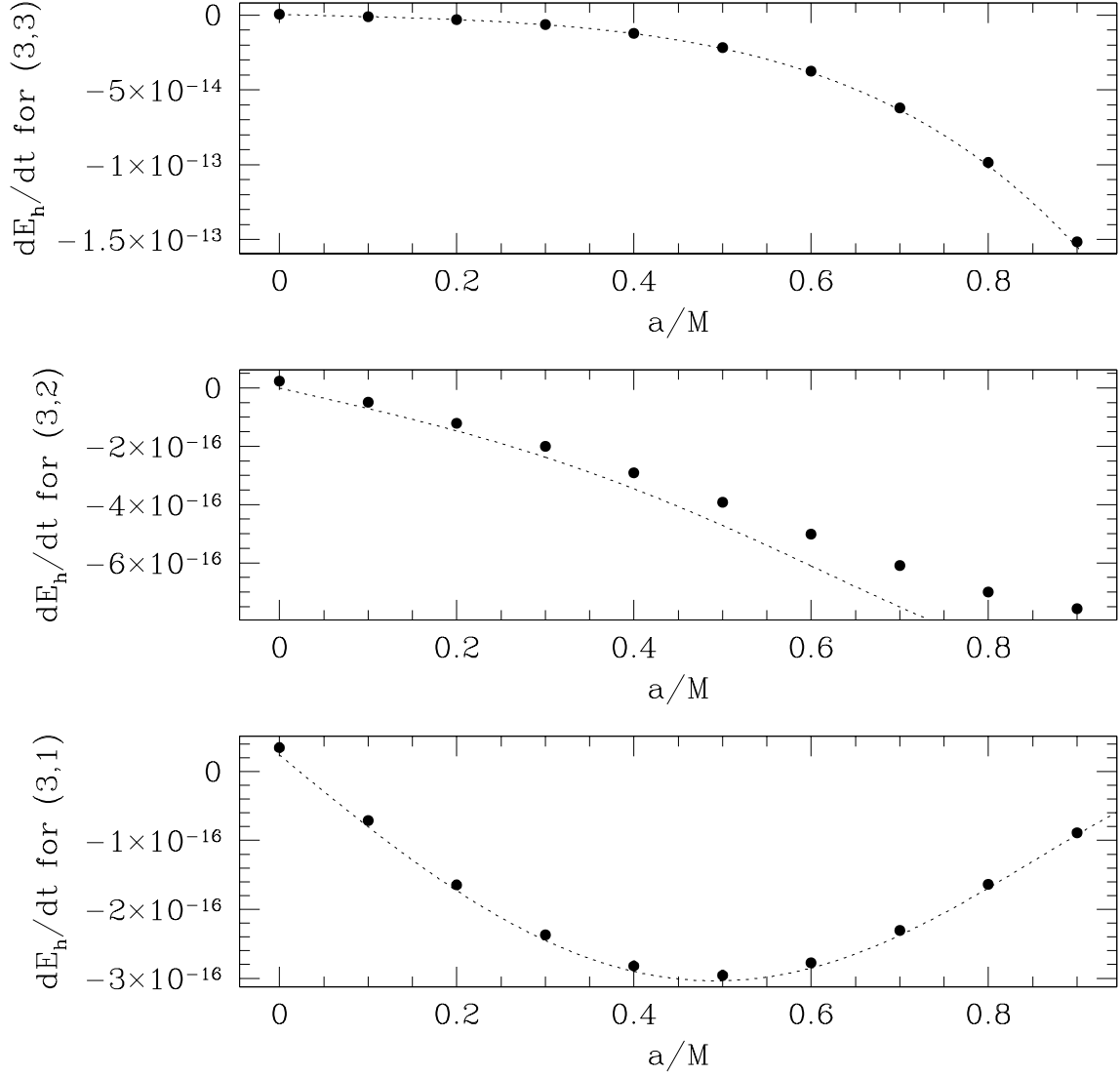, width = 16cm}
\caption{\label{fig:fluxdownholel=3}
Comparison of the flux down the event horizon for orbits at $r = 25 M$
as a function of black hole spin $a/M$ for $l = 3$ modes.  Agreement
between the numerical and post-Newtonian fluxes is quite good, except
for $m=2$; this is because the post-Newtonian expansions contain many
terms, and usually are quite robust.  The case $m=2$ is an example
where the expansion is not as robust.  The interesting upturn in the
down-horizon flux for $m = 1$ and $a \ge 0.5 M$ is due to superradiant
scattering --- some incoming radiation gets scattered by the black
hole's ergosphere out to infinity.}
\end{center}
\vskip -0.5cm
\end{figure}

\begin{figure}[ht]
\begin{center}
\epsfig{file = 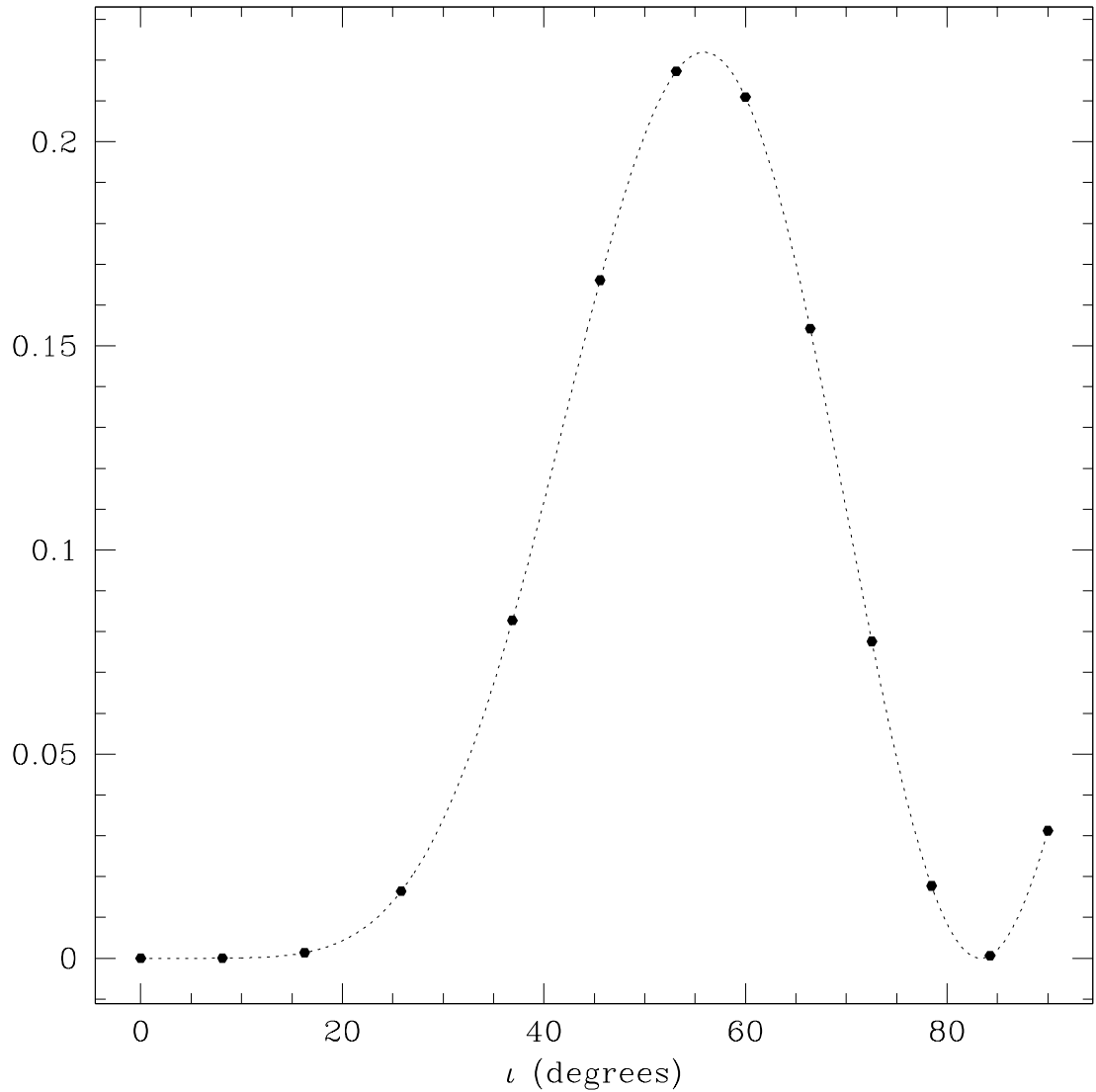, width = 16cm}
\caption{\label{fig:compareinclined}
Typical result of validation test for inclined, Schwarzschild orbits.
The dotted line is the modulus squared of the Wigner D-function for $l
= 4$, $m = 2$, $k = 1$ as a function of inclination angle $\iota$; the
large black points are the ratio ${\dot E}_{lmk}(\iota)/{\dot E}^{\rm
eq}_{lm}$.  The numerical data for the fluxes agrees with the
analytical formula for the D-function to within $10^{-6} - 10^{-7}$.
(For this plot, the numerical fluxes are evaluated at infinity; the
results are identical to within the error when examining fluxes down
the horizon.  Also, these results are invariant --- within the error
bounds --- as a function of orbital radius; this plot is generated for
a particle orbiting at $r = 15 M$.)}
\end{center}
\vskip -0.5cm
\end{figure}

\begin{figure}[ht]
\begin{center}
\epsfig{file = 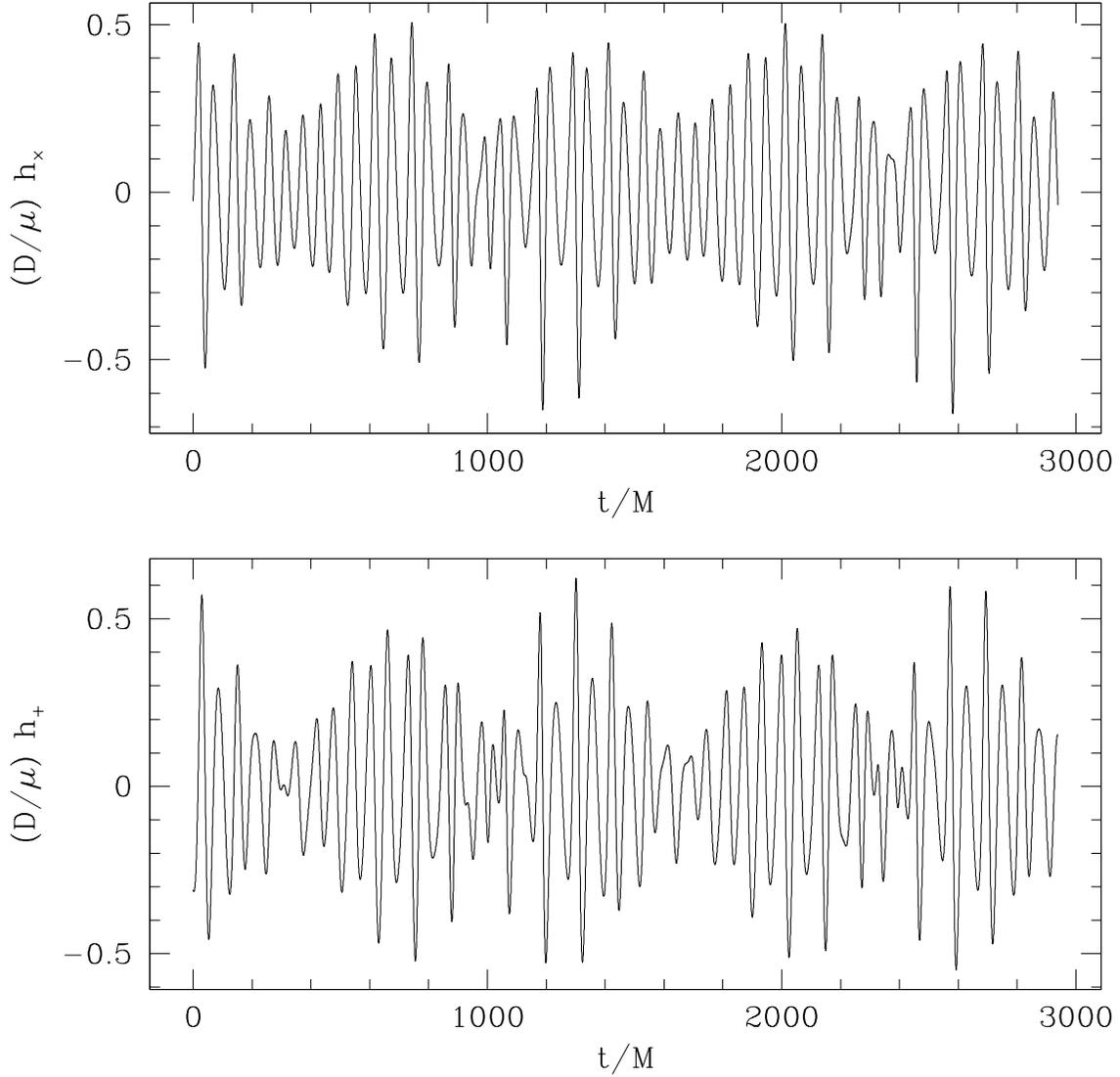, width = 16cm}
\caption{\label{fig:wave_r7_a0.95} The gravitational waveform produced
by orbits with $r = 7 M$, $\iota = 62.43^\circ$ about a black hole
with $a = 0.95 M$.  The observer is in the hole's equatorial plane,
$\theta = 90^\circ$.  The distance to the source is $D$.  Notice that
there are many sharp features in this waveform, indicating the strong
presence of relatively large harmonics of the fundamental frequencies
$\Omega_\phi$ and $\Omega_\theta$.  This is consistent with the rather
broad emission spectra produced by this orbit (cf.\ Fig.\
{\ref{fig:dEinfdt_r7_a0.95_l2}}).  The low frequency modulation is due
to Lense-Thirring precession ({\it i.e.}, the precession of the
orbital plane due to dragging of inertial frames by the black hole's
spin).}
\end{center}
\vskip -0.5cm
\end{figure}

\begin{figure}[ht]
\begin{center}
\epsfig{file = 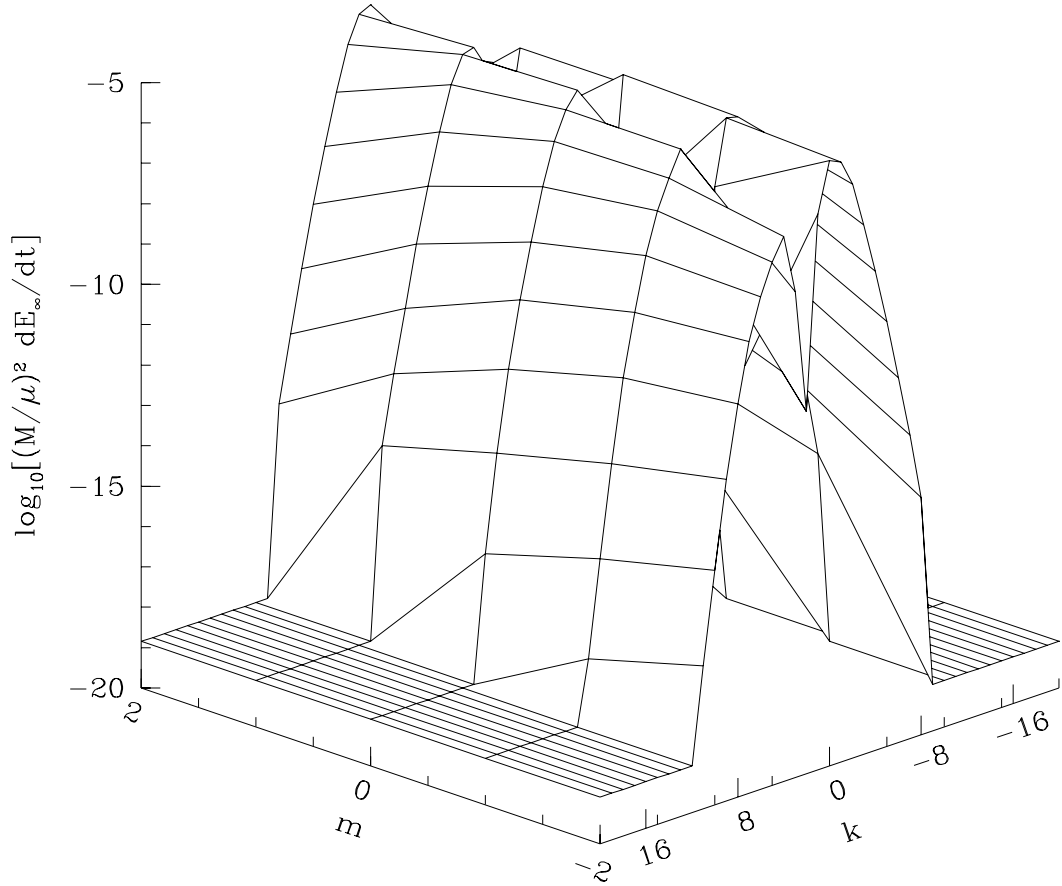, width = 16 cm}
\caption{\label{fig:dEinfdt_r7_a0.95_l2}
The spectrum of energy for $l = 2$ modes radiated to infinity by
orbits with $r = 7 M$, $\iota = 62.43^\circ$ about a black hole with
$a = 0.95 M$.  Of particular note in this case is that the
distribution is rather broad with respect to $k$.  This is primarily
due to the fact that for very large spin, the Teukolsky potential
[cf.\ Eq.\ ({\ref{eq:teukpotential}})] is fairly transmissive to high
frequency modes.  Notice that it is more transmissive to corotating
modes ($m \omega > 0$) than it is to counterrotating modes: corotating
modes are more readily scattered by the hole.}
\end{center}
\vskip -0.5cm
\end{figure}

\begin{figure}[ht]
\begin{center}
\epsfig{file = 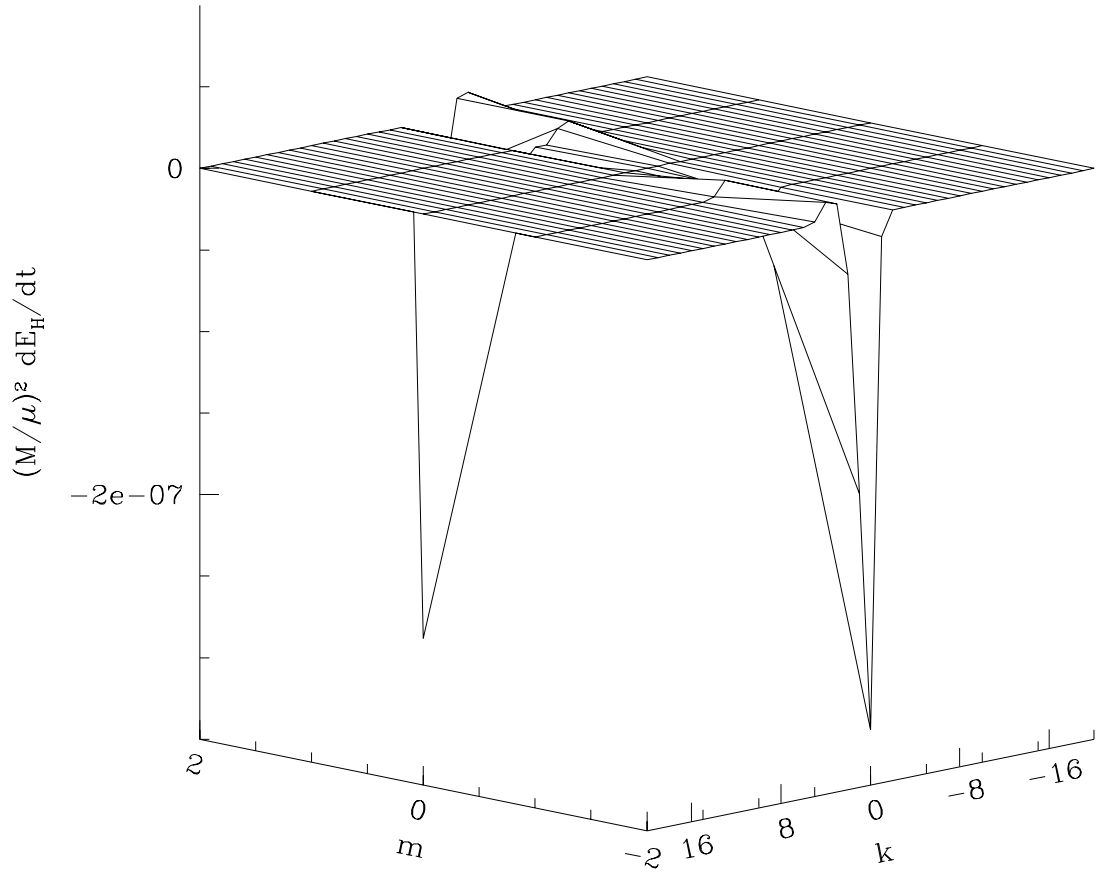, width = 16 cm}
\caption{\label{fig:dEHdt_r7_a0.95_l2}
The spectrum of energy radiated down the event horizon for $l = 2$
modes for orbits with $r = 7 M$, $\iota = 62.43^\circ$ about a black
hole with $a = 0.95 M$.  The distribution is, for the most part,
sharply negative (particularly for corotating modes, $m\omega > 0$),
indicating superradiant scattering.  This is essentially a
manifestation of the Penrose process --- radiation extracts energy
from the black hole's ergosphere.}
\end{center}
\vskip -0.5cm
\end{figure}

\begin{figure}[ht]
\begin{center}
\epsfig{file = 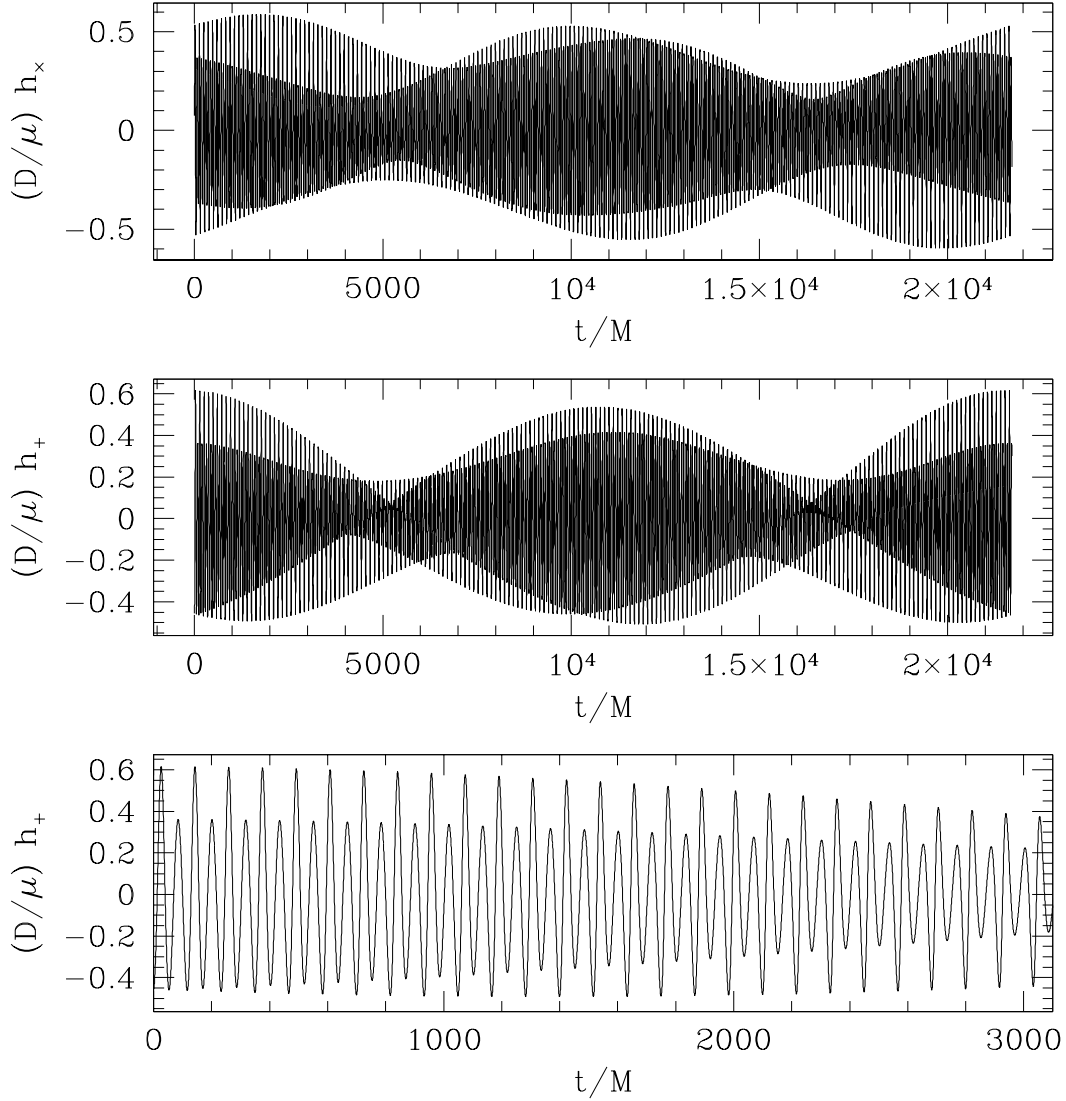, width = 16cm}
\caption{\label{fig:wave_r7_a0.05}
The gravitational waveform produced by orbits with $r = 7 M$, $\iota =
60.14^\circ$ about a black hole with $a = 0.05 M$.  The observer is in
the hole's equatorial plane, $\theta = 90^\circ$.  Although not much
detail regarding the waveform is visible in this figure, the low
frequency modulation is markedly slower than in the case $a = 0.95$
(cf.\ Fig.\ {\ref{fig:wave_r7_a0.95}}).  This is not surprising: the
Lense-Thirring precession frequency is much smaller in this case since
the spin is so low.  The lowest panel is a zoom on $h_+$.  The time
shown is chosen so that comparison can be easily made with the
waveform for $a = 0.95 M$, Fig.\ {\ref{fig:wave_r7_a0.95}}.  In
contrast to the $a = 0.95 M$ case, the waveform is quite a bit
simpler, lacking the many sharp features seen when there is large
spin.  This is primarily because the Teukolsky potential
(\ref{eq:teukpotential}) is not nearly as transmissive to
high-frequency modes when $a$ is small.}
\end{center}
\vskip -0.5cm
\end{figure}

\begin{figure}[ht]
\begin{center}
\epsfig{file = 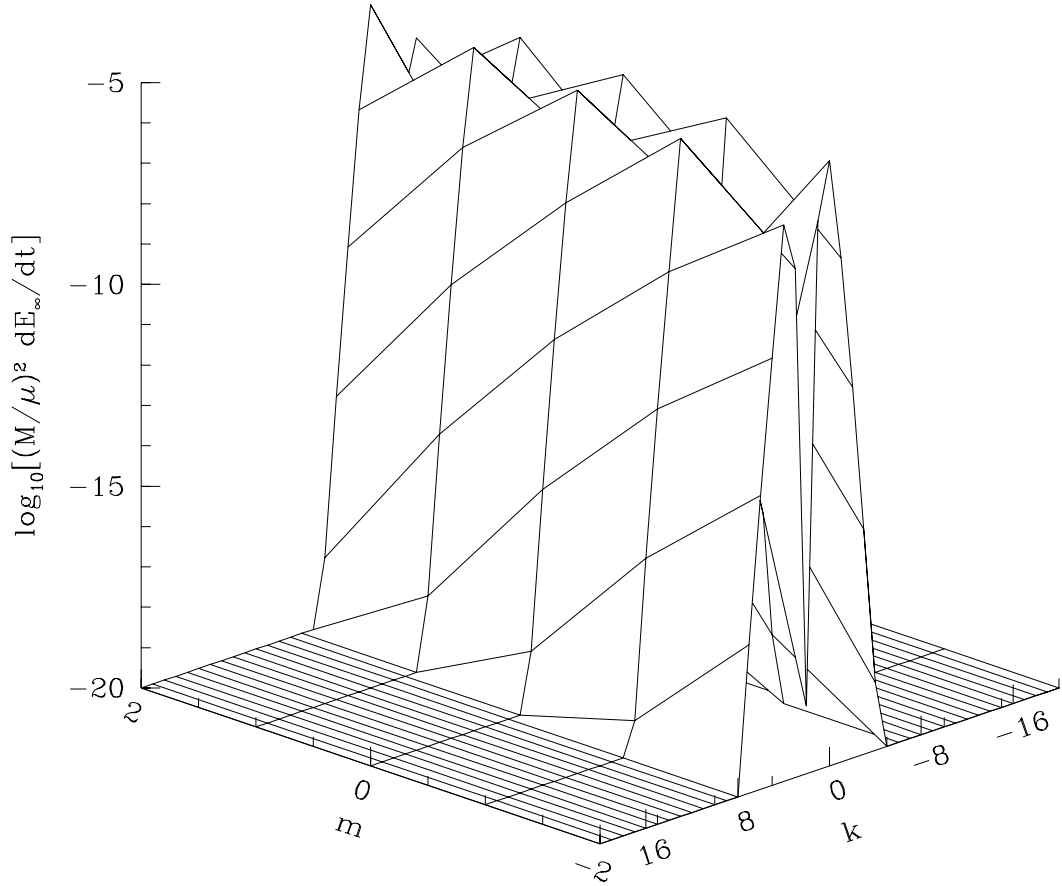, width = 16cm}
\caption{\label{fig:dEinfdt_r7_a0.05_l2}
The spectrum of energy for $l = 2$ modes radiated to infinity by
orbits with $r = 7 M$, $\iota = 60.14^\circ$ about a black hole with
$a = 0.05 M$.  The distribution is fairly narrow with respect to $k$,
particularly when compared with the distribution for $a = 0.95 M$
(Fig.\ {\ref{fig:dEinfdt_r7_a0.95_l2}}).  This is because the
Teukolsky potential is not very transmissive to high-frequency modes
for small $a$.  Note, though, that the distribution shows the
potential is more transmissive to corotating modes ($m \omega > 0$)
than to counterrotating modes, just as in the case of large $a$.}
\end{center}
\vskip -0.5cm
\end{figure}

\begin{figure}[ht]
\begin{center}
\epsfig{file = 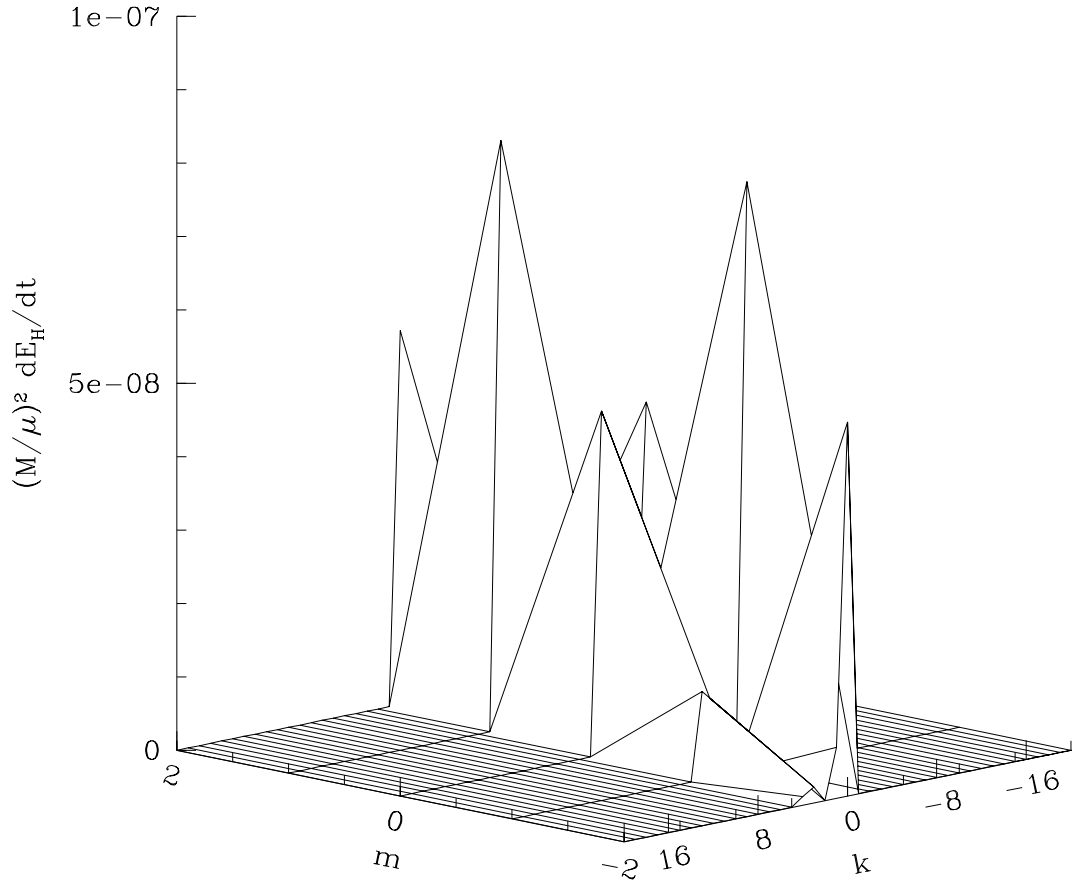, width = 16cm}
\caption{\label{fig:dEHdt_r7_a0.05_l2}
The spectrum of energy radiated down the event horizon for $l = 2$
modes for orbits with $r = 7 M$, $\iota = 60.14^\circ$ about a black
hole with $a = 0.05 M$.  In this case, the distribution is nowhere
negative: the ergosphere for such a slowly rotating hole is
practically irrelevant, and as a consequence we never see any
superradiant scattering.}
\end{center}
\vskip -0.5cm
\end{figure}

\begin{figure}[ht]
\begin{center}
\epsfig{file = 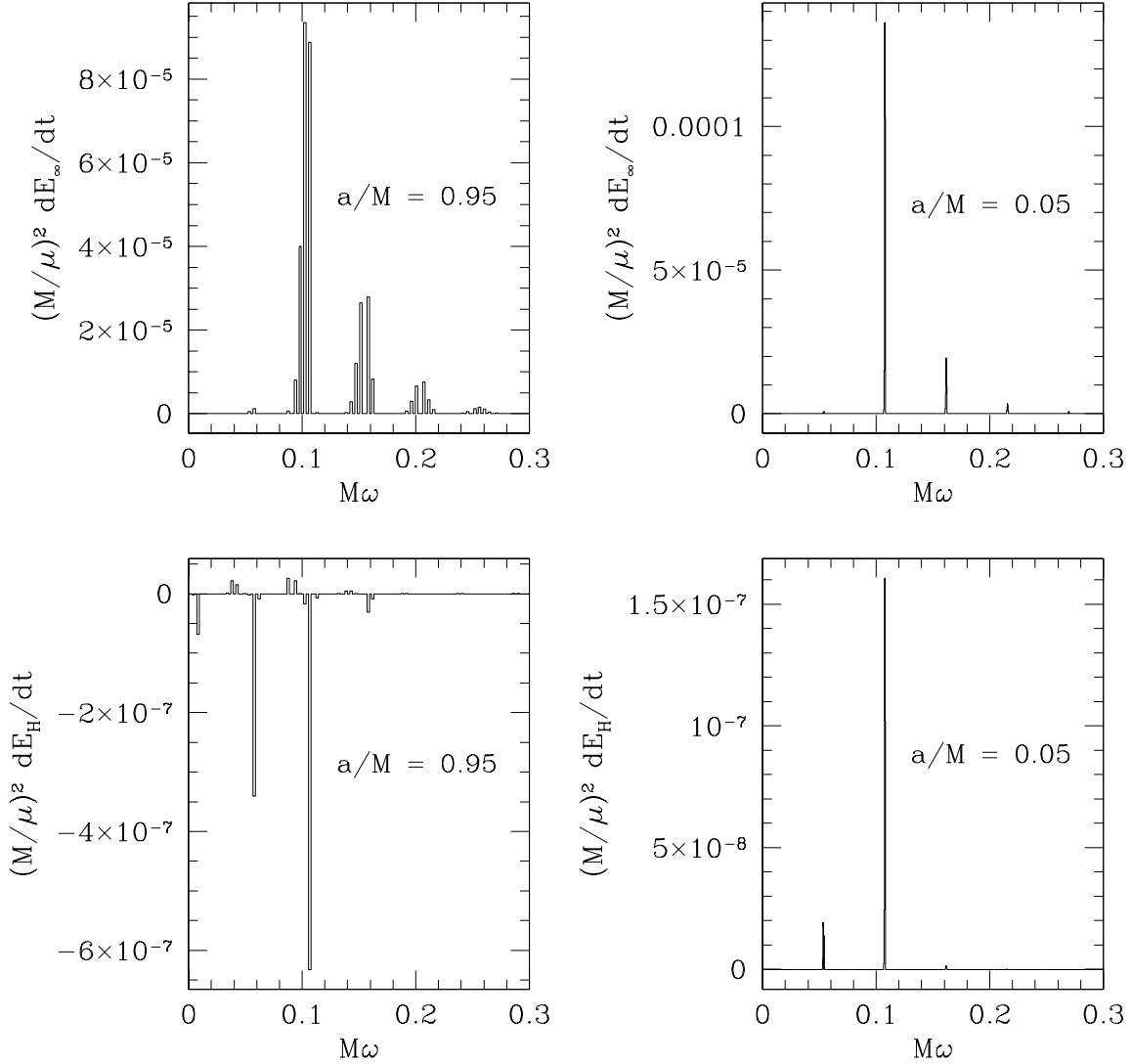, width = 16cm}
\caption{\label{fig:dEsumdt_r7}
Energy spectra $dE/dt$ as a function of frequency $\omega$ for orbits
at $r = 7 M$.  The spectra on the left are for $a = 0.95 M$, $\iota =
62.43^\circ$; the spectra on the right are for $a = 0.05 M$, $\iota =
60.14^\circ$.  The top two spectra are $dE_\infty/dt$, the energy
radiated to infinity; the bottom two are $dE_H/dt$, the energy down
the horizon.  All spectra have been summed over $l$.  In all cases,
the greatest amount of radiation comes out at $\omega \sim 0.1/M$;
this corresponds to $\omega \sim 2\Omega_\phi \sim 2\Omega_\theta$.
For $a = 0.95 M$, the power is smeared over several frequency bins
near each peak.  By contrast, the power is well-confined near a single
bin for $a = 0.05 M$.  In both cases, there is significant power at
several sum and difference harmonics of $\Omega_\phi$ and
$\Omega_\theta$.  For $a = 0.95 M$, these frequencies are different
enough (relative difference $\sim10\%$) that the effect of these
harmonics is quite marked.  For $a = 0.05M$, the frequencies are
practically identical (relative difference $\sim 0.5\%$), so the
harmonics are barely distinguishible from the main peak.}
\end{center}
\vskip -0.5cm
\end{figure}

\begin{figure}[ht]
\begin{center}
\epsfig{file = 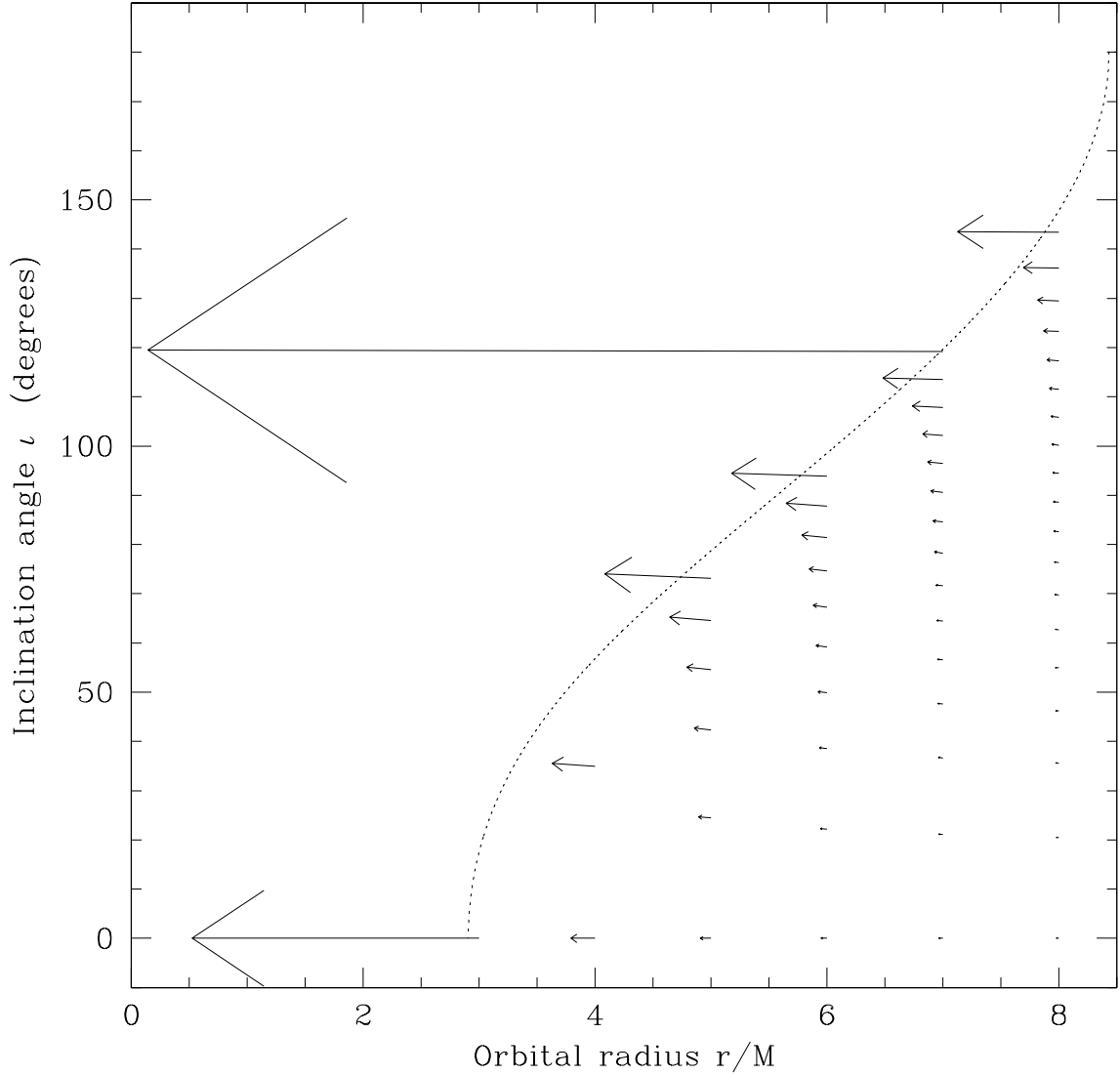, width = 16cm}
\caption{\label{fig:innerseq_a0.8}
The effect of radiation reaction on orbits around a black hole with $a
= 0.8 M$.  The dotted line is the maximum allowed inclination angle;
orbits tilted beyond that line are dynamically unstable and rapidly
plunge into the black hole.  Each arrow is proportional to the vector
$[(M/\mu)\dot r,(M^2/\mu)\dot\iota]$.  Thus, the orientation of the
arrow indicates the direction in phase space to which radiation
reaction drives the orbit; the length of the vector indicates how
strongly it is so driven.  In all cases, the vectors point inwards and
upwards --- radiation reaction drives circular orbits to smaller radii
and larger inclination angles (except when $\iota = 0^\circ$ or
$180^\circ$, in which case the inclination angle does not change).
The rate at which the inclination angle changes is rather slow,
especially compared to the rate at which the radius changes.  Note the
very long vector (indicating extremely rapid orbital evolution) at
$\iota \simeq 120^\circ$, $r = 7M$.  This orbit evolves so quickly
because it happens to lie extremely close to the maximum allowed angle
--- it is barely dynamically stable, so a small push from radiation
reaction has marked effects.}
\end{center}
\vskip -0.5cm
\end{figure}

\begin{figure}[ht]
\begin{center}
\epsfig{file = 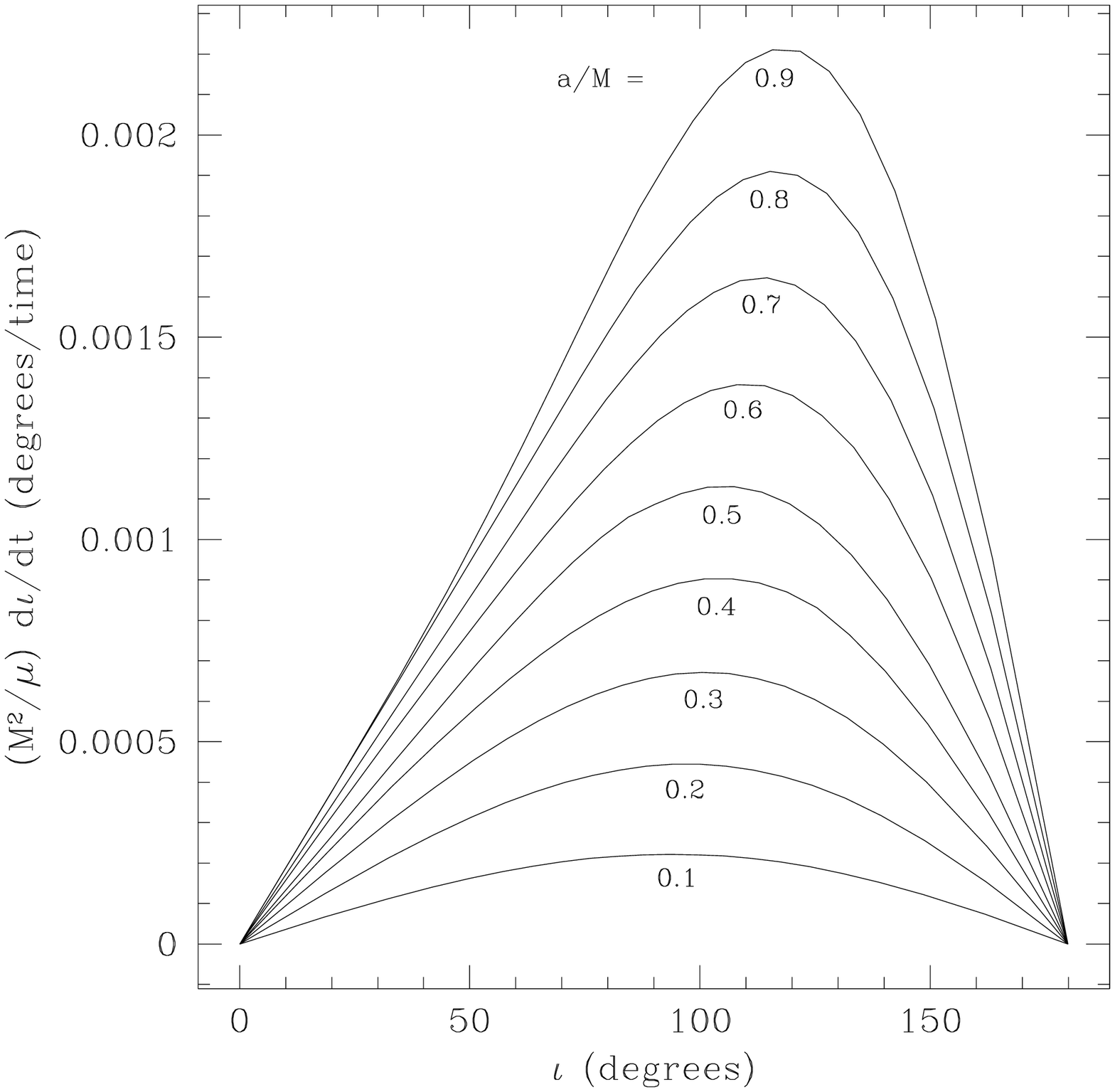, width = 16cm}
\caption{\label{fig:iotadot_vs_iota}
The rate of change of the inclination angle $\dot\iota$ versus
$\iota$, parameterized by black hole spin.  All curves are for orbits
at $r = 10 M$.  For large spin, the maximum rate of change of the
angle occurs for $\iota > 90^\circ$, in contrast to the post-Newtonian
prediction, which yields $\dot\iota\propto\sin\iota$ [cf.\ Eq.\
(\ref{eq:pn_rdotiotadot})].}
\end{center}
\vskip -0.5cm
\end{figure}


\begin{references}
\bibitem{lxlycomment} Shibata's analysis {\cite{shibata93}} only
	includes radiation to infinity.  In this nearly spherically
	symmetric asymptotically-flat region, notions of $L_x$ and
	$L_y$ components to the angular momentum can be made rigorous.
	In the highly non-spherical region near the event horizon,
	such notions cannot be made rigorous.  It would probably not
	be possible to generalize the notion of $L_x$ and $L_y$ to
	radiation carried down the event horizon.
\bibitem{spindownhole} Any further evolution of the system would be
	driven by the spindown of the massive black hole as radiative
	energy losses bled away its rotational energy.  The timescale
	for a such a change would be extremely long, scaling as
	$M^{-3}$.
\bibitem{nooverlaphere} Strictly speaking, the overlap between any
	two waveforms with different parameters presented in this
	paper is {\it zero}.  This is because the waveforms last for
	infinite time and have strictly different frequencies.  It is
	only when the waveforms are computed from a radiation reaction
	sequence that they last for finite time, and overlap with each
	other's frequency domains.
\end{references}
\end{document}